\documentclass{aa}  

\usepackage{graphicx}
\usepackage{rotating}         
\usepackage{txfonts}
\usepackage{subfigure}
\usepackage{subfig}
\usepackage{amssymb}
\usepackage{longtable}
\usepackage{pdflscape}
\usepackage{natbib}
\bibpunct{(}{)}{;}{a}{}{,} 
\usepackage{url}

\def\la{\mathrel{\mathchoice {\vcenter{\offinterlineskip\halign{\hfil
$\displaystyle##$\hfil\cr<\cr\sim\cr}}}
{\vcenter{\offinterlineskip\halign{\hfil$\textstyle##$\hfil\cr
<\cr\sim\cr}}}
{\vcenter{\offinterlineskip\halign{\hfil$\scriptstyle##$\hfil\cr
<\cr\sim\cr}}}
{\vcenter{\offinterlineskip\halign{\hfil$\scriptscriptstyle##$\hfil\cr
<\cr\sim\cr}}}}}
\def\ga{\mathrel{\mathchoice {\vcenter{\offinterlineskip\halign{\hfil
$\displaystyle##$\hfil\cr>\cr\sim\cr}}}
{\vcenter{\offinterlineskip\halign{\hfil$\textstyle##$\hfil\cr
>\cr\sim\cr}}}
{\vcenter{\offinterlineskip\halign{\hfil$\scriptstyle##$\hfil\cr
>\cr\sim\cr}}}
{\vcenter{\offinterlineskip\halign{\hfil$\scriptscriptstyle##$\hfil\cr
>\cr\sim\cr}}}}}
\def\deg{\hbox{$^\circ$}}

\newcommand{\beq}{\begin{equation}}
\newcommand{\eeq}{\end{equation}}
\newcommand{\bdi}{\begin{displaymath}}
\newcommand{\edi}{\end{displaymath}}

\begin{document} 

\titlerunning{High-mass starless and proto-stellar clumps identified by the Hi-GAL survey} 
%

\title{
{\emph{Herschel}}\thanks{\emph{Herschel} is an ESA space observatory with science instruments provided by European-led  
Principal Investigator consortia and with important participation from NASA.}-HIFI  observations of 
H$_2$O, NH$_3$ and N$_2$H$^+$ toward high-mass starless and proto-stellar clumps identified by the Hi-GAL survey
}

\authorrunning {L.~Olmi et.~al.}  

\author{
L. Olmi \inst{\ref{inst1},\ref{inst2}} \and
C.M. Persson \inst{3} \and
C. Codella \inst{1}
}

\institute{
          INAF, Osservatorio Astrofisico di Arcetri, Largo E. Fermi 5,
          I-50125 Firenze, Italy,  \email{olmi.luca@gmail.com} \label{inst1}
\and
          University of Puerto Rico, Rio Piedras Campus, Physics Dept., Box 23343,
          UPR station, San Juan, Puerto Rico, USA  \label{inst2}
\and
         Chalmers University of Technology, Department of Earth and Space Sciences,
         Onsala Space Observatory,  SE-439 92 Onsala, Sweden.
          }

\date{Received Month Date, 2015; accepted Month Date, 2015}

\abstract
{Our present understanding of high-mass star formation still remains very schematic.
In particular, it is not yet clear how much of the difference between low-mass and high-mass 
star formation  occurs during the earliest star formation phases.
}
{
The chemical characteristics of massive cold clumps, and the comparison with
those of their low-mass counterparts, could provide crucial clues about the exact 
role that chemistry plays in differentiating the early phases of low-mass and high-mass
star formation.  Water, in particular, is a unique probe of physical and chemical 
conditions in star-forming regions.
}
{
Using the HIFI instrument of \emph{Herschel} we have observed the 
ortho$-$NH$_3\,(1_0-0_0)$ (572\,GHz),  ortho$-$H$_2$O$\,(1_{10}-1_{01})$ (557\,GHz) and
N$_2$H$^+\,(6-5)$ (559\,GHz) lines toward a sample of high-mass starless and proto-stellar 
clumps selected from the  ``{\it Herschel} Infrared Galactic Plane Survey'' (Hi-GAL).
We compare our results to previous studies of low-mass and high-mass proto-stellar objects.
%
}
{
At least one of the three molecular lines was detected in 4 (out of 35) and 7 (out of 17) objects 
in the $\ell=59\deg$ and $\ell=30\deg$ galactic regions, respectively.  All detected sources
are {\it proto}-stellar.  The water spectra are complex 
and consist of several kinematic components, identified through a
Gaussian decomposition, and in a few sources inverse and regular P-Cygni profiles have been detected.  
All water line profiles of the $\ell=59\deg$ region are dominated by a 
broad Gaussian emission feature, indicating that the bulk of the water emission arises 
in outflows. No such broad emission is detected toward the $\ell=30\deg$ objects.
The ammonia line in some cases also shows line wings and an inverse P-Cygni profile, thus confirming that
NH$_3$ rotational transitions can be used to probe the dynamics of high-mass star forming regions.
Both bolometric and water line luminosity increase with the continuum temperature. 
}
{
The higher water abundance toward the $\ell=59\deg$ sources, 
characterized by the presence of outflows and shocks, supports a scenario where 
the abundance of this molecule is linked to the shocked gas.
Various indicators suggest that the detected sources toward
the $\ell=30\deg$ region are in a somewhat later evolutionary phase compared to the
$\ell=59\deg$ field, although a firm conclusion is limited by the small number of
observed sources. 
We find many similarities with studies carried out 
toward low-mass proto-stellar objects, but there are indications that the level
of infall and turbulence in the high-mass protostars studied here
could be significantly higher.
} 
 


   \maketitle
%

%
%
\begin{table*}[\!htb]
\centering
\caption{Observed  transitions.
}
\begin{tabular} {lccccc}
\hline\hline
\noalign{\smallskip}
Species & Frequency\tablefootmark{a}    & Transition\tablefootmark{b}   & $E_{\mathrm{l}}$      & $E_{\mathrm{u}}$  & ${\rm log}(A_{\rm ul})$ \\
\noalign{\smallskip}
&[GHz]  &       & [K]   & [K]   & [s$^{-1}$]    \\
\noalign{\smallskip}
\hline
\noalign{\smallskip}
%
%
ortho-NH$_3$                            &572.4982   &   $J_{K}=1_{0}$ -- 0$_{0}$    &   0   &  27   &       $-$2.80    \\  
ortho-H$_2$O                            &556.9359   &    $J_{K_a,K_c}=1_{1,0}-1_{0,1}$      &   34  &  61   &$-$2.46   \\  
N$_2$H$^+$                          &558.9665   &   $J =6-5$    &  67   &  94   & $-$1.96                          \\  
%
%
%
%
%
%

\noalign{\smallskip}
\hline

\label{Table:transitions}
\end{tabular}
\tablefoot{
\tablefoottext{a}{Frequencies from {\tt http://www.cv.nrao.edu/php/splat/}}
\tablefoottext{b}{Critical densities are $n_{\rm cr} \ga 10^7 - 10^8$\,cm$^{-3}$ for all three lines.}
}
\end{table*}
%

\section{Introduction}

High-mass stars (OB spectral type, $M \ga 8\,M_\odot$), though much rarer than low-mass
and intermediate-mass stars, are fundamental in the evolution of galaxies.
But, while a reasonably robust evolutionary sequence has been established for the
formation and early evolution of individual,
our present understanding of high-mass star formation still remains very schematic
(\citealp{zinnecker2007}, \citealp{tan2014}).  In particular, some of the main
issues regarding high-mass star formation that still need to be studied are:
{\it (i)} what is the physical and dynamical evolutionary
path of massive starless\footnote{As identified through the absence of Mid-IR emission
and/or the strength of the PACS $70\,\mu$m emission.}
clumps\footnote{In the literature, the term ``core'' or ``fragment'' usually refers
to small-scale objects ($\la 0.1\,$pc or even $\ll 0.1\,$pc), possibly corresponding to a later stage of
fragmentation, while the term ``clump'' is generally used for a somewhat larger ($\ga 0.1-1\,$pc),
unresolved object, possibly composed of several cores. The term clump is thus more appropriate
to refer to the compact objects described in this work.}
toward the proto-stellar and stellar phases? And,
{\it (ii)} how much of the difference between low-mass and high-mass star formation occurs during the
earliest star formation phases?

The high-mass analogues of low-mass Class 0 protostars and starless clumps
(e.g., \citealp{motte2007}, \citealp{bontemps2010}, \citealp{olmi2010a}, \citealp{olmi2013})
constitute a fundamental link between the
global processes that regulate star formation on large scales and the properties of newly formed
(proto)stars and clusters.  Therefore, to better define the evolutionary path from
molecular clumps to proto-stars it is crucial to determine the physical and dynamical
(and, possibly, chemical) properties of these early phases of high-mass star formation.

The formation of both low-mass and high-mass stars from the initial dense clumps
is accompanied by large changes
in the temperature and density structure in their surrounding envelopes.
These changes in turn open up different chemical routes.
Current observational evidence indicates that significant amounts of volatile
molecules, for instance CO, H$_2$O and NH$_3$,
reside on the surfaces of dust grains in dense clumps. These
surface molecules can be formed either in the gas phase via ion-molecule reactions
followed by accretion onto the dust particles, or via direct production on the
grain surfaces.
Different physical conditions, for example during the early formation phases
of low-mass and high-mass stars, could thus determine which of these
different chemical routes to molecule formation are active and/or dominant.

\begin{landscape}    
  
%
%
%
\begin{table*}    
\centering                      
\caption{ Selected observational and physical properties of the source sample: $1\sigma$ noise level,  
SSB continuum intensity ($T_\mathrm{c}$), beam-filling factor ($\eta_{\rm bf}$), 
evolutionary flag, molecular hydrogen volume density ($n(\mathrm{H_2})$), 
and column density ($N(\mathrm{H_2})$).
Luminosity ($L$), mass ($M$), dust temperature $T_\mathrm{d}$, source diameter 
(Full Width at Half Maximum, or FWHM)  and distance ($d$) 
are all derived from the Hi-GAL observations.
Sources with their name in boldface have at least one detection and thus they could be assigned 
a $V_{\rm lsr}$.
}
\hspace*{-7cm}
\begin{tabular}{lcccccccccccccc} 
\hline\hline
\noalign{\smallskip}
Source  & Gal. Long.  & Gal. Lat  & $L$ & $M$ & $T_\mathrm{d}$  & Source FWHM   & $d$  & $T_\mathrm{c}$\tablefootmark{a}  & $1\sigma$\tablefootmark{b}  & $V_\mathrm{lsr}$\tablefootmark{c}  & $\eta_{\rm bf}$\tablefootmark{d} &   $n(\mathrm{H_2})$\tablefootmark{e} & $N(\mathrm{H_2})$\tablefootmark{e} & Evol.Flag\tablefootmark{f} \\ 
&  [deg]  &  [deg]   & [L$_\odot$]  & [M$_\odot$]   & [K]  & [$\arcsec$]  & [kpc]  & [mK]   & [mK]  & [km s$^{-1}$] &  & [$\times 10^4\,$cm$^{-3}$]  & [$\times 10^{22}\,$cm$^{-2}$]  & \\
\noalign{\smallskip}
\hline
\noalign{\smallskip} \noalign{\smallskip} 
%
l59-251  &  57.984803  &  0.0927620 &     27.2 &    165.0 &  10.1 &   17.3 &      7.0 &    0 &   9  &    \ldots   &   0.2  &   2.7  &   4.82  &      1  \\
l59-139  &  58.048693  &  0.1889100 &    117.5 &   2118.0 &   9.2 &   27.8 &      9.1 &    0 &   8  &    \ldots   &   0.4  &   3.7  &   14.0  &      1  \\
l59-376  &  58.171557  &  0.7804730 &      7.7 &    118.0 &   8.9 &   25.3 &      2.6 &    0 &   8  &    \ldots   &   0.3  &   12.0 &   12.0  &      1  \\
l59-199  &  58.256292  &  0.1431680 &     67.5 &    323.0 &  11.5 &   31.5 &      7.2 &    0 &   8  &    \ldots   &   0.4  &   0.81 &   2.7   &      1  \\
l59-203  &  58.464407  &  0.2996740 &     11.5 &     81.0 &  10.8 &   20.0 &      3.5 &    0 &   8  &    \ldots   &   0.2  &   6.5  &   6.9   &      1  \\
l59-129  &  58.464903  &  0.4929190 &  \ldots &  \ldots &  12.6 &   17.9 &  \ldots &    0 &   8  &    \ldots   &   0.2                 &   \ldots   &    \ldots   &      1  \\
l59-185  &  58.512225  &  0.5060260 &      5.2 &    132.0 &   8.3 &   25.3 &      2.8 &    0 &   8  &    \ldots   &   0.3  &   11.0 &   11.0  &      1  \\
l59-102  &  58.517332  &  0.4718170 &      8.7 &    100.0 &   9.2 &   25.3 &      2.8 &    0 &   9  &    \ldots   &   0.3  &   \ldots   &   \ldots    &      1  \\
l59-78  &  58.558761  &  0.3525380 &     33.7 &   3296.0 &   7.0 &   25.3 &      6.2 &    0 &   9  &    \ldots   &   0.3  &    25.0 &   58.0  &      1  \\
l59-314  &  58.569252  &  0.0207140 &     27.6 &    100.0 &  11.1 &   24.8 &      5.7 &    0 &  10  &    \ldots   &   0.3  &   0.53 &   1.1  &      1  \\
l59-391  &  58.606871  &  0.7818820 &     25.0 &    109.0 &  10.7 &   25.3 &      4.7 &    0 &   9  &    \ldots   &   0.4  &   \ldots   &   \ldots   &      1  \\
l59-80  &  58.646263  &  0.5462850 &     77.4 &   1027.0 &   9.1 &   25.3 &      6.1 &   24 &  10  &    \ldots   &   0.3  &    8.0  &   19.0 &      1  \\
l59-345  &  58.693246  &  0.7685080 &     25.7 &    149.0 &  10.3 &   25.3 &      4.9 &    0 &   8  &    \ldots   &   0.3  &   4.5  &   10.0 &      1  \\
l59-136  &  58.754808  &  0.6835960 &     23.1 &    446.0 &   8.6 &   25.3 &      5.5 &    0 &   8  &    \ldots   &   0.3  &   4.8  &   9.9  &      1  \\
l59-231  &  58.755031  &  0.7204550 &     88.4 &    137.0 &  13.8 &   33.8 &      5.6 &    0 &   9  &    \ldots   &   0.5  &   0.58 &   1.7  &      1  \\
l59-293  &  58.810992  &  0.6553700 &     14.5 &      3.0 &  19.6 &   17.8 &      5.2 &    0 &   9  &    \ldots   &   0.2  &   0.11 &   0.15 &      0  \\
l59-667  &  58.865456  &  -0.212801 &    653.6 &    799.0 &  12.5 &   17.0 &     13.7 &    0 &   8  &    \ldots   &   0.2  &   1.8  &   6.3  &      2  \\
l59-230  &  59.051623  &  -0.229814 &     23.0 &     79.0 &  12.1 &   27.0 &      2.6 &    0 &   8  &    \ldots   &   0.4  &   6.8  &   7.1  &      1  \\
l59-57  &  59.151275  &  -0.268629 &    185.7 &    281.0 &  13.9 &   24.2 &      6.2 &   22 &   9  &    \ldots   &   0.3  &    2.4  &   5.4  &      1  \\
l59-191  &  59.208976  &  0.1322060 &     62.8 &    239.0 &  11.0 &   25.3 &      6.7 &    0 &   8  &    \ldots   &   0.4  &   1.1  &   1.0  &      1  \\
l59-263  &  59.212680  &  0.1586920 &     14.0 &    170.0 &   9.2 &   25.3 &      4.0 &    0 &   9  &    \ldots   &   0.5  &   0.66 &   2.1  &      1  \\
l59-150  &  59.300967  &  0.2429840 &     47.3 &    210.0 &  10.7 &   25.3 &      5.9 &    0 &   8  &    \ldots   &   0.2  &   13.0 &   20.0 &      1  \\
l59-104  &  59.304277  &  0.2083220 &     17.0 &      2.0 &  23.3 &   25.3 &      2.7 &    0 &   9  &    \ldots   &   0.3  &   0.18 &   0.18 &      1  \\
l59-371  &  59.309467  &  -0.397915 &     15.8 &    227.0 &   9.6 &   19.3 &      6.2 &    0 &   8  &    \ldots   &   0.2  &   3.8  &   6.8  &      1  \\
l59-31  &  59.541533  &  -0.101984 &     20.7 &      5.0 &  19.4 &   25.3 &      2.3 &    0 &   9  &    \ldots   &   0.2  &    60.0 &   30.0 &      1  \\
{\bf l59-441}  &  59.633104  &  -0.191683 &   2259.6 &    294.0 &  21.6 &   15.3 &      2.2 &  110 &   8  &   27.8  &   0.1  &  230.0 &   110.0  &      2  \\
{\bf l59-445}  &  59.635570  &  -0.186242 &    787.9 &     82.0 &  20.9 &   23.0 &      2.3 &   75 &   8  &   27.4  &   0.3  &   16.0 &   13.0  &      2  \\
l59-282  &  59.707751  &  0.5772220 &     30.1 &    112.0 &  10.9 &   25.3 &      4.3 &    0 &   8  &    \ldots   &   0.1        &   \ldots   &   \ldots  &      1  \\
l59-409  &  59.737176  &  0.5263750 &     60.7 &    123.0 &  13.2 &   27.8 &      6.2 &    0 &   8  &    \ldots   &   0.4  &   0.7  &   1.8  &      1  \\
{\bf l59-444}  &  59.789451  &  0.6307310 &  \ldots &  \ldots &  21.8 &    9.0 &  \ldots &   37 &   8  &   34.3  &   0.1           &   \ldots   &   \ldots   &      2  \\
{\bf l59-442}  &  59.832526  &  0.6718730 &  \ldots &  \ldots &  25.3 &   13.9 &  \ldots &   28 &   8  &   34.2  &   0.1           &   \ldots   &   \ldots   &      2  \\
l59-359  &  59.862726  &  -0.210802 &     55.3 &    429.0 &   9.8 &   25.3 &      6.4 &    0 &  11  &    \ldots   &   0.5  &   5.1  &   18.0 &      1  \\
l59-101  &  59.880982  &  -0.102392 &     85.9 &   2814.0 &   8.4 &   25.9 &      6.6 &    0 &   8  &    \ldots   &   0.3  &   16.0 &   41.0 &      1  \\
l59-69  &  59.891888  &  -0.112074 &     27.8 &     92.0 &  12.2 &   36.2 &      1.9 &    0 &   8  &    \ldots   &   0.5  &    7.5  &   8.0  &      1  \\
l59-198  &  59.946964  &  0.0610780 &      5.9 &      5.0 &  15.3 &   13.5 &      1.9 &   32 &   8  &    \ldots   &   0.1  &   8.3  &   3.2  &      1  \\
\noalign{\smallskip} \noalign{\smallskip}
\hline
\label{Table:sources}
\end{tabular}
\tablefoot{
\tablefoottext{a}{SSB continuum intensity at 557~GHz. }
\tablefoottext{b}{Rms noise  at a resolution of 1~km~s$^{-1}$ with subtracted baseline.}
\tablefoottext{c}{Velocity is specified only for sources with at least a detection.}
\tablefoottext{d}{Beam-filling factor, see Section~\ref{sec:radex}. }
\tablefoottext{e}{See Section~\ref{sec:radex} for a description of these parameters.}
\tablefoottext{f}{0: no detection at 70$\,\mu$m, Larson's criterion \citep{larson1981} not fulfilled (starless, unbound).
1: no detection at 70$\,\mu$m, Larson's criterium fulfilled (pre-stellar). 2: detection
at 70$\,\mu$m (proto-stellar).}
}
\end{table*}

\addtocounter{table}{-1}

\begin{table*}
\centering
\caption[]{
continued
}
\hspace*{-7cm}
\begin{tabular}{lcccccccccccccc} 
\hline\hline
\noalign{\smallskip}
Source  & Gal. Long.  & Gal. Lat  & $L$ & $M$ & $T_\mathrm{d}$  & Source FWHM   & $d$  & $T_\mathrm{c}$\tablefootmark{a}  & $1\sigma$\tablefootmark{b}  & $V_\mathrm{lsr}$\tablefootmark{c}  & $\eta_{\rm bf}$\tablefootmark{d} &   $n(\mathrm{H_2})$\tablefootmark{e} & $N(\mathrm{H_2})$\tablefootmark{e} & Evol.Flag\tablefootmark{f} \\
&  [deg]  &  [deg]   & [L$_\odot$]  & [M$_\odot$]   & [K]  & [$\arcsec$]  & [kpc]  & [mK]   & [mK]  & [km s$^{-1}$] &  & [$\times 10^4\,$cm$^{-3}$]  & [$\times 10^{22}\,$cm$^{-2}$]  & \\
\noalign{\smallskip}
\hline
\noalign{\smallskip} \noalign{\smallskip}
{\bf l30-376}  &  29.002507  &  0.0689530 &   9328.5 &   1072.0 &  19.8 &   21.8 &      8.9 &   23 &   8  &   70.5  &   0.3  &   4.1  &   12.0  &      2  \\
{\bf l30-512}  &  29.235020  &  -0.046735 &   1312.0 &   1860.0 &  13.8 &   12.6 &      9.9 &   63 &   8  &   61.4  &   0.1  &   27.0 &   51.0  &      2  \\
{\bf l30-313}  &  29.862299  &  -0.043707 &  53790.2 &   1866.0 &  22.9 &   12.2 &      8.5 &   68 &   8  &  101.5  &   0.1  &   47.0 &   73.0  &      2  \\
l30-110  &  29.868288  &  -0.007673 &    313.8 &   1311.0 &  10.6 &   14.7 &      8.5 &   27 &   8  &    \ldots   &   0.3        &   10.0 &   20.0  &      1  \\
{\bf l30-327}  &  30.009612  &  -0.273200 &  12371.5 &    964.0 &  23.5 &    9.2 &      6.7 &   98 &   8  &  102.6  &   0.1  &  120.0 &   110   &      2  \\
l30-281  &  30.121077  &  -0.548844 &    435.5 &    628.0 &  12.8 &   25.3 &      8.1 &   44 &   8  &    \ldots   &   0.3        &   2.1  &   6.4   &      1  \\
l30-375  &  30.197099  &  0.3101250 &  15852.5 &   2921.0 &  17.7 &   12.7 &     14.2 &    0 &   9  &    \ldots   &   0.1        &   14.0 &   39.0  &      2  \\
l30-432  &  30.447046  &  -0.024628 &   9848.3 &    271.0 &  26.7 &   28.0 &     11.6 &   44 &   8  &    \ldots   &   0.2        &   6.5  &   22.0  &      2  \\
l30-103  &  30.575773  &  0.1188560 &    186.4 &    878.0 &  10.6 &   25.3 &      5.7 &   44 &   8  &   55.6  &   0.3  &   8.4  &   18.0  &      1  \\
l30-403  &  30.627182  &  -0.063627 &   1739.5 &    711.0 &  15.8 &   28.0 &      5.6 &   38 &   8  &    \ldots   &   0.4        &   5.3  &   12.0  &      2  \\
l30-73  &  30.658809  &  0.0450200 &    143.7 &    962.0 &  11.1 &   17.7 &      5.6 &   34 &   8  &    \ldots   &   0.2         &   28.0 &   42.0  &      1  \\
l30-111  &  30.661139  &  0.1419250 &    392.9 &    249.0 &  14.8 &   25.3 &      5.6 &   35 &   10  &    \ldots   &   0.4        &   7.9  &   20.0  &      1  \\
l30-132  &  30.675761  &  -0.166004 &   1223.9 &   1559.0 &  14.3 &   27.4 &      9.0 &    0 &   8  &    \ldots   &   0.4        &   3.0  &   11.0  &      1  \\
{\bf l30-42}  &  30.693536  &  -0.148593 &   1082.2 &    723.0 &  16.7 &   17.4 &      5.7 &   50 &   9  &   90.4  &   0.2   &   22.0 &   32.0  &   2  \\
{\bf l30-304}  &  30.703241  &  -0.068204 &  49112.4 &   3612.0 &  22.5 &   20.4 &      6.0 &  350 &  12 &   90.7  &   0.2        &   57.0 &   100.0 &      2  \\
{\bf l30-43}\tablefootmark{g}  &  30.822869  &  0.1343020 &   2405.6 &  16437.0 &  10.2 &   32.6 &      8.8 &   50 &   8  &  108.4  &   0.5   &   20.0 &   86.0  &      2  \\
l30-230  &  30.846821  &  0.3426510 &    186.5 &    425.0 &  12.0 &   25.3 &      6.2 &    0 &   8  &    \ldots   &   0.4        &   5.7  &   21.0  &      1  \\
\noalign{\smallskip} \noalign{\smallskip}
\hline 
\end{tabular}
\tablefoot{
\tablefoottext{g}{Tentative detection (see Section~\ref{sec:parspec}).}
} 
\end{table*}
%

\end{landscape}    

The chemical characteristics of massive cold clumps, and the comparison with
those of their low-mass counterparts, are only starting to emerge (\citealp{russeil2010},
\citealp{marseille2010}). While governed by the same processes as lower mass objects,
the much larger luminosity of massive stars during all their evolutionary phases will
greatly affect the outcome.
Therefore, the comparison of chemical models (e.g., with different weights of grain
vs. gas-chemistry) with aimed spectral line observations will provide crucial
clues about the exact role that chemistry
plays in differentiating the early phases of high-mass star formation from those of low-mass star formation.

The $(1_0-0_0)$ ground-state rotational transition of {\it o-}NH$_3$  (e.g., \citealp{ho1983})
has an upper state energy of 27\,K (see Table~\ref{Table:transitions}), comparable to the rotation-inversion transitions
at cm wavelengths.
The critical density is, however, four orders of magnitude higher and is thus a better probe of
the physical conditions of the dense high-mass starless clumps detected by the Hi-GAL survey.
Moreover, its {\it ortho} symmetry form makes it
a complementary tool to the commonly used  rotation-inversion transitions
of para$-$NH$_3$ observable from the ground.
Both NH$_3$ and N$_2$H$^+$ are well known tracers of low-mass pre- and proto-stellar clumps.

Water is another important component in the chemistry of regions of star formation.
In cold gas, water is mainly produced on dust grains, and released to the gas phase when the
temperature increases ($\ga 100$\,K). Water is thus not expected to
to be found with high abundances in cold, starless clumps
(e.g., \citealp{bergin2007}, \citealp{caselli2012}, \citealp{wirstrom2014}). In fact,
recent results from the Herschel Key Program ``Water in star-forming regions with {\it Herschel}''
(WISH; \citealp{vandishoeck2011}) have shown that during the quiescent phases of star formation,
the water abundance is very low. The embedded stage of star formation, however, shows very
broad and complex line profiles consisting of many dynamical components
(\citealp{kristensen2011}, \citealp{kristensen2012}).
The WISH team also found, comparing typical low-, intermediate-, and high-mass young stellar objects, that
their spectra were remarkably similar, characterized by broad, complex line profiles
and consisting of multiple components. Our observations mainly consists of high-mass clumps
in various evolutionary stages, and thus we will also try to identify possible evolutionary differences.

The outline of the paper is as follows.
In Section~\ref{sec:obs} we discuss how our sources were selected and observed with {\it Herschel}.
In Section~\ref{section: results} we describe how the main kinematical and physical parameters of the
detected sources were derived. The gas kinematics is further discussed in Section~\ref{section: discussion}
and we draw our conclusions in Section~\ref{section:conclusions}.

\section{Observations and data reduction}
\label{sec:obs}

\subsection{Source selection}

We have selected our targets from the preliminary source catalog obtained
toward the two Galactic fields, at $\ell=30\deg$ and $\ell=59\deg$,
observed for the ``{\it Herschel} Infrared Galactic Plane Survey'' (Hi-GAL,
\citealp{molinari2010PASP}) during the Science Demonstration Phase of {\it Herschel}
(\citealp{elia2010}).
The Hi-GAL (and Hi-GAL360) Project used SPIRE and PACS in parallel mode to carry out an
unbiased imaging survey of the Galactic Plane, uniformly sampling a $2\deg$-wide strip,
with $-60\deg < \ell < 60\deg$,
in the 70, 170, 250, 350 and 500\,$\mu$m photometric bands.
The source-identification, flux-extraction and spectral energy density-building in these two Hi-GAL maps
were carried out by adapting to the {\it Herschel} SPIRE/PACS wavebands the
methods and techniques described by \citet{netterfield2009} and \citet{olmi2009}.
From the preliminary catalogs of the $\ell=30\deg$ and $\ell=59\deg$ fields,
we have extracted both high-mass starless clumps 
and high-mass proto-stellar objects.

Candidate high-mass starless clumps have been selected by imposing that
$M_{\rm clump} > 100 \,M_\odot$ ($M_{\rm clump} > 200\, M_\odot$) and
$L_{\rm fir} < 100\, L_\odot$ ($L_{\rm fir} < 500\, L_\odot$) for the $\ell=59\deg$
($\ell=30\deg$) field. Moreover, we have checked that no compact MIPS
(``Multiband Imaging Photometer for Spitzer'', \citealp{rieke2004})
$24\,\mu$m emission were present at the position of the candidate high-mass starless clump.
Candidate high-mass proto-stellar objects do have instead a compact MIPS $24\,\mu$m counterpart.
Finally, the sources have been visually inspected to reject ambiguous detections.
This procedure resulted in a total of $81$ clumps, including the
four candidate high-mass starless clumps already observed with the VLA and the Effelsberg 100-m telescope
in the NH$_3(1,1)$ and (2,2) inversion lines by \citet{olmi2010a}.
Out of the initial sample, 52 sources were finally observed with {\it Herschel}: 17 objects in
the $\ell=30\deg$ region and 35 in the $\ell=59\deg$ region. The observed sources
and their basic physical parameters are listed in Table~\ref{Table:sources}.    
Sources with a detection in at least one of the three observed lines have been
highlighted in boldface.


%
%
%
\begin{figure}
\centering
%
\hspace{-0.55cm}
\includegraphics[width=9.0cm,angle=0]{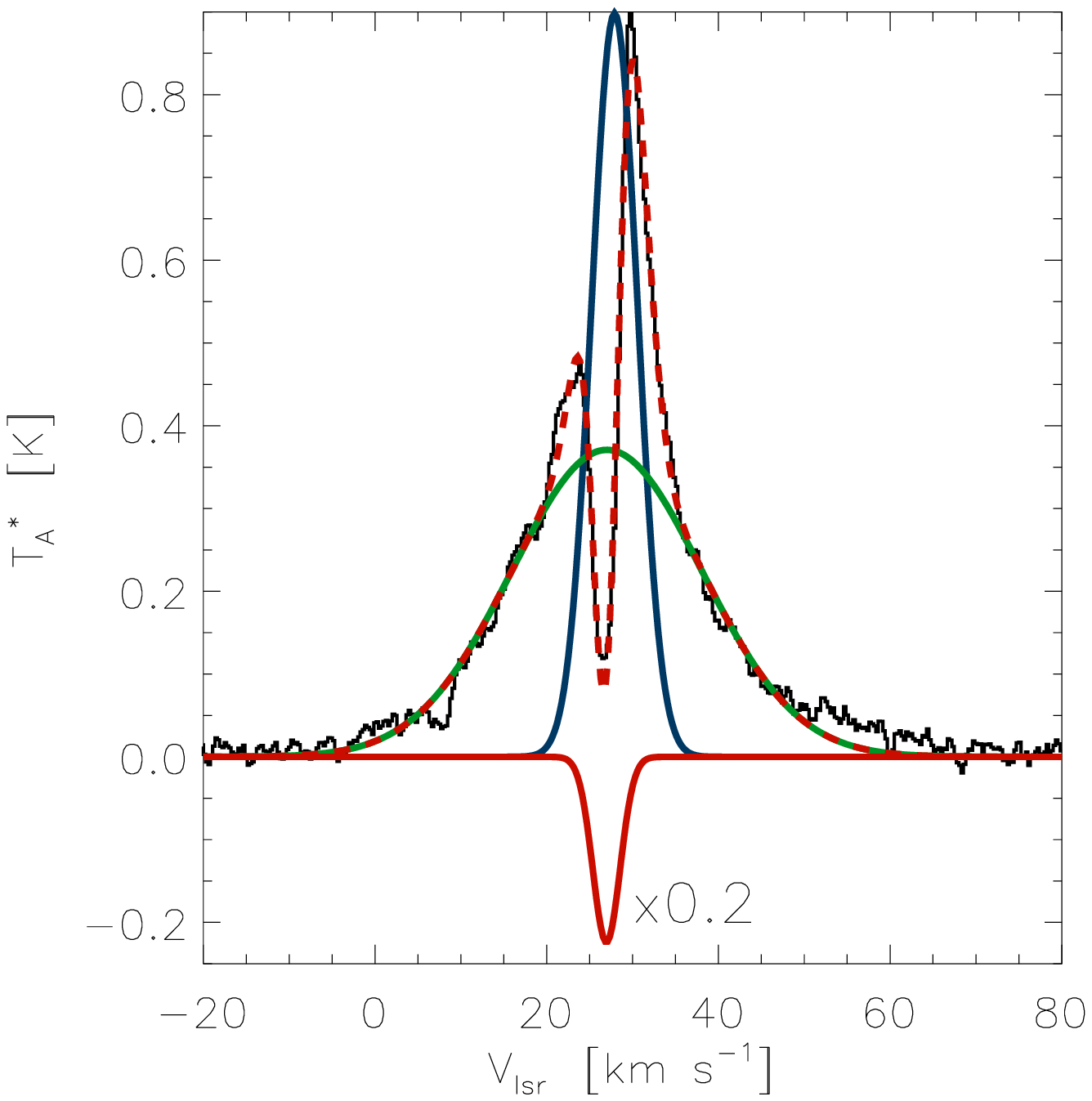}          
\hspace{0.0cm}
\includegraphics[width=7.8cm,angle=0]{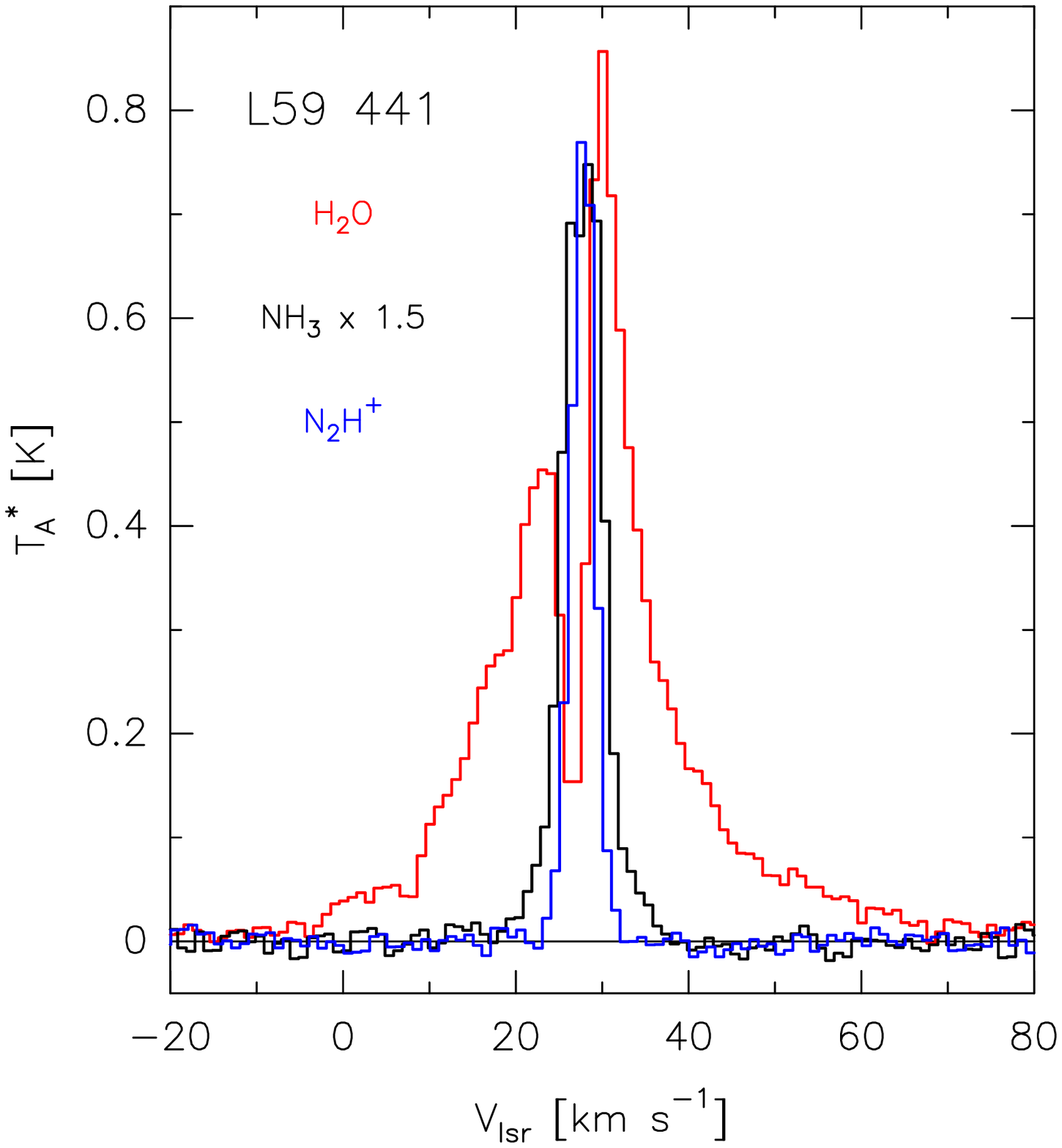}
\caption{
{\it Top panel}.
Gaussian decomposition of the {\it o}-H$_2$O$(1_{10}-1_{01})$
spectrum towards the source l59-441.
The individual Gaussian components are shown as solid color lines
overlaid on the (continuum-subtracted) spectrum. The overall Gaussian fit is shown as a red
dashed line, and the residual deviation at a velocity of $\simeq 55$\,km\,s$^{-1}$ is visible.
The intensity of the component in absorption has been multiplied by a 0.2 factor.
{\it Bottom panel}. Overlay of the {\it o}-H$_2$O$\,(1_{10}-1_{01})$, {\it o}-NH$_3\,(1_0-0_0)$
and N$_2$H$^+\,(6-5)$ lines towards the source l59-441.
}
\label{fig:l59_441}
\end{figure}
%

\subsection{{\it Herschel} observations}

Using the HIFI instrument (\citealp{degraauw2010}) of {\it Herschel} (\citealp{pilbratt2010})
we have observed the {\it o}-NH$_3\,(1_0-0_0)$ line at 572\,GHz in the upper side band of band 1b, and
simultaneously in the lower side band both the {\it o}-H$_2$O\,$(1_{10}-1_{01})$ line at 557\,GHz,  and
the N$_2$H$^+\,(6-5)$ line at 559\,GHz (see Table~\ref{Table:transitions}). We used the wide band spectrometer 
with a bandwidth of 4\,GHz and an effective spectral resolution of 1.1\,MHz ($\Delta V =0.27$\,km\,s$^{-1}$).
The half-power beam width and the main-beam efficiency at 572\,GHz are 36\,\arcsec and 0.62, respectively
(Mueller et al., 2014\footnote{\tt http://herschel.esac.esa.int/twiki/pub/Public/ \\ /HifiCalibrationWeb/HifiBeamReleaseNote\_Sep2014.pdf}). ”
The total calibration uncertainties are $\simeq 9\,$\%
for band 1\footnote{\tt http://herschel.esac.esa.int/Docs/HIFI/html/ch5.html}
(\citealp{roelfsema2012}).
We have  used the dual beam switching (DBS) observing mode.  The
reference beams  in these observations were located within 3\arcmin~on either side of the source.
 The observations were performed during the {\it Herschel} Open Time (OT2) on April 19 and 20, 2013, 
and the corresponding observation identifier (OBSID) range is [1342270449, 1342270604].  

The data were processed using the standard \emph{Herschel} Interactive Processing Environment\footnote{
{\tt http://herschel.esac.esa.int/HIPE\_download.shtml}},
version 11.1,  up to level 2, providing fully calibrated  double side band  spectra    in
the $T_\mathrm{A}^*$  antenna temperature   intensity scale where the lines are
calibrated on a single side band (SSB) scale. Because HIFI is
intrinsically a double side band  instrument, the continuum has to be
divided by two to be properly scaled.
%
%
The data quality is excellent with very low intensity ripples in most cases,
typically below a few percent of the double side band continuum.
However, in some cases the baselines are bad enough that the large rms ($\ga 20-30\,$mK) 
does not allow a reliable estimate of the continuum temperature to be obtained (see Section~\ref{sec:corr}).
Therefore, for these sources we set the continuum temperature to 0.
The FITS  files were exported to the spectral line analysis software packages
{\tt xs}\footnote{ {\tt http://www.chalmers.se/rss/oso-en/observations/\\data-reduction-software}}
and CLASS\footnote{CLASS is part of the GILDAS software package developed by IRAM.}.
Both polarisations and all three LO-settings were included in the averaged noise-weighted  
spectra. The resulting averages were convolved to a resolution of 1~km\,s$^{-1}$
in sources with a low signal-to-noise ratio.  
All HIFI spectra, in $T_\mathrm{A}^*$ units, are shown in Appendices~\ref{sect:app1} ($\ell=30\deg$ region) 
and \ref{sect:app2} ($\ell=59\deg$ region).

\section{Results}
\label{section: results}

\subsection{Distance determination}
\label{sec:dist}

Assigning a distance to sources detected with a photometer is a
crucial step in giving physical significance to all information
extracted from the Hi-GAL data. While reliable distance estimates are
available for a limited  number of known objects (e.g., H{\sc ii} regions,
see \citealp{russeil2003}, and masers, see e.g. \citealp{green2011}),
this information does not exist for the majority of Hi-GAL sources.
We therefore adopted the scheme presented by \citet{russeil2011} to assign
kinematic distances to large lists of sources. A $^{12}$CO (or
$^{13}$CO) spectrum (e.g., from the BU-FCRAO Galactic Ring Survey,
or GRS, \citealp{jackson2006}) is extracted in the line of sight of every
Hi-GAL source. Then, the velocity, $V_\mathrm{lsr}$, of the brightest spectral
component is assigned to that specific source thereby allowing calculation of a
kinematical distance.  By using extinction maps (derived from the {\it 2MASS}
point source catalog, see e.g. \citealp{schneider2011})
and a catalog of sources with known distances, such as H{\sc ii}
regions, masers, and others, the distance ambiguity is resolved and a
recommendation given. In this way it is possible to produce a
``distance map'' having the same pixel size of the CO cube as used to
extract the spectra for every target, where the value of the pixel is
the assigned distance of the Hi-GAL source(s) falling in that pixel.
The typical error on the kinematic distance\footnote{This estimate assumes
that the distance ambiguity has been correctly solved.} (column 8 in Table~\ref{Table:sources})
is estimated to vary between $\sim 0.6-0.9\,$kpc in the range of longitude
$\ell \sim 30^{\circ} - 60^{\circ}$.

\subsection{Profile decomposition}
\label{sec:parspec}


The observed line profiles, and in particular those of the water spectral lines, show complex
shapes, and likely represent  contributions from proto-stellar envelopes,
molecular outflows, foreground clouds as well as infall motions. These contributions,
both in emission and in absorption, are disentangled and parameterized by fitting
multiple Gaussians to the line profiles. We show an example in Fig.~\ref{fig:l59_441}
and list all the results in Table~\ref{Table:gaussian}.
%
%
The Gaussian fits have been performed using up to four
velocity components and all line parameters were allowed to vary.
A similar decomposition of the H$_2$O$(1_{10}-1_{01})$ (557\,GHz) line has
also been used by \citet{kristensen2012} and \citet{herpin2012}.

%
%
\begin{figure}
\centering
\includegraphics[width=9.3cm,angle=0]{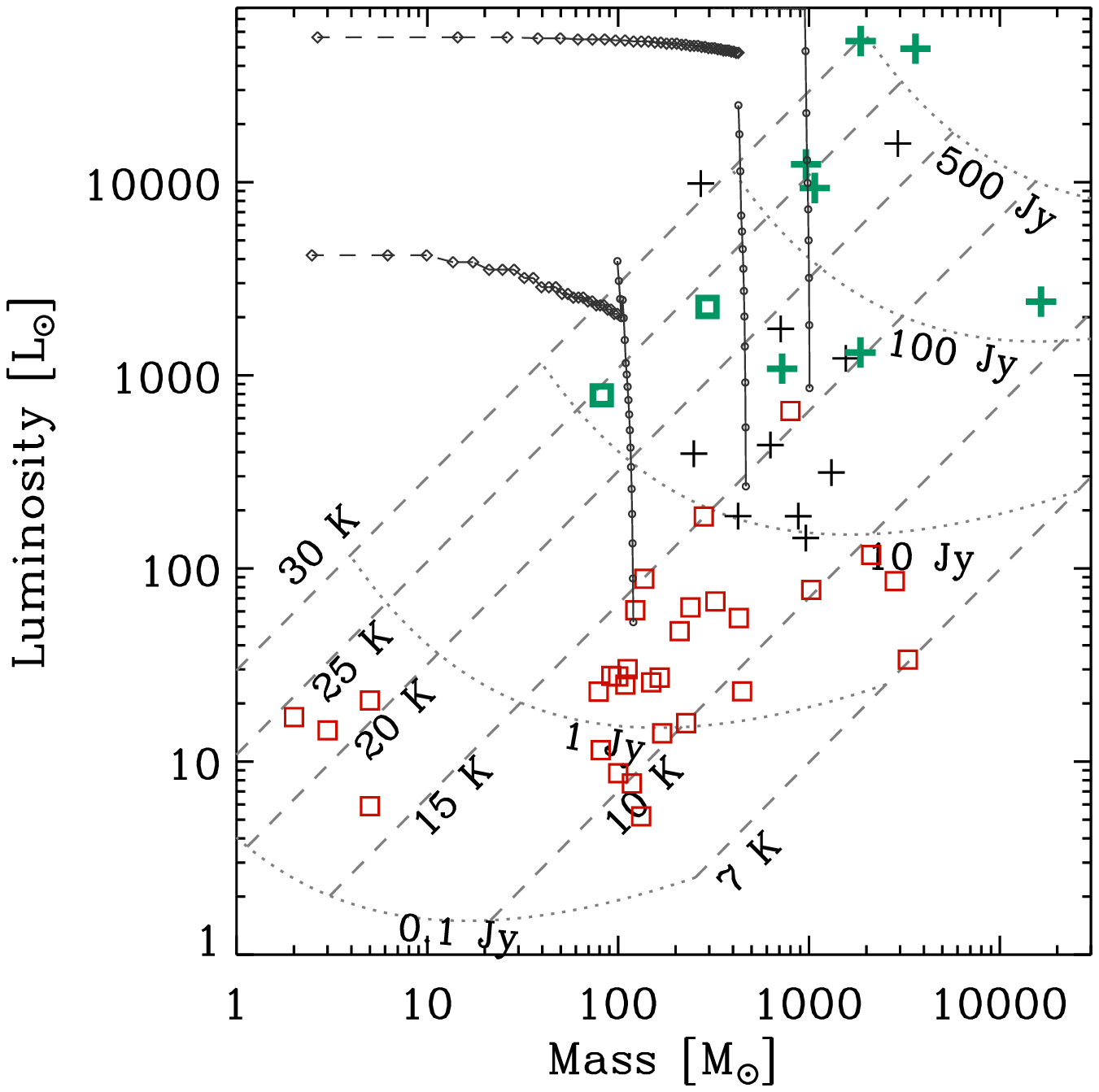}
\caption{
Clump bolometric luminosity vs. mass, as derived from the Hi-GAL observations for all sources
where a distance could be determined. Sources in the $\ell=30\deg$ and $\ell=59\deg$ regions
are represented by the (black) ``+'' sign and (red) open squares, respectively.
Sources with a detection in at least one of the three observed spectral lines are shown
with thicker (green) symbols.
The dashed lines are loci at constant temperature.
Roughly orthogonal to these are loci (dotted lines) of constant 250$\,\mu$m
flux density, ranging from 0.1 to 500\,Jy, assuming a modified blackbody spectral energy distribution
with $\beta = 1.5$ and a fixed distance of $\sim 6\,$kpc, equal to the median of the distances
of the observed sources.
The black dots and the lines joining them represent the evolutionary tracks of clumps
with starting envelope mass equal to 120, 470 and $10^3\,$M$_\odot$ (see Section~\ref{sec:evolution}).
}
\label{fig:LvsM}
\end{figure}

%
%
\begin{figure}
\centering
\includegraphics[width=8.0cm,angle=0]{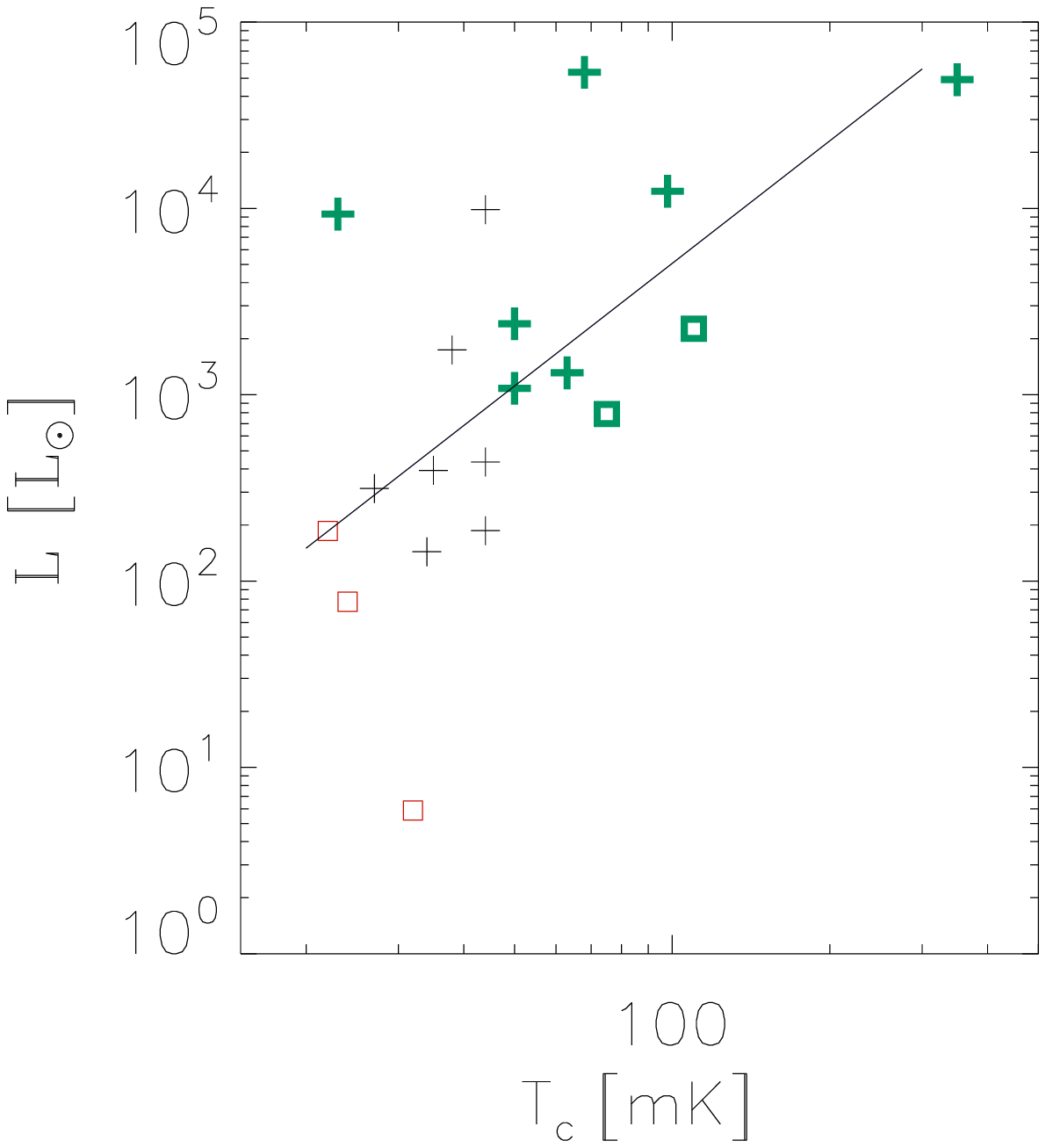}
\includegraphics[width=8.0cm,angle=0]{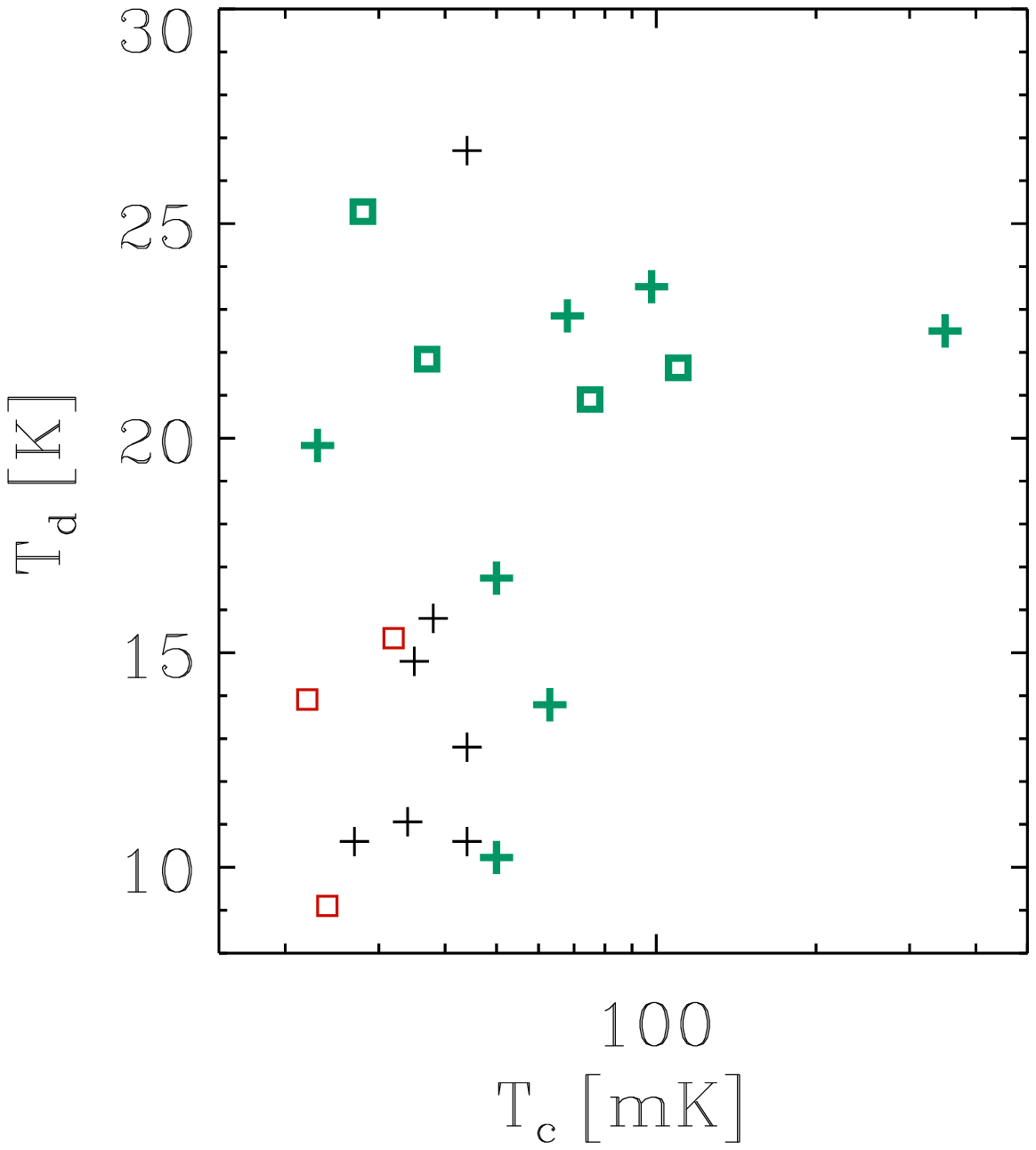}
\caption{
{\it Top panel}.
Clump bolometric luminosity vs. continuum temperature, $T_{\rm c}$. 
The solid line represents the linear fit
to all points (see text); the Spearman rank coefficient is 0.65.
{\it Bottom panel}.
Dust-derived temperature, $T_{\rm d}$, vs. continuum temperature. 
Symbols are as in Fig.~\ref{fig:LvsM}.
}
\label{fig:corrTc}
\end{figure}

%
%
\begin{table*}[\!htb] 
\centering
\caption{Gaussian fits of the \emph{Herschel}-HIFI observations.
}
\begin{tabular} {lccccccccccc } 
\hline\hline
\noalign{\smallskip} \noalign{\smallskip}
& \multicolumn{5}{c}{H$_2$O}  &  \multicolumn{3}{c}{$\mathrm{N_2H^+}$}  & \multicolumn{3}{c}{NH$_3$}\\
%

Source & H$_2$O Comp.\tablefootmark{a}  & E/A\tablefootmark{b}  & $V_\mathrm{lsr}$  &  $T_\mathrm{A}^*$  &  $\Delta V$   & $V_\mathrm{lsr}$ &   $T_\mathrm{A}^*$  &  $\Delta V$    &    $V_\mathrm{lsr}$ &  $T_\mathrm{A}^*$  &  $\Delta V$   \\


\noalign{\smallskip}
&   &   & [km~s$^{-1}$]  & [K]  & [km~s$^{-1}$] & [km~s$^{-1}$]  & [K]  & [km~s$^{-1}$]  & [km~s$^{-1}$]  & [K]  & [km~s$^{-1}$] \\

\noalign{\smallskip}
\hline
\noalign{\smallskip}



l59-441  &M   & E  & 27.9 &0.89         &6.3   &27.8    & 0.83    & 3.1      &  27.7  &0.81     &3.6 \\
         &B   & E  & 27.0 &0.37         &26.0  & \ldots & \ldots  & \ldots   &  28.2  & 0.09    & 10.1\\
         &N   & A  & 26.9 &-1.1         &3.7   & \ldots & \ldots  & \ldots   &27.7    & -0.46   & 2.0\\

\noalign{\smallskip}
	  
l59-442 &M   & E  & 36.7  &0.24   & 6.3   & 34.2    & 0.14     & 3.6     & 34.8   & 0.13   & 6.4  \\ 
	&B   & E  & 36.7  &0.12   & 23.6  & \ldots  &  \ldots  & \ldots  & \ldots & \ldots  & \ldots \\   
        &N   & A  & 34.9  &-0.30  & 4.8   & \ldots  &  \ldots  & \ldots  &35.1    & -0.05  & 2.5       \\
\noalign{\smallskip}

l59-444  &N  & E  & 32.0 & 0.21   &2.9    & 34.3    &0.20     &4.6     & 34.4   &0.21     &5.8   \\ 
	 &B  & E  &31.2  & 0.08   & 20.7  & \ldots  & \ldots  & \ldots & \ldots & \ldots  & \ldots \\
	 &N  & A  &27.7  & -0.08  & 1.0   & \ldots  & \ldots  & \ldots & \ldots & \ldots  & \ldots\\ 
	 &N  & A  &33.9  & -0.14  & 4.8   & \ldots  & \ldots  & \ldots &35.4    & -0.18   &3.4 \\ 
\noalign{\smallskip}

l59-445 &M  & E  &28.0 &0.67   &7.2  & 27.4    &0.67    & 3.1     & 27.4   & 0.64   & 4.2 \\
        &B  & E  &28.4 &0.24   &28.2 & \ldots  & \ldots  & \ldots  & \ldots & \ldots  & \ldots \\
        &N  & A  &27.0 &-0.80  &3.7  & \ldots  & \ldots  & \ldots  & 27.7   & -0.25  & 1.7\\


 \noalign{\smallskip} \noalign{\smallskip} \noalign{\smallskip}

l30-42 & \ldots  & \ldots & \ldots  & \ldots & \ldots   & 90.4   &0.03   &1.9 &91.3   &0.08   &4.2  \\  
\noalign{\smallskip}


l30-43\tablefootmark{d} &N  & E  & 108.4  & 0.07  & 2.7 & \ldots & $\lesssim0.02$\tablefootmark{c}  & \ldots & \ldots& $\lesssim0.02$\tablefootmark{c}& \ldots  \\ 
\noalign{\smallskip}




l30-304 & \ldots  & \ldots  & \ldots &  \ldots& \ldots & 	90.7 & 0.86 & 4.1 & 87.8 & 0.33 & 3.1 \\ 
        &M  & A  &92.9  &-0.35 &8.5 &\ldots &  \ldots& \ldots& 93.7 & -0.31 & 5.7  \\
\noalign{\smallskip}
l30-313 &M  & E    & 101.2& 0.07  & 6.5  & 101.5  &0.08 & 2.7  & 101.5  &0.14 & 5.8  	\\ 
\noalign{\smallskip}
l30-327 &\ldots &\ldots     &  \ldots &  \ldots& \ldots& 102.6 & 0.07  & 3.6 &\ldots &  \ldots& \ldots	\\  
        &M  & A   &  103.8 & -0.11 & 5.7 & \ldots &  \ldots  & \ldots &104.6 & -0.08 & 4.6	\\ 
\noalign{\smallskip}


l30-376 &N  & E  &70.5 &0.07 &3.2 & \ldots & $\lesssim0.02$\tablefootmark{c}  & \ldots & 70.5  &0.05  &2.7 \\ 
\noalign{\smallskip}


l30-512 &M  & E  & 63.2  &0.11   &5.5  &61.4    &0.06      &2.9      &61.8    &0.15      &2.5	\\ 
	&N  & A  & 59.5  &-0.06  &2.4  & \ldots & \ldots   & \ldots  & \ldots &  \ldots  & \ldots  \\ 
	&N  & A  & 64.0  &-0.06  &1.6  & \ldots & \ldots   & \ldots  & \ldots &  \ldots  & \ldots  \\ 
\noalign{\smallskip}
\noalign{\smallskip} \noalign{\smallskip}
\hline 
\label{Table:gaussian}
\end{tabular}
\tablefoot{
\tablefoottext{a}{Definition of the Gaussian components of the water line:
N=narrow, M=medium, B=broad. } 
\tablefoottext{b}{E=emission, A=absorption.}
\tablefoottext{c}{3$\sigma$ limit.} 
\tablefoottext{d}{Tentative detection (see Section~\ref{sec:parspec}).} 
} 
\end{table*}

Table~\ref{Table:gaussian} shows that the H$_2$O line profiles of the $\ell=59\deg$ sources
are very similar. Almost all water spectra can be described as the sum of a narrow
(FWHM$ \sim 3-5$\,km\,s$^{-1}$), a medium (FWHM$ \sim 6-10$\,km\,s$^{-1}$) and a broad
(up to $\simeq 30$\,km\,s$^{-1}$) velocity component. The narrow component is mostly
seen in absorption,  whereas the medium and broad components have positive intensities.
Source l59-444 shows an inverse P-Cygni profile overlaid on a broad component.  
In source l59-441 and, to a lesser extent, in l59-445 (please note that these
two sources are only $21.5\arcsec$ apart) if three Gaussian components are used in the fit the
red-shifted wing of the spectrum still shows a residual deviation
($\ga 3\,\sigma$, see Fig.~\ref{fig:l59_441}) from the overall
fit at a velocity of $\simeq 55$\,km\,s$^{-1}$. This residual can be eliminated with an additional
Gaussian component, which may either have a high velocity ($\ga 50\,$km\,s$^{-1}$), or
a velocity comparable to the other components but with a wide line width ($\ga 40\,$km\,s$^{-1}$).
In the first case the additional velocity component could be interpreted as an extremely high velocity
(EHV) component. Similar EHV components have previously been reported, e.g., by \citet{kristensen2012}
toward several low-mass protostars and by \citet{leurini2014} toward the massive protostar
IRAS\,17233-3606, and are usually associated with shocked gas. However, we do not have further evidence
(such as spectra of other water lines and/or spectra at different positions)
to confirm that an EHV component indeed exists in source l59-441. In the second scenario, an additional broad
velocity component must be invoked, which may be caused by the fact that molecular outflows are not
Guassian shaped. Due to this ambiguity, we choose to show only three Gaussian
components for this source.
Section~\ref{sec:kin} discusses the water line profile components in detail.

%
%
%

The situation is very different in the $\ell=30\deg$ region, with no two profiles being identical.
We have no clear detection of a broad component in the water spectra, and the line profile
can be fit by fewer Gaussian components.
%
%
Ammonia also displays complex line shapes with mixed emission and absorption, and only weak
signs of outflows. In contrast to water and ammonia, the N$_2$H$^+$ spectra do not show any line
asymmetries and we observe only emission.  In addition,
two sources also show a regular or inverse P-Cygni profile, either in the water
line (l30-512) or in the NH$_3$ spectrum (l30-327) (see Section~\ref{sec:PCygni}).
Tables~\ref{Table:sources}, \ref{Table:gaussian} and \ref{Table:Columns} mark source l30-43
as a tentative detection because recent (June 2015) follow-up observations with the 20-m 
telescope of the Onsala Space Observatory have detected the NH$_3$(1,1) line toward
this source at a velocity of $\simeq 95.3$\,km\,s$^{-1}$, instead of 108.4\,km\,s$^{-1}$
as shown in Table~\ref{Table:gaussian} and Appendix~\ref{sect:app1}. Therefore, either 
the water line in source l30-43 is an artefact of some type, or both lines are actually real
and thus the issue remains as to whether or not they belong to the same object.
With our current data we cannot answer this question and therefore in this work we will
assume that the water line detection is real.

%
%
%
\begin{figure}
\centering
\includegraphics[width=7.0cm,angle=0]{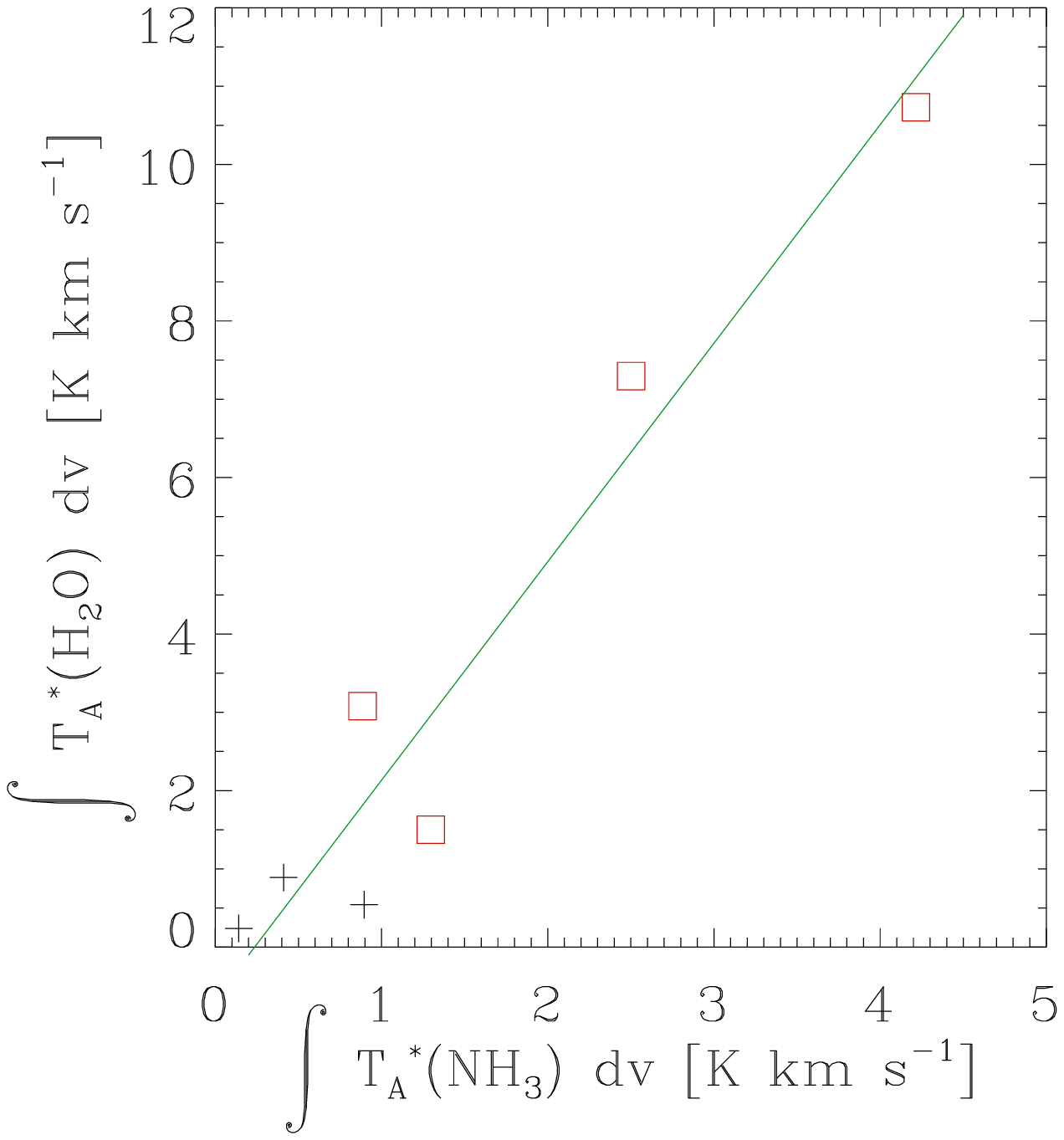}
\includegraphics[width=7.0cm,angle=0]{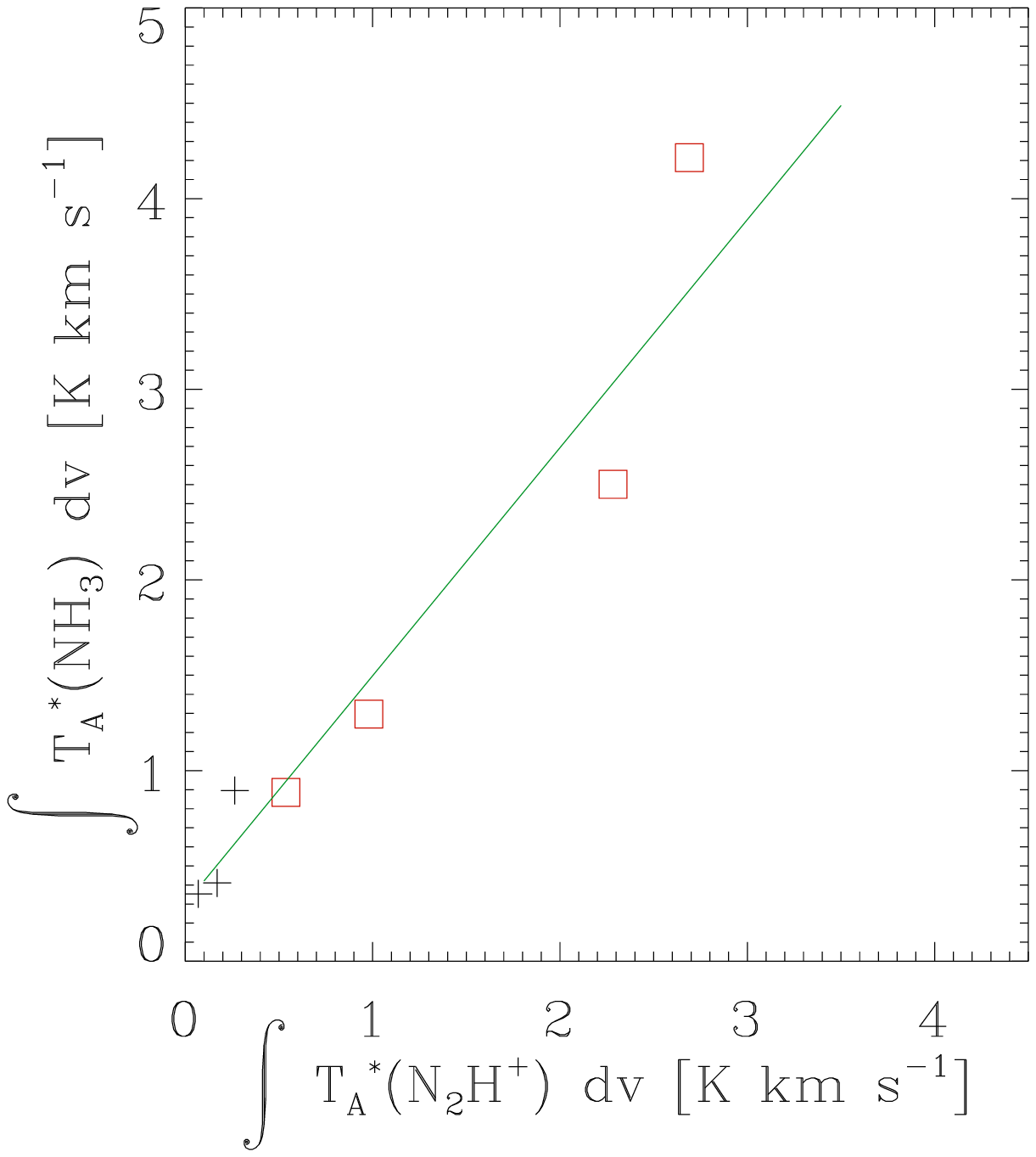}
\includegraphics[width=7.0cm,angle=0]{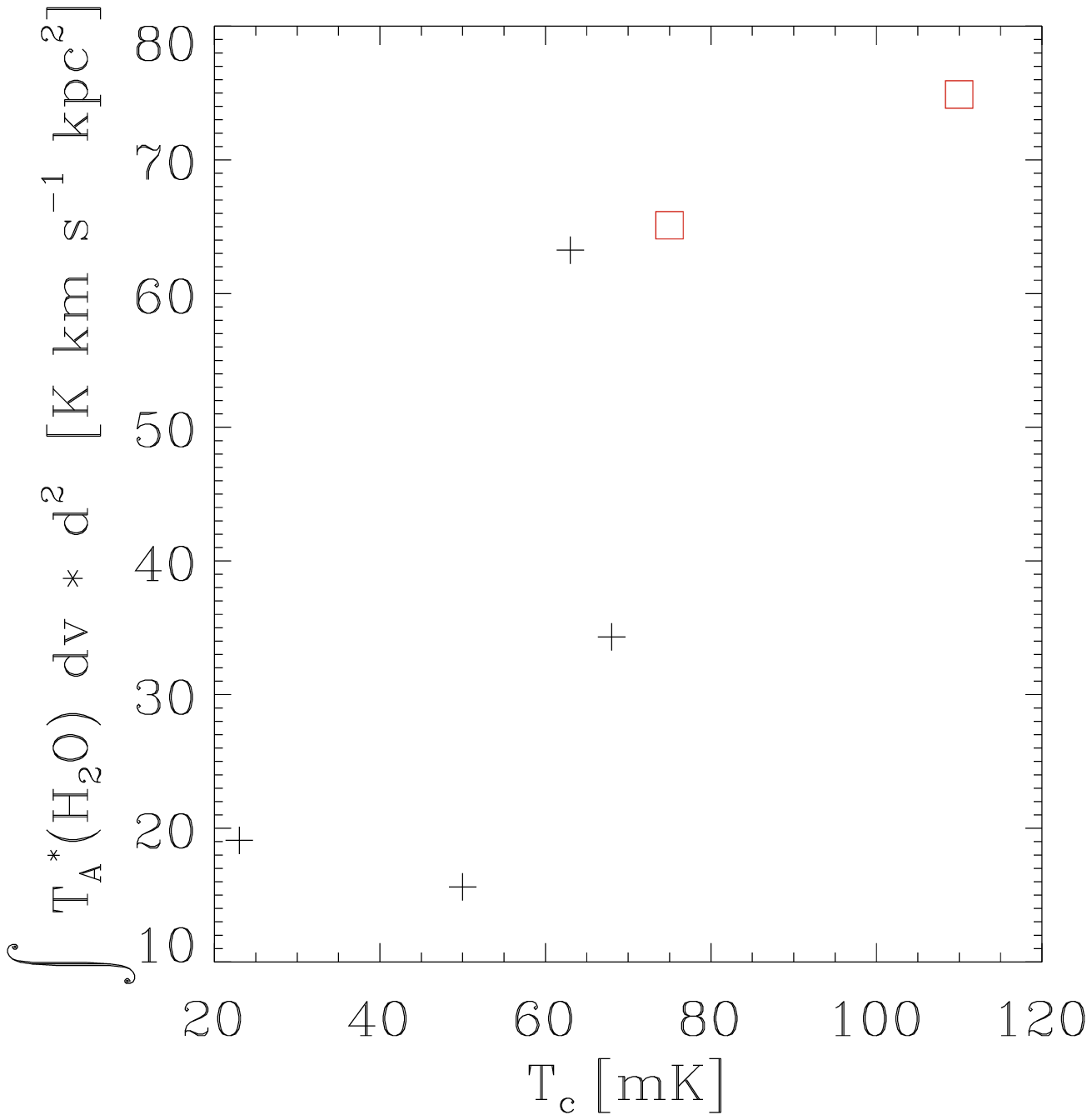}
\caption{
{\it Top panel}.
Integrated H$_2$O$\,(1_{10}-1_{01})$ line intensity plotted
vs. the integrated NH$_3\,(1_0-0_0)$ line intensity.
Symbols are as in Fig.~\ref{fig:LvsM}. The solid line represents the linear fit
to all points. The Spearman rank coefficient is 0.94.
{\it Middle panel}.
Integrated NH$_3\,(1_0-0_0)$ line intensity vs. the integrated
N$_2$H$^+\,(6-5)$ line intensity.  
Source l30-304 has not been included in the plot (see text).
The Spearman rank coefficient is 0.96.
{\it Bottom panel}.
$\int T_{\rm A}^*[{\rm H_2O}] \, {\rm d}v \times d^2$, proportional to the H$_2$O line luminosity,
vs. continuum temperature, $T_c$. 
Symbols are as in Fig.~\ref{fig:LvsM}.
}
\label{fig:IntegIntens}
\end{figure}

\subsection{Correlations between dust- and gas-derived physical parameters}
\label{sec:corr}

In order to better understand the nature of the detected sources in the $\ell=30\deg$
and $\ell=59\deg$ regions, as well as to determine the origin of the H$_2$O emission,
we have performed correlations between several parameters.
We begin with a comparison which includes both dust-derived physical parameters,
as determined from our previous Hi-GAL observations, and HIFI-derived parameters,
such as the continuum temperature, $T_{\rm c}$. Then, the total H$_2$O integrated line intensity
is compared to different parameters.

In Fig.~\ref{fig:LvsM} we plot the clump luminosity vs. mass, as derived from
our previous Hi-GAL observations.  Since both are distance-dependent quantities, only
sources where a distance could be determined are shown. The plot shows that
the clumps found in the $\ell=59\deg$ region are characterized by lower luminosity
and mass compared to those in the $\ell=30\deg$ region (see the detailed statistical analysis
by \citealp{olmi2013} and \citealp{olmi2014a}). Sources with a detection in at least one of the three
observed spectral lines are clearly found at higher luminosity ($L \ga 10^3\,$L$_\odot$),
as also confirmed by Table~\ref{tab:median}.
%
In Fig.~\ref{fig:LvsM} we also plot a series of evolutionary tracks which we discuss
in Section~\ref{sec:evolution}.

In Fig.~\ref{fig:corrTc} we plot the dust-derived temperatures 
($T_{\rm d}$, column 6 in Table~\ref{Table:sources}) and luminosities of the clumps in the $\ell=30\deg$ 
and $\ell=59\deg$ regions vs. $T_{\rm c}$. 
The continuum flux density associated with $T_{\rm c}$ is likely to have a contribution
from both free-free emission and warm dust thermal emission. We do not have an alternative
method to determine the contribution of free-free emission in our sources. Furthermore, because 
we have only measured $T_{\rm c}$ at one frequency, we cannot determine how $T_{\rm c}$
varies with frequency, which could otherwise be used to discriminate between dust and free-free
emission. 
\citet{vandertak2013}, however, found that the free-free emission in their massive proto-stellar
clumps is negligible compared to warm dust emission. If this is true also for our sources,
then $T_{\rm c}$ should correlate with the source bolometric luminosity.
This is indeed shown in the top panel of Fig.~\ref{fig:corrTc}, where we also 
used the Bayesian IDL routine {\tt LINMIX\_ERR} to perform a linear
regression to find the slope of the best fit line.  The source luminosity is expected 
to increase with time, as the clump evolves from the starless to the proto-stellar 
phase (see, e.g., \citealp{molinari2008}). Therefore, the top panel of Fig.~\ref{fig:corrTc}
is also likely to represent an evolutionary sequence.
However, the bottom panel of Fig.~\ref{fig:corrTc} shows that there is only a marginal 
correlation between $T_{\rm d}$ and $T_{\rm c}$, which are both distance-independent.
The large scatter is certainly due in part to the large uncertainties in $T_{\rm c}$,
since the baseline rms was in general quite large ($\ga 20-30\,$mK). 


%
%
%
In Fig.~\ref{fig:IntegIntens} we plot various line integrated intensities.
The integrated area of the profile has been
determined by summing all components in emission.
For a Gaussian line profile,
the integrated emission is $T_{\rm A}^{\rm pk} \, \Delta V \, \sqrt{\pi /(4 \ln 2)}$, where
$T_{\rm A}^{\rm pk}$ is the peak antenna temperature and $\Delta V$ is the FWHM of the line.
Sources with pure absorption spectra in any of the plotted spectral lines have
not been considered.
Despite the low number of data points, the top and middle panels show an excellent
correlation between the integrated intensities of the H$_2$O and NH$_3$ lines,
and also between those of the two N-bearing molecules. In the middle panel
source l30-304 has not been included in the plot because it has a completely self-absorbed 
water line profile, and the NH$_3$ line has an inverse P-Cygni profile.  Therefore, the 
integrated intensity (in emission) of ammonia is very uncertain.
The reasonable agreement between the integrated
intensities of ammonia and N$_2$H$^+$ suggests that they are emitted from similar
volumes of gas, assuming similar excitation conditions.

Finally, in the bottom panel of Fig.~\ref{fig:IntegIntens} we plot the quantity
$\int T_{\rm A}^*[{\rm H_2O}] \, {\rm d}v \times d^2$, proportional to the
H$_2$O line luminosity, vs. the continuum temperature, $T_{\rm c}$.
Although we do not show any linear fit in this case, there seems to be a general trend of
increasing line luminosity with $T_{\rm c}$. The l30 source with the highest water line
luminosity ($\ga 60\,$K\,km\,s$^{-1}$\,kpc$^2$) is l30-512 which, however, has a P-Cygni profile
(see Section~\ref{sec:PCygni}) and thus the actual line luminosity is uncertain.
Except for this point, the plot shows that in these
objects the ``typical'' water line luminosity toward the $\ell=30\deg$ sources is lower
compared to that of the $\ell=59\deg$ region.

%
%

 \begin{table*}[\!htb] 
\centering
\caption{Column densities and abundances from {\tt RADEX} modelling\tablefootmark{a}. 
}
\begin{tabular} {lcccccccccc } 
 \hline\hline
     \noalign{\smallskip} \noalign{\smallskip}
Source& Comp.\tablefootmark{b}  &$V_\mathrm{lsr}$  &       $N$(o-H$_2$O)\tablefootmark{c}    &       $X$(o-H$_2$O)\tablefootmark{d}& $V_\mathrm{lsr}$ &     $N$(N$_2$H$^+$)\tablefootmark{c}  
&  $X(\mathrm{N_2H^+})$\tablefootmark{d} & $V_\mathrm{lsr}$ &      $N$(o-NH$_3$)\tablefootmark{c} &      $X$(o-NH$_3$)\tablefootmark{d}\\
 \noalign{\smallskip}
  &  &[km~s$^{-1}$] &  [$\times 10^{14}\,$cm$^{-2}$]  &[$\times 10^{-9}$]  &[km~s$^{-1}$] &  [$\times 10^{14}\,$cm$^{-2}$]  &[$\times 10^{-9}$]  & [km~s$^{-1}$]  & [$\times 10^{14}\,$cm$^{-2}$]  & [$\times 10^{-9}$] \\
     \noalign{\smallskip}
     \hline
     \noalign{\smallskip}

l59-441  & M   & 27.9 &	450     & 41   &27.8  & 0.34  & 6.2      &27.7  & 37  & 3.4  \\
	 & B   & 27.0 & 67	&\ldots  & \ldots &\ldots  &\ldots & 28.2 & 1.3 &\ldots \\
	 & N   & 26.9 & 0.34	& 0.06   & \ldots &\ldots  &\ldots &27.7  & 0.05 & 0.01 \\  \noalign{\smallskip}
	  
l59-442 &M  & 36.7 &\ldots &\ldots &34.2    &\ldots &\ldots   & 34.8   &\ldots & \ldots  \\ 
	&B  & 36.7 &\ldots &\ldots & \ldots &\ldots  &\ldots  &\ldots  &\ldots & \ldots  \\   
	&N  & 34.9 &\ldots &\ldots &\ldots  &\ldots &\ldots   &35.1    &\ldots &\ldots     \\\noalign{\smallskip}

l59-444 &N  & 32.0&\ldots &\ldots &34.3    &\ldots &\ldots  &34.4     &\ldots &\ldots \\ 
        &B  & 31.2&\ldots &\ldots &\ldots  &\ldots &\ldots  & \ldots  &\ldots &\ldots   \\
	&N  & 27.7&\ldots &\ldots &\ldots  &\ldots &\ldots  & \ldots  &\ldots &\ldots    \\ 
	&N  & 33.9&\ldots &\ldots &\ldots  &\ldots &\ldots  & 35.4 &\ldots  &\ldots  \\ \noalign{\smallskip}

l59-445  &M &28.0 & 350  & 270   &27.4 & 32  & 25  & 27.4 & 60  & 46  \\
	 &B &28.4 & 300  &\ldots      &\ldots     &\ldots   &\ldots  &\ldots  &\ldots  &\ldots   \\
	 &N &27.0 & 0.36 & 0.6   &\ldots     &\ldots &\ldots & 27.7 & 0.03  & 0.04 \\
 \noalign{\smallskip} \noalign{\smallskip} \noalign{\smallskip}

l30-42 &\ldots   &\ldots &\ldots & \ldots  &90.4  & 4.3  & 1.3  & 91.3  &  6.0	 & 1.9  \\  \noalign{\smallskip}

l30-43\tablefootmark{g} & N	 & 108.4  & 5  & 0.58   & \ldots  & $\lesssim$30   & $\lesssim$4  & \ldots  & $\lesssim$0.6\tablefootmark{e}  & $\lesssim$0.06\tablefootmark{e}   \\ \noalign{\smallskip}



l30-304 &\ldots     & \ldots  &  \ldots  & \ldots  & 90.7 & 33   & 3.3  &87.8  & 4.9  & 0.49 \\ 
	&M          &92.9     & $\gtrsim$2\tablefootmark{f} &$\gtrsim$0.4\tablefootmark{f} &\ldots&\ldots &\ldots&  93.7  & 0.52  & 0.1\\\noalign{\smallskip}

l30-313 &M    &101.2& 15  & 2.19 &101.5  & 3.1  &0.42  & 101.5  & 10  & 1.4	\\ \noalign{\smallskip}

l30-327 &\ldots    & \ldots & \ldots  & \ldots &102.6 & 2.4  & 0.22  & \ldots  & \ldots  & \ldots 	\\  
        &M         &103.8   & 0.73    & 0.1  & \ldots  & \ldots  & \ldots  &104.6  & 0.21  & 0.04\\ \noalign{\smallskip}

l30-376 &N  &70.5& 39   & 33  & \ldots  & $\lesssim$8  & $\lesssim$7  & 70.5  & 6.1  & 5.1 \\ \noalign{\smallskip}

l30-512 &M  & 63.2  & 110 & 22   &61.4     &26  & 5.1  &61.8  & 30  & 5.9 	\\ 
	&N  & 59.5  & 0.3 & 0.1  & \ldots  & \ldots& \ldots& \ldots&  \ldots& \ldots \\ 
	&N  & 64.0  & 0.2 & 0.08 &  \ldots&  \ldots& \ldots& \ldots&  \ldots& \ldots \\ \noalign{\smallskip}
    \noalign{\smallskip} \noalign{\smallskip}
\hline 
\label{Table:Columns}
\end{tabular}
\tablefoot{
\tablefoottext{a}{Using our derived $n(\mathrm{H_2})$ and $T_\mathrm{k} = T_\mathrm{dust}$ (Table~\ref{Table:sources}).} 
\tablefoottext{b}{N=narrow component, M=medium, B=broad. }
\tablefoottext{c}{Column density corrected for main beam efficiency and beam-filling factor (Table~\ref{Table:sources}).} 
\tablefoottext{d}{Abundance with respect to molecular hydrogen (Table~\ref{Table:sources}). In the absorption
components we have used $N(\mathrm{H_2})/2$  assuming that the absorption is probing approximately half of the total column density
in front of the background continuum. }
\tablefoottext{e}{3$\sigma$ limit.} 
\tablefoottext{f}{Optically thick; $\tau$ is assumed to be 5.} 
\tablefoottext{g}{Tentative detection (see Section~\ref{sec:parspec}).}
} 
\end{table*}
%

\subsection{ Column densities and molecular abundances}
\label{sec:radex}

\subsubsection{{\tt RADEX} modelling}
\label{sec:radexmod}


Since we lack direct information about the excitation temperatures and opacities of
the observed transitions, we use the non-equilibrium homogeneous radiative transfer code
{\tt RADEX}\footnote{\tt{http://www.sron.rug.nl/~vdtak/radex/radex.php}}
\citep{vanderTak2007}  to correct for possible populations of molecules in unobserved
excited levels, in order to relate the observed emission and absorption components to
the column densities of H$_2$O, NH$_3$ and N$_2$H$^+$.

The principal parameters needed to specify a model are the kinetic temperature, $T_{\rm k}$,
the average number density of molecular hydrogen $n_{\rm H2}$, and the line width
(see for example, \citealp{persson2009}).
The density of each source is computed from the Hi-GAL observed size and mass, assuming a
homogeneous, spherical cloud. The kinetic temperature is assumed to be equal to the dust
temperature obtained from our Hi-GAL observations.  All parameters are listed in Table~\ref{Table:sources}.
The  background radiation field is an average of the Galactic background radiation in the
solar neighbourhood and of the cosmic microwave background radiation.


The column density obtained from {\tt RADEX} is varied until the model line
brightness temperature,  $T_{\rm b}$, matches the amplitude of the Gaussian fit
(Tab.~\ref{Table:gaussian}). The line brighness temperature is obtained from the
antenna temperature after correcting for main-beam
efficiency, $\eta_{\rm mb}$, and beam-filling factor, $\eta_{\rm bf}$, i.e.:
\beq
T_\mathrm{A}^* = T_{\rm b} \, \eta_{\rm mb} \, \eta_{\rm bf} = T_{\rm b} \, \eta_{\rm mb} \,
\frac{ \theta_{\rm mb}^2 }  { \theta_{\rm mb}^2 + \theta_{\rm s}^2 }
\eeq
with $\theta_{\rm mb}$ and $\theta_{\rm s}$ being the FWHM of the main-beam and source,
respectively. The values of $\eta_{\rm bf}$ are also listed in Table~\ref{Table:sources}.
For the components in absorption, we match the observed and modelled integrated opacity
instead of the brightness temperature (see Section~\ref{sec:narrow}).

The computed {\tt RADEX} results are listed in Table~\ref{Table:Columns}. The results
are {\it not} very sensitive to changes in density because of the high critical densities
($n_{\rm cr} \ga 10^7 - 10^8$\,cm$^{-3}$) of all three molecular transitions.
However, the column density is sensitive to changes in kinetic temperature if $T_{\rm k} \la 15\,$K.
For sources with non-detections in one (or two) of the molecular transitions, the upper limit
to the column density has been estimated by using the $3\, \sigma$ limits for $T_{\rm A}^*$
listed in Table~\ref{Table:gaussian}.
%
%
%

Abundances with respect to molecular hydrogen, $X({\rm mol}) = N({\rm mol})/N({\rm H}_2)$,
are estimated using the derived volume densities and observed source sizes to compute $N({\rm H}_2)$.
For the components in absorption we divide the estimated $N({\rm H}_2)$ by a factor of two since 
the absorption is probing half of the total column density that is in front of the continuum source. 
For the outflow components we cannot estimate the abundance since $N({\rm H}_2)$ is likely orders 
of magnitude lower than that of the whole cloud.  

Table~\ref{Table:Columns} indicates that the water molecular abundance is possibly
higher toward sources in the $\ell=59\deg$ region. Our results may be affected 
by the uncertainties in the derived Hi-GAL physical parameters and other assumptions. 
However, combined with the result that the typical water line luminosity toward the
$\ell=30\deg$ sources (where no broad component is detected) is lower compared 
to that of the $\ell=59\deg$ region (see bottom panel of Fig.~\ref{fig:IntegIntens}), 
we conclude that our observations support a scenario where the water abundance is 
linked to shocked gas.

%
%
%
\begin{table*}
\caption{
Mean values of bolometric luminosity, mass and continuum temperature.
}
\label{tab:median}
\centering\begin{tabular}{lcccccccr}
\hline\hline
%
             Region                      &    & \multicolumn{3}{c}{Sources with detection }    &
& \multicolumn{3}{c}{Sources with no detection\tablefootmark{a}}  \\
%
\cline{3-5}
\cline{7-9}
                   &   & Mean $L$           & Mean $M$           & Mean $T_{\rm c}$   &    
                       & Mean $L$           & Mean $M$           & Mean $T_{\rm c}$  \\
                   &   & $[{\rm L}_\odot]$  & $[{\rm M}_\odot]$  & [K]                &    & $[{\rm L}_\odot]$  & $[{\rm M}_\odot]$ & [K]              \\
\hline
    \noalign{\smallskip}
        $\ell=30$      &   & $16200\pm 17600$    & $3430\pm 3300$  & $93\pm 65$      &    & $2145\pm 2570$   & $690\pm 305$     & $37\pm 5$ \\
        $\ell=59$      &   & $1520\pm 735$       & $190\pm 105$    & $63\pm 30$      &    & $90\pm 64$       & $440\pm 390$     & $26\pm 4$ \\
    \noalign{\smallskip} 
\hline
\end{tabular}
\tablefoot{
\tablefoottext{a}{
Only sources where all three parameters could be measured were included. }
}
\end{table*}

\subsubsection{Column density for the absorption component}
\label{sec:absorptioncd}

We also used an alternative approach to compute the {\it o}-H$_2$O column density of the
envelope gas responsible for the observed absorption in the water lines, which 
we also use to double-check the {\tt{RADEX}} results.
We first derived the optical depth at the dip of the absorption profile as:
%
\beq
\tau_{\rm dip} = - \ln \left ( \frac{T_{\rm A}^{\rm em} - T_{\rm A}^{\rm dip} }
{T_{\rm A}^{\rm em} }  \right)
\eeq
where, in order to include the line contribution to the absorption background,
the antenna temperature $T_{\rm A}^{\rm em}$ represents the sum of all
Gaussian components in emission (at the velocity of the dip in the absorption component)
and of the continuum temperature $T_{\rm c}$. This procedure assumes that the absorber
is located in front of both the continuum and the line emitting source.
$T_{\rm A}^{\rm dip}$ represents the
peak absolute value of the antenna temperature of the Gaussian component in absorption.
The integrated optical depth of the absorption conponent is then calculated as
$ \tau_{\rm abs} = \int \tau \, {\rm d}\nu =
1.06 \, \tau_{\rm dip} \, \Delta V_{\rm abs} $    
where $\Delta V_{\rm abs} $ represents the line FWHM of the Gaussian component in absorption.
Finally, the column density can be estimated as:
\beq
N_{\rm tot} \simeq N_l = \frac{8 \pi \nu_{\rm ul}^3 g_l} {c^3 A_{\rm ul} g_u }
\left [ 1 - \exp \left (-\frac{h \nu_{\rm ul}}{k T_{\rm ex}} \right ) \right ]^{-1}
\tau_{\rm abs}
\label{eq:coldens}
\eeq
where $A_{\rm ul}$ and $\nu_{\rm ul}$ represent the Einstein $A$ coefficient and frequency
associated with the transition between the lower and upper level, indicated
by the subscripts $l$ and $u$, respectively, and with degeneracies $g_l$ and $g_u$.
Physical constants are indicated with standard symbols and $T_{\rm ex}$ represents the
excitation temperature. Since we do not know $T_{\rm ex}$ we will assume for the moment
that $T_{\rm ex} \ll h \nu_{\rm ul}/k = 26.6$\,K for the 557\,GHz water line. Therefore,
the $\exp$ factor in Eq.(\ref{eq:coldens}) can be neglected.

We thus obtain remarkably similar {\it o}-H$_2$O column densities, varying from about
$10^{13}\,$cm$^{-2}$ to $\simeq 5 \times 10^{13}\,$cm$^{-2}$ (consistent with 
the {\tt{RADEX}} results), with the highest value
obtained toward the source l30-304, where in fact the absorption appears to be almost
saturated, since the line profile is nearly down to the zero-temperature level.
The column densities of the outer envelope are thus quite similar to those
estimated for both low- \citep{kristensen2012} and high-mass \citep{vandertak2013}
star forming regions.

\section{Discussion} 
\label{section: discussion}

\subsection{Average physical conditions in the $\ell=30\deg$ and $\ell=59\deg$ regions}
\label{sec:overview}

As shown by Tables~\ref{Table:sources} and \ref{Table:gaussian}, in the $\ell=30\deg$ region
about 41\% of the sources have a detection in at least one of the observed spectral
lines. This percentage drops to about 11\% in the $\ell=59\deg$ region.  We also note that
all of the sources with at least one detection 
are {\it proto}-stellar in both galactic regions.
The significant difference in detections between the two regions
may be a consequence of the different fraction of proto-stellar sources
observed in each region, corresponding to 59\% and 14\% in $\ell=30\deg$ and $\ell=59\deg$,
respectively. In fact, the fraction of {\it proto}-stellar sources
with at least one detection corresponds to 70\% and 80\% in the $\ell=30\deg$ and $\ell=59\deg$
regions, respectively, making the difference in the detection rate less significant.

In Table~\ref{tab:median} we list the mean values of the clump mass and bolometric luminosity,
as derived from the Hi-GAL observations, for the observed sources in the two galactic regions.
We note that the mean values of the luminosity for sources with a detection in one of the
observed spectral lines are much higher than the luminosity of sources with no detection
at all. The scatter around the mean value is quite high and thus our conclusion is
only tentative and will need more data to be confirmed. Furthermore, the mean values in the
$\ell=59\deg$ region are derived from only two sources, since in sources l59-442 and l59-444
the distance could not be determined reliably.

As with luminosity, Table~\ref{tab:median} shows that $T_{\rm c}$ has a higher mean value
(by more than a factor of 2) in sources with at least one detected spectral line, for both observed regions.
Given that all of these sources are proto-stellar, the mean values of $T_{\rm c}$
for sources with and without detection  are in fact very similar to the mean values of $T_{\rm c}$
for proto-stellar and starless  sources, respectively.  Compared to luminosity and mass,
the mean values of $T_{\rm c}$ show a lower scatter, which is still quite high only in the case
of sources with detection in the $\ell=30\deg$ region. In fact, the scatter
for the case of sources with no detection is quite low in both regions, despite the low number
of sources.
Given the marginal correlation between luminosity and $T_{\rm c}$ (see Section~\ref{sec:corr})
this result is not totally unexpected. However, given that the continuum temperature is
a distance-independent quantity this result is more significant in charaterizing the sources.
Finally, from Table~\ref{tab:median} we also note that the
mean values of the mass for sources with and without a detection 
do not show any specific relation.
Thus, if we assume tentatively that the difference in the mean value of luminosity 
(and $T_{\rm c}$) between sources with and without detections is indeed real, then this might 
be a consequence of evolutionary effects rather then an
effect associated with more massive clumps (see also Section~\ref{sec:evolution}).

\subsection{Gas kinematics}
\label{sec:kin}

Our decomposition of the H$_2$O line profiles (see Section~\ref{sec:parspec}) is similar to that
of \citet{kristensen2012} toward low-mass protostars, and to the description of the H$_2$O
profiles of \citet{vandertak2013} toward high-mass protostars. \citet{herpin2012} found
a similar profile decomposition toward the massive protostar W43-MM1.
In the following we first describe the general properties of the line shapes in the 
$\ell=30\deg$ and $\ell=59\deg$ regions, and then we discuss the possible physical origin 
of each component.

\subsubsection{Line profile components in the $\ell=59\deg$ region}
\label{sec:profile59}

Here we summarize the main observational properties of the line profiles of the
three transitions, H$_2$O$(1_{10}-1_{01})$, NH$_3(1_0-0_0)$, and 
N$_2$H$^+(6-5)$ toward the $\ell=59\deg$ region. \\

%
{\tt H$_2$O.} 
{\it All} detected sources of the $\ell=59\deg$ region have a strikingly
similar water line-profile, which  is always charaterized by an absorption dip at the 
source velocity (as given by N$_2$H$^+$). This absorption dip does not appear to be
saturated.  As shown by Table~\ref{Table:gaussian}, the offset between these absorption 
components and the source velocity does not appear to be significant.
The broad and medium Gaussian components have 
line widths exceeding those of N$_2$H$^+$, which is tracing the more quiescent ambient gas,
and are thus likely originating in proto-stellar outflows.
The water line profile in source l59-444 is somewhat different from that of the other 
three detected sources in the $\ell=59\deg$ region since it is charaterized by an 
inverse P-Cygni profile overlaid on a broad component (see Section~\ref{sec:PCygni}). \\

{\tt N$_2$H$^+$ and NH$_3$.} 
While the N$_2$H$^+$ line-profile looks quite Gaussian, and thus used 
to set the source velocity, NH$_3$ shows an asymmetric line-profile which exhibits 
self-absorption (less evident in l59-442), though not at the exact same velocity as 
absorption in the water line (see Table~\ref{Table:gaussian}). 
Absorption is also present in the NH$_3$ spectrum of l59-444.
Finally, we note that one of the Gaussian components of the NH$_3$ line 
toward sources l59-441 and l59-442 has a line width of $10.1\,$km~s$^{-1}$ and 
$6.4\,$km~s$^{-1}$ (see Table~\ref{Table:gaussian}). This suggests that the outflows
observed in the water line are also partially detected in the ammonia line, at least
in source l59-441.
%

%
%
\begin{figure}
\centering
\includegraphics[width=7cm,angle=0]{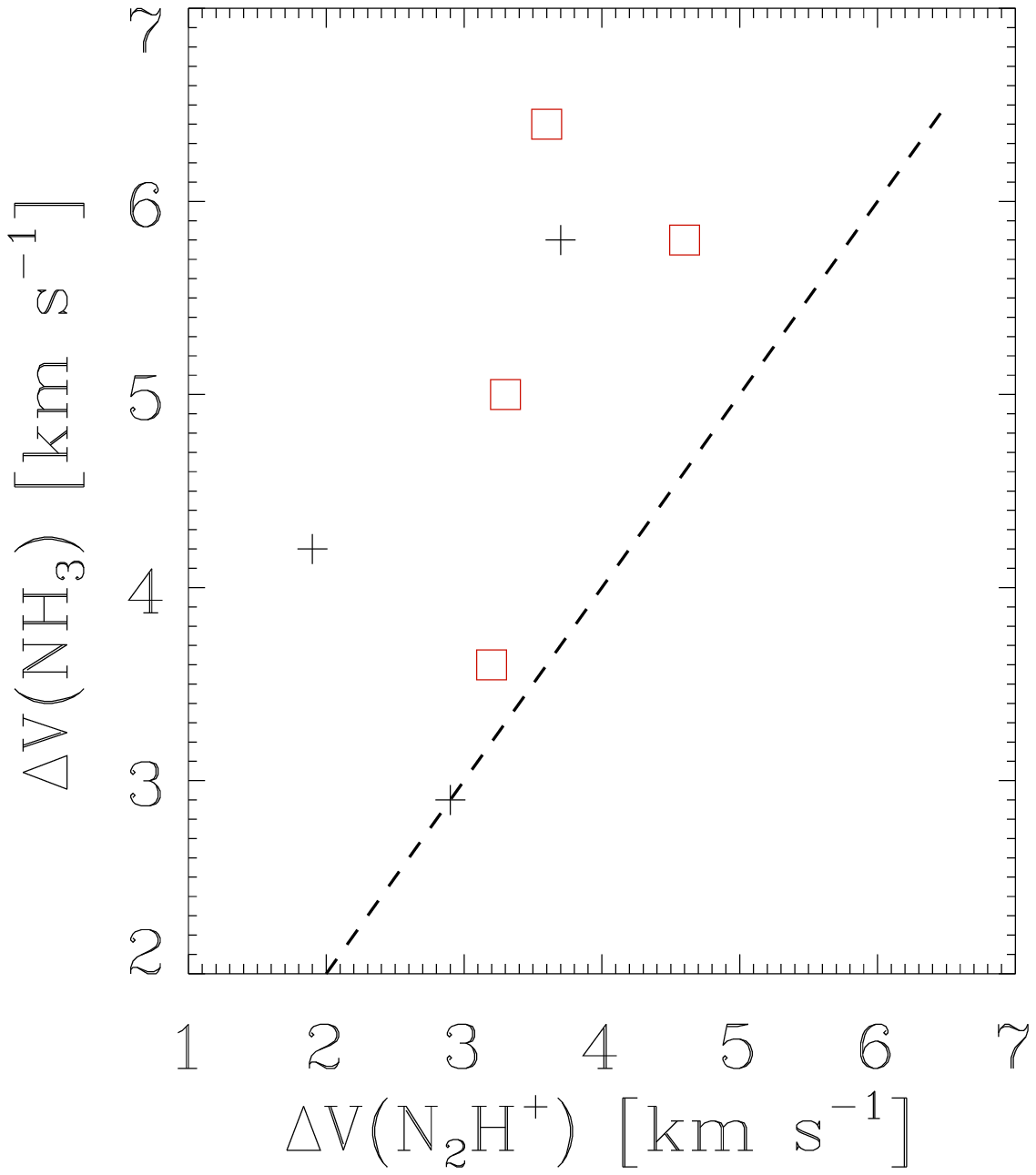}
\includegraphics[width=7cm,angle=0]{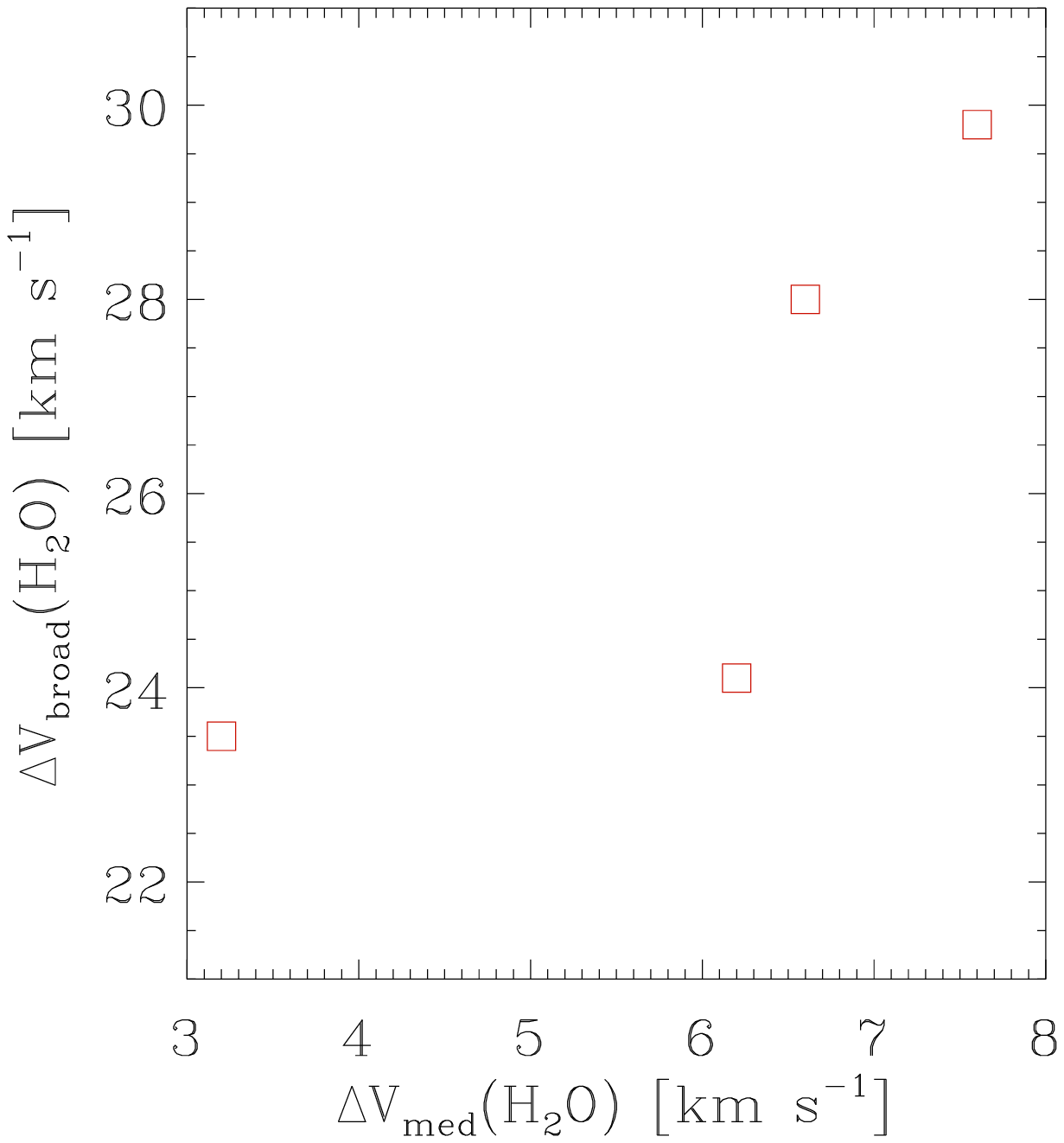}
\caption{
{\it Top.} Line width of the NH$_3$ line vs. line width of the N$_2$H$^+$ line, when both
are detected in emission.
The dashed line represents the $\Delta V$(NH$_3$) = $\Delta V$(N$_2$H$^+$) locus.
{\it Bottom.} Line width of the broad component of the water line profile vs. the
line width of the medium component for  the $\ell=59\deg$ sources.
Symbols are as in Fig.~\ref{fig:LvsM}.
}
\label{fig:Dv}
\end{figure}

\subsubsection{Line profile components in the $\ell=30\deg$ region and comparison of line widths}
\label{sec:profile30}

The nature of the observed lines toward this region is significantly different from that
of the sources in the $\ell=59\deg$ region. The line intensities are in general lower 
compared to those measured toward the $\ell=59\deg$ region, which is likely a consequence 
of the average larger distance to the $\ell=30\deg$ sources. \\

%
{\tt H$_2$O.}
 The water line is detected both in emission and absorption, but the 
line-profiles are much more varied. 
As already mentioned, in the sources l30-304 and l30-327 the Gaussian component in absorption appears
to be almost saturated.  No line wings are detected 
above the noise level. 
In several sources (e.g., l30-304 and l30-327) the presence of
multiple water absorption lines is evidently tracing the diffuse interstellar gas 
(which will be the subject of a separate paper, 
Persson et al., in prep.), that are likely a consequence of the larger amount of diffuse 
gas encountered along the $\ell=30\deg$ direction.  
In Fig.~\ref{fig:IntegIntens} we have already shown that the water line luminosity toward the 
$\ell=30\deg$ sources is lower compared to that of the $\ell=59\deg$ region.
\\

{\tt N$_2$H$^+$ and NH$_3$.}
 As for the $\ell=59\deg$ region, the N$_2$H$^+$ line-profile looks Gaussian, 
and (when detected) it is also used to set the source velocity. In most cases the 
line-profiles of water and ammonia are consistent, i.e., they are either both in 
emission or absorption. Toward source l30-304 the NH$_3$ spectrum exhibits an inverse 
P-Cygni profile and the component in absorption looks almost saturated as in the 
water line. Toward source l30-327 (and possibly l30-42), ammonia also shows absorption 
lines along the line of sight, as already observed in the water spectra.  The maximum 
velocity difference between  NH$_3$ and N$_2$H$^+$, when both lines  
are seen in emission, is $\simeq 0.9\,$km~s$^{-1}$ in source l30-42. \\
%

We have compared the line widths of the NH$_3$ and N$_2$H$^+$ lines, as shown in
Fig.~\ref{fig:Dv}. Except for source l30-304, which shows an inverse P-Cygni profile
in the ammonia spectrum, in all other sources we note that 
$\Delta V$(NH$_3$)$\geq \Delta V$(N$_2$H$^+$) including all four detected sources
in the $\ell=59\deg$ region. 
We also note that $\Delta V$(NH$_3$) tends to be somewhat higher in the $\ell=59\deg$ sources.
Since molecular outflows are the most distinctive feature
of the detected $\ell=59\deg$ sources, this is an indication that the line width of 
NH$_3$ is being broadened by these outflows. Likewise, the N$_2$H$^+$ emission must
then originate in a volume where the molecular gas is not significantly affected by 
the molecular outflows.

%
%
%
\begin{figure}
\centering
\includegraphics[width=6.0cm,angle=0]{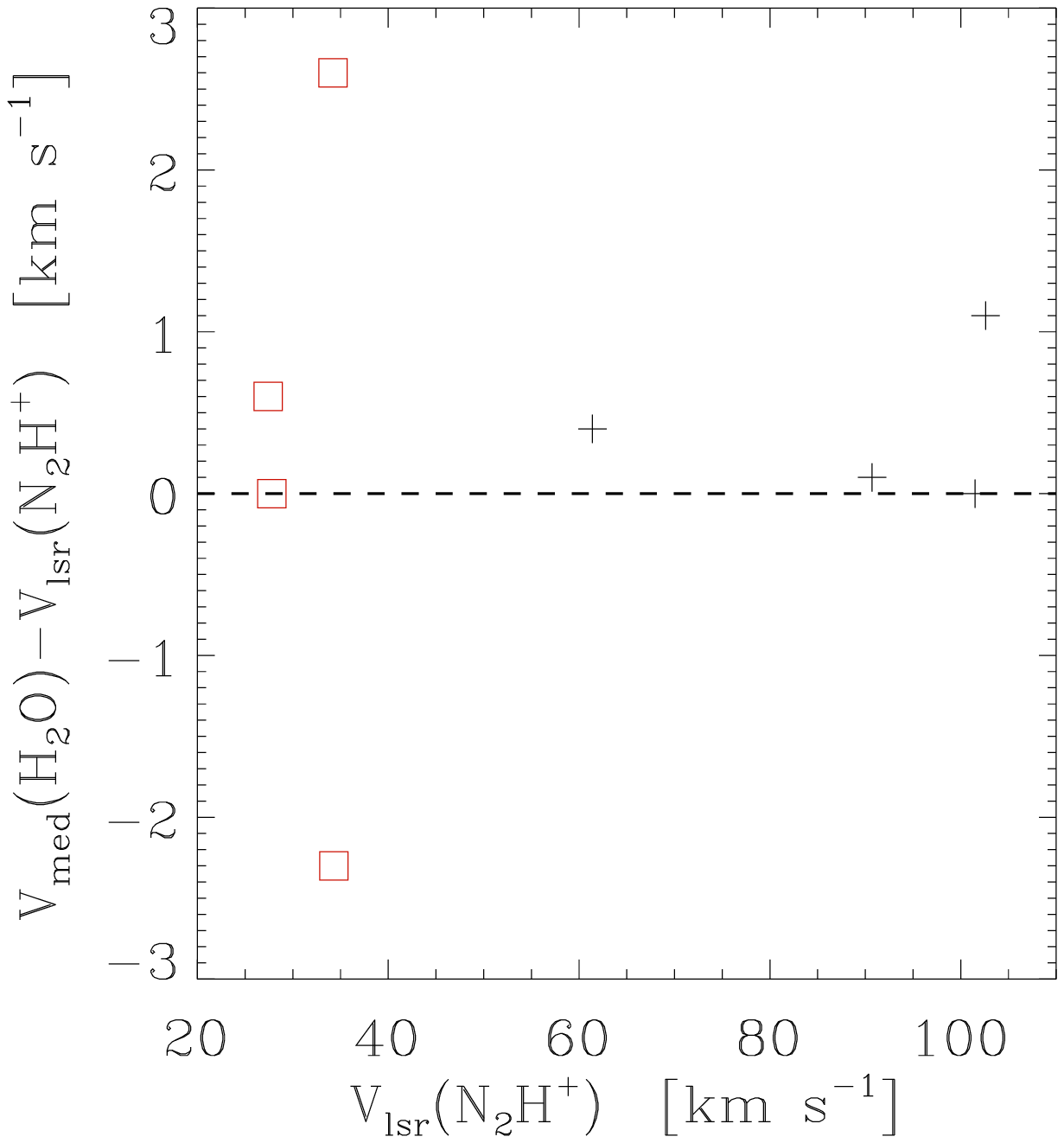}
\includegraphics[width=6.0cm,angle=0]{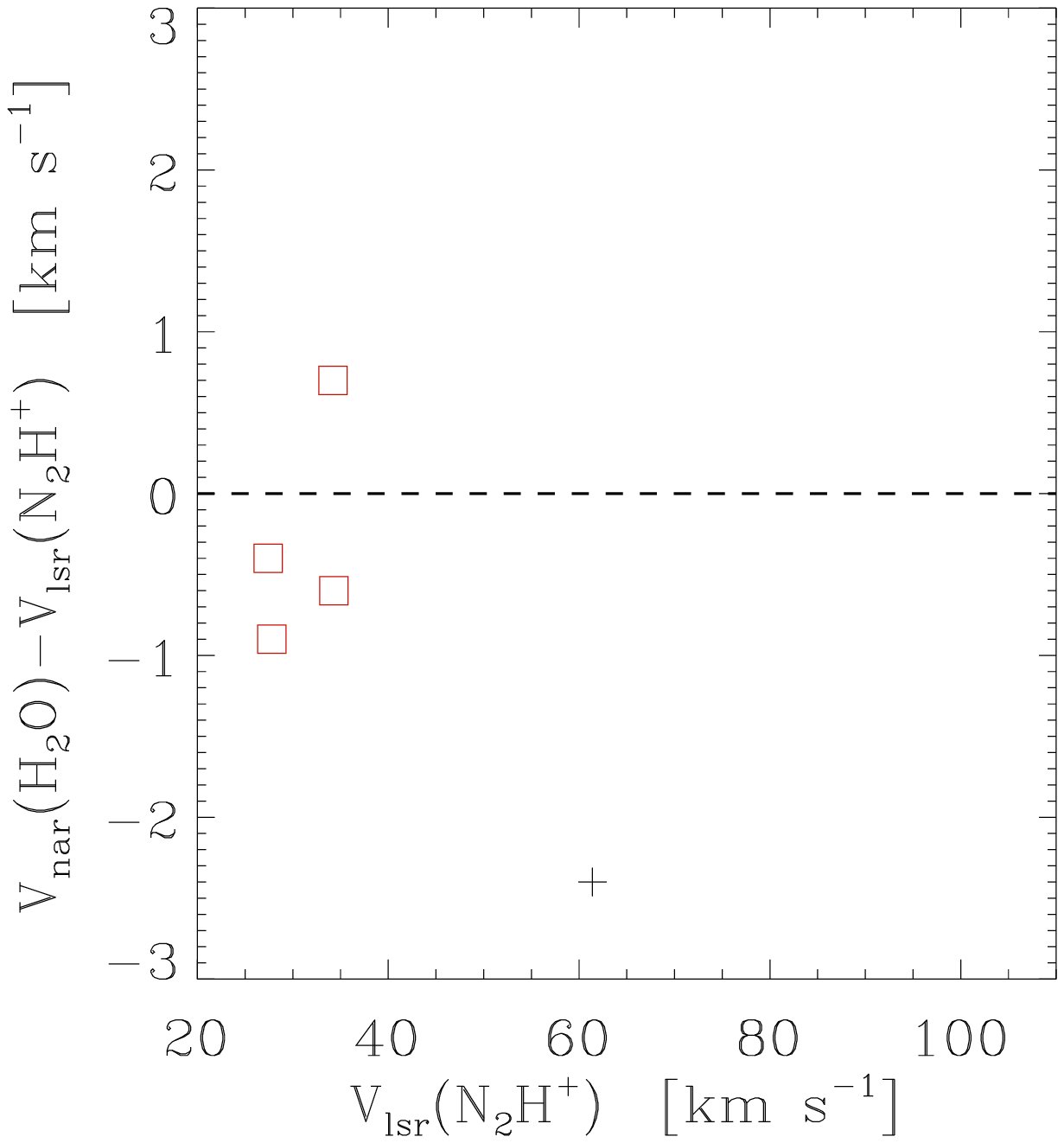}
\includegraphics[width=6.0cm,angle=0]{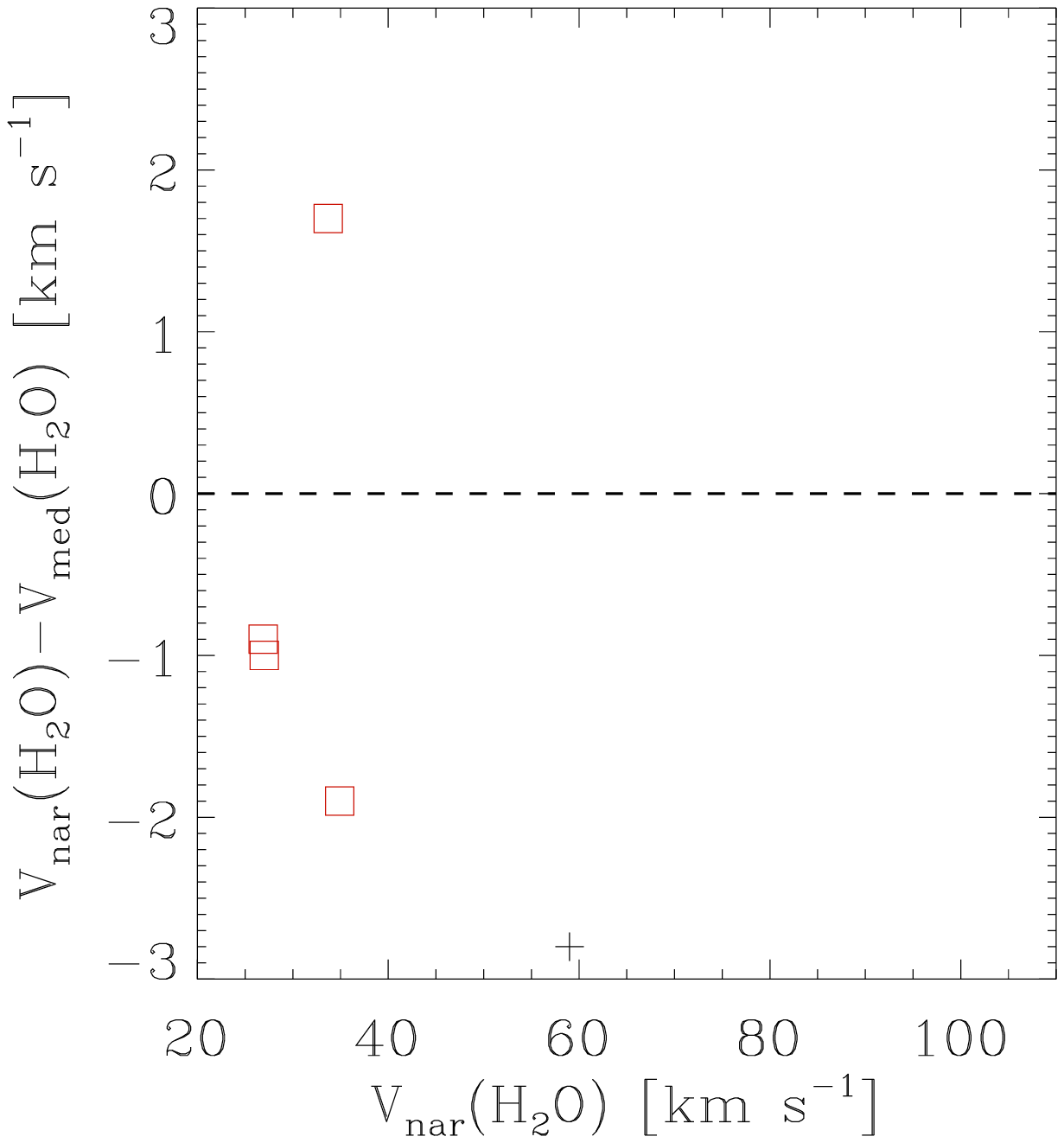}
\caption{
Central velocities of the narrow ($V_{\rm nar}$) and  medium ($V_{\rm med}$) Gaussian
components of the observed H$_2$O lines compared with each other and with the
systemic velocity from N$_2$H$^+$.
Symbols are as in Fig.~\ref{fig:LvsM}.
}
\label{fig:VelGaussComp}
\end{figure}

\subsubsection{Narrow component}
\label{sec:narrow}

The general appearance of the narrow component, mostly in absorption, indicates a high H$_2$O 
column density and a low excitation temperature, suggesting that the absorption
is being caused by the outer envelope and ambient cloud. Source l30-376 is the 
only one showing a clear narrow component in emission, which could be related to temperature, 
as appearance in emission requires an excitation temperature above the continuum level. 
We note, in fact, that l30-376 has the {\it lowest} continuum temperature 
(23\,mK, see Table~\ref{Table:sources}) among sources in the $\ell=30\deg$ region with a 
detectable continuum emission.
As mentioned earlier, in the other $\ell=30\deg$ sources the water line, either in absorption 
or emission, is broader than $\simeq 5\,$km\,s$^{-1}$ and has thus been classified as a 
medium component. 

As already noted by \citet{kristensen2012}, because the absorption of the narrow component 
is seen against both the outflow and continuum emission, the absorbing layer must be located 
in front of both the emitting layers, i.e., the outflows must also be embedded.
Figure~\ref{fig:VelGaussComp} shows that the narrow component is either centred at the source 
velocity (offset $\la 1\,$km\,s$^{-1}$) or blueshifted  (l30-512) relative to the systemic
velocity, as given by N$_2$H$^+$. In most $\ell=59\deg$ sources, however, the narrow
component is also slightly blushifted relative to N$_2$H$^+$ emission, suggesting that
the envelope, where the absorbed narrow component originates, is expanding.
In addition, Fig.~\ref{fig:VelGaussComp} shows that the narrow component is also 
blushifted relative to both the medium and broad (not shown) components.
The only source where $V_{\rm nar}$(H$_2$O) is red-shifted compared to both $V_{\rm med}$(H$_2$O)
and $V_{\rm broad}$(H$_2$O) is l59-444, which has an inverse P-Cygni profile, which is instead
typical of an infalling envelope.

Contrary to \citet{vandertak2013} we do not find any correlation between the source
mass (which is mostly due to the envelope contribution, being derived from the Hi-GAL observations
of cold dust emission) and the line width of the narrow component.
In Section~\ref{sec:radex} using {\tt{RADEX}} we have estimated the {\it o}-H$_2$O column density of the
envelope gas responsible for the observed absorption toward all $\ell=59\deg$ sources
and toward the $\ell=30\deg$ sources l30-304, l30-327  and l30-512 
(see also Table~\ref{Table:gaussian}).

\subsubsection{Outflows and shocked gas}
\label{sec:outflow}

The H$_2$O line profiles are characterized by two additional and broader components:
a broad  emission component, always in emission and visible only towards the $\ell=59\deg$ sources, 
and a medium absorption or emission component seen in almost all sources.
The broad and medium line components have widths exceeding the line width typical 
of the more quiescent envelope (e.g., \citealp{jorgensen2002}) and are thus 
likely related to a proto-stellar outflow.

In the $\ell=59\deg$ sources, both Gaussian components are required to fit the line profile.
The medium component has already been observed by \citet{kristensen2012},
toward a sample of low-mass proto-stellar objects, by \citet{vandertak2013} toward
a sample of high-mass star forming regions,
and also by \citet{herpin2012} toward the massive protostar W43-MM1.
Figure~\ref{fig:VelGaussComp} shows that the medium component in most cases is centred at the source
velocity (offset $\la 1\,$km\,s$^{-1}$), with $V_{\rm med}({\rm H_2O}) - V_{\rm lsr}({\rm N_2H^+})$
having positive values in almost all cases.
We already noted in the previous section that the narrow component is
blueshifted relative to the medium component. \citet{vandertak2013} found that
their medium component (which they call ``narrow outflow component'') usually appears in
absorption and is instead blueshifted relative to the narrow component  (which they call
``envelope component''). However, it should be noted that their decomposition of the
line profiles is based on higher-excitation water lines.

Both \citet{kristensen2012} and \citet{vandertak2013} associate the origin of the medium
component to the region, closer to the central protostar, where the gas is currently
being shocked, whereas they associate the broader component with the swept-up gas of
the molecular outflow. However, \citet{herpin2012} report that their
medium velocity component is likely due to a combination of turbulence and infall.
A more detailed analysis of the different velocity components in shocked molecular gas 
has been carried out, for example, by \citet{busquet2014} in the L1157-B1 shock region. 
They performed a multi-line H$_2$O survey 
probing a wide range of excitation conditions. In addition to the component in absorption
(probably due to the photodesorption of icy mantles at the edge of the cloud), two gas components, 
connected with a J-type shock associated with a jet and with different
excitation conditions, were found. The first component is a relatively extended ($\sim$ 10$\arcsec$ 
corresponding to $\sim 2\times 10^4$\,AU), 
warm ($\sim$ 200 K), and dense ($\sim$ 10$^{6}$ cm$^{-3}$)
region, and the second is a more compact (less than 5$\arcsec$ or $\sim 10^4$\,AU), hot ($\sim$ 1000 K) 
and tenuous ($\sim 10^3 - 10^4$\,cm$^{-3}$) component.  The water abundance is very high 
in the hot gas ($\sim 10^{-4}$ ) and it decreases down to $\sim 10^{-6}$ in the warm component.
Both warm and hot components have their peak at low-velocities, with the hot region dominating the 
total flux at the high end of the velocity range.  

In our work, and with only the data presented here, it is not possible to infer the exact
physical origin of the medium component of the H$_2$O$(1_{10}-1_{01})$ transition.
However, the correlation between the line widths of the broad and medium components
in the $\ell=59\deg$ sources, shown in the bottom panel of Fig.~\ref{fig:Dv}, suggests
an outflow or shock origin for the medium component as well.
We also note that \citet{kristensen2012} speculate that the phenomenon producing the
masers should also have a thermal component associated with it, which then produces the
medium component. We analyze the association with masers in Section~\ref{sec:maser}.
%

As far as sources in the $\ell=30\deg$ region are concerned, no broad component has been
detected. Observationally, this may be a consequence of the larger distances of
these sources compared to the $\ell=59\deg$ region. Alternatively, the viewing angle
of the molecular outflows and/or shocks may also determine whether a source shows
a broad, or both a broad and medium component. Because
of the small sample size of $\ell=30\deg$
sources, such an effect could be statistically significant, although in the
$\ell=59\deg$ region, with a comparable sample size, {\it all} sources show a broad
component. An alternative, more physical reason for the lack of a broad component
toward the $\ell=30\deg$ region could be a different evolutionary phase
(see Section~\ref{sec:evolution}).

As far as NH$_3$ and N$_2$H$^+$ are concerned, 
the line width of the ammonia line is likely affected by the presence of molecular
outflows. We note, in particular, that source l59-441 is the only one where the NH$_3$  profile
shows the same three components as the water line. While the water narrow component is generally
blueshifted relative to the medium and broad components, all ammonia components have remarkably
similar velocities. Three more sources show a NH$_3$ narrow and medium component 
(or two narrow components): l59-442, l59-444, and
l30-304. The two ammonia components have very similar velocities also in l59-442. In the other
two sources, l59-444, and l30-304, however, inverse P-Cygni profiles are observed and will
be discussed in the next Section.

%
%
\begin{figure}
\centering
\includegraphics[width=6.2cm,angle=0]{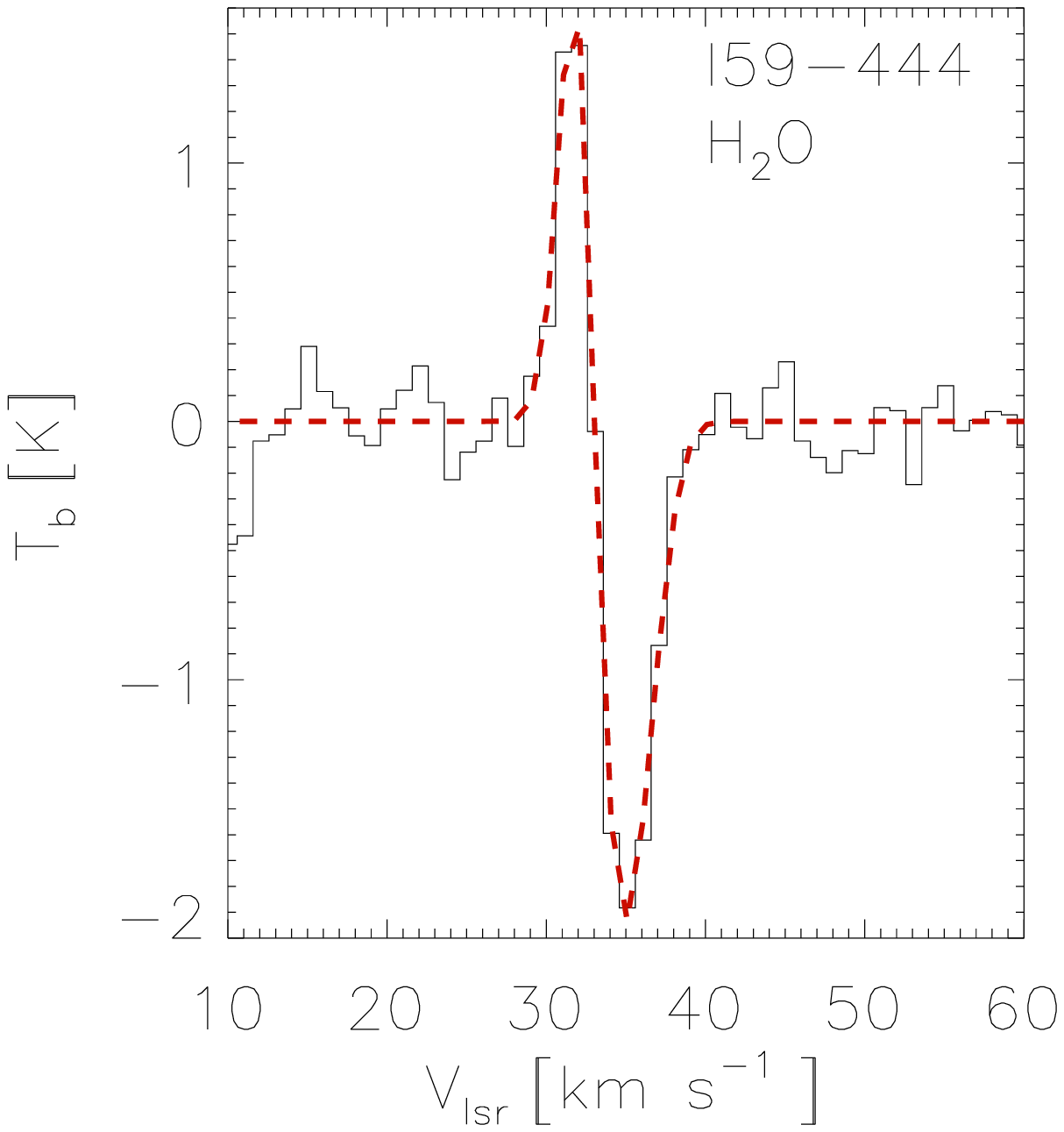}
\includegraphics[width=6.2cm,angle=0]{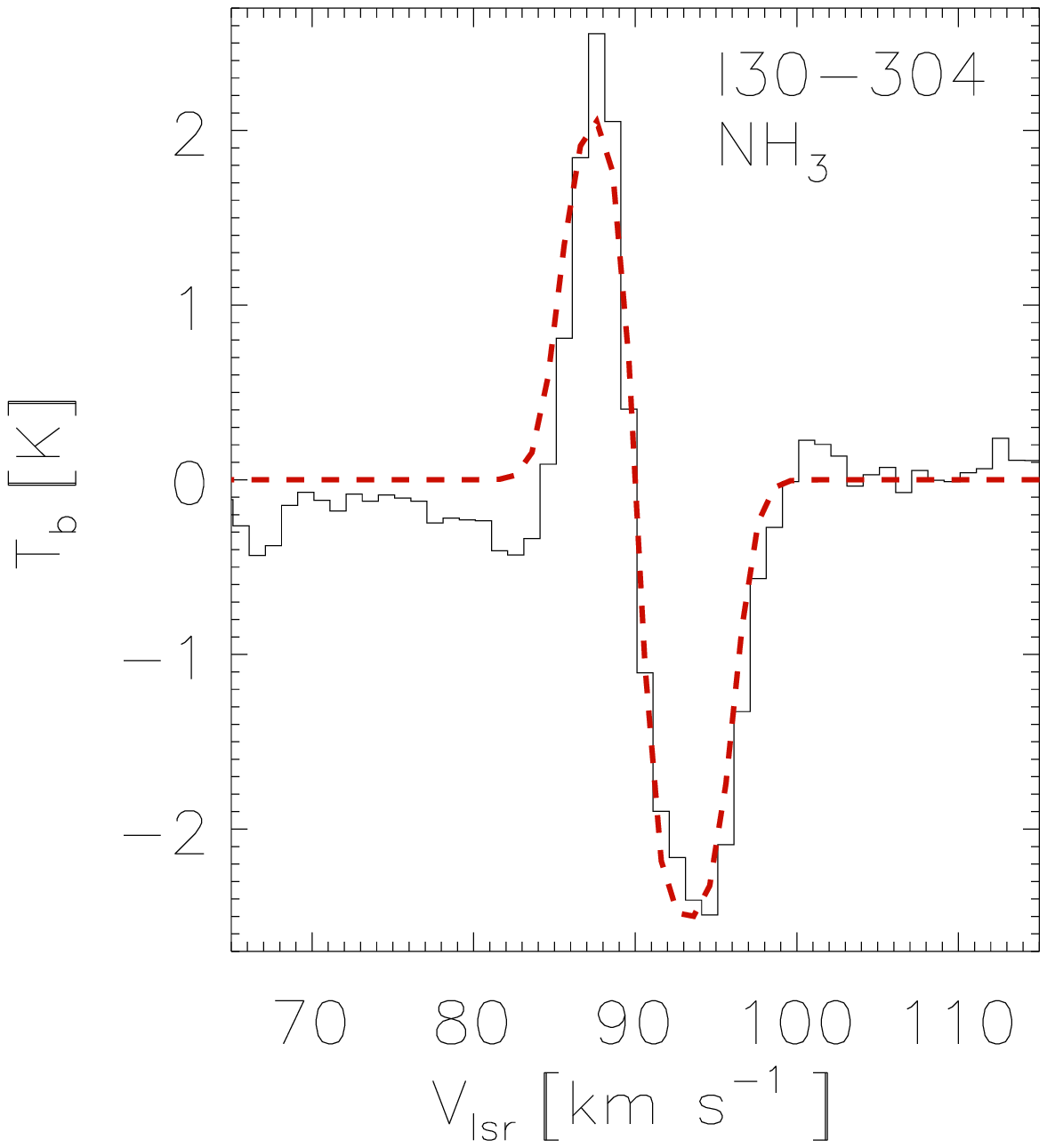}
\includegraphics[width=6.2cm,angle=0]{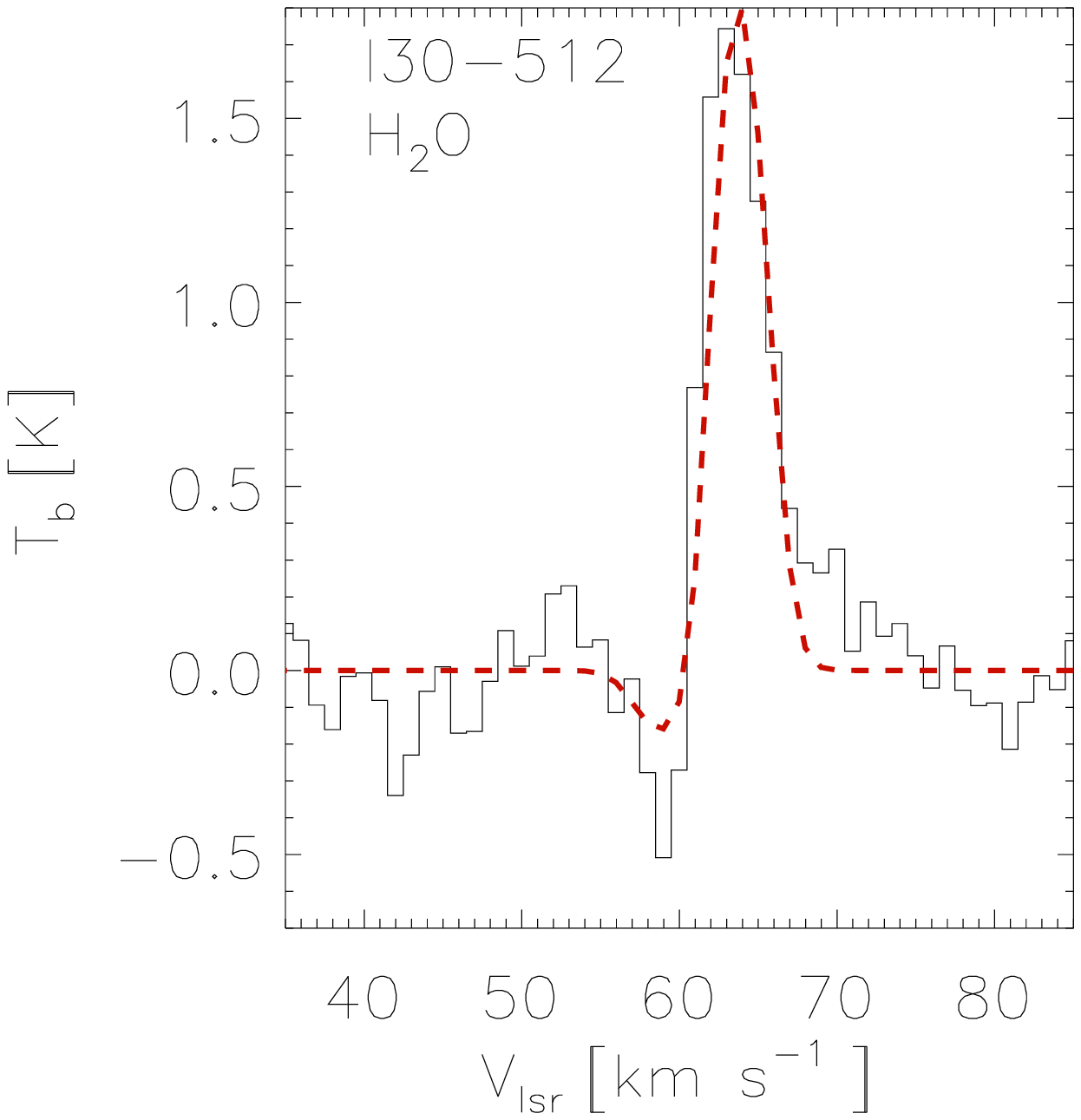}
\caption{
From top to bottom, inverse P-Cygni profiles detected toward sources l59-444 (H$_2$O)
and l30-304 (NH$_3$), and regular P-Cygni profile detected toward l30-512 (H$_2$O).
The infall (or expansion, in the case of l30-512) model is shown on top (red, dashed line).
The broad outflow component has been subtracted from the spectrum of source l59-444, whereas
a narrow absorption component has been added back to the spectrum of l30-512.
}
\label{fig:pcygni}
\end{figure}

%
%
\begin{table*}[\!htb]
\centering
\caption{Best-fit parameters of the infall/outflow modelling.  See text for a description
of the parameters used in the fit.}
\begin{tabular} {lccccccccc}
\hline\hline
\noalign{\smallskip}
Source  & Species  & $V_{\rm in}$   & $\sigma$        & $V_{\rm lsr}$\tablefootmark{a}   & $\tau_o$  & $\Phi$   & $T_c$\tablefootmark{a}  & $T_f$\tablefootmark{a}  & $T_r$  \\
\noalign{\smallskip}
        &          &[km~s$^{-1}$]  & [km~s$^{-1}$]  & [km~s$^{-1}$]                      &           &          & [K]                     & [K]                     & [K]    \\
\noalign{\smallskip}
\hline
\noalign{\smallskip}
l59-444  & H$_2$O  & 0.5  & 1.5  & 34.3  & 1.7  & 0.14  & 37.0  & 4.0  & 17.2  \\
l30-304  & NH$_3$  & 2.5  & 1.7  & 90.7  & 3.1  & 0.01  & 350.0 & 4.0  & 10.5  \\
l30-512\tablefootmark{b}  & H$_2$O  & -2.1 & 1.6  & 61.4  & 1.3  & 0.01  & 63.0  & 4.0  & 11.0\tablefootmark{a}  \\
\noalign{\smallskip}
\hline
\label{Table:pcygni}
\end{tabular}
\tablefoot{
\tablefoottext{a}{Fixed parameter, see text.}
\tablefoottext{b}{Tentative result.}
}
\end{table*}
%

\subsubsection{Regular and inverse P-Cygni profiles}
\label{sec:PCygni}


In a total of three cases, we have observed a regular or inverse P-Cygni profile, which are usually 
associated to envelope expansion and infall, respectively.  
An inverse P-Cygni profile is characterised by an absorption feature which is red-shifted relative 
to an emission feature, and is typically caused by a velocity and excitation gradient in the gas 
falling in toward a central object.  Two sources have inverse P-Cygni profiles, l30-304 and l59-444. 
In the case of source l59-444, the inverse P-Cygni profile is visible in the spectrum of water as 
superposed on the broad outflow component. In this source a red-shifted absorption is also visible
in the ammonia spectrum.
Toward source l30-304 the inverse P-Cygni profile is only detected in the NH$_3$ spectrum, whereas
the water spectrum shows multiple absorption lines from the diffuse interstellar gas, 
and at the source systemic velocity the water line
profile appears almost completely self-absorbed (Section~\ref{sec:absorptioncd}).
A recent example of  multiple ammonia lines, observed with {\it Herschel} 
toward the massive star forming region G34.3+0.15, 
showing inverse P-Cygni profiles can be found in Hajigholi et al. (submitted).
%
The only regular P-Cygni profile we have detected, toward source l30-512, is also superposed on 
what looks like an outflow/shock component. These three profiles are thus all different at some
level and suggest a great variety of kinematic and excitation conditions.

In order to give the simplest possible description of a contracting or expanding system that can 
reproduce the observed line profiles, \citet{difrancesco2001} used a simple model consisting 
of two slabs moving towards each other with a continuum source between them. This model was
a modification of a previous  ``two-layer'' code used by \citet{myers1996}. 
This model is appealing because of its simplicity, but on the other hand 
it gives only the crudest picture of the optical depth and excitation temperature of 
the absorbing and emitting gas. In addition, the model uses a relatively large number of 
free parameters (eight in total, if none is held fixed), including the
excitation temperature of the front/rear layers and the continuum source 
($T_f$, $T_r$, $T_c$) the filling factor of the absorbing source with respect to 
the telescope beam ($\Phi$), the peak optical depth of the front and rear layers 
($\tau_o$), the turbulent velocity dispersion in each layer ($\sigma$), the infall 
velocity ($V_{\rm in}$) and the systemic velocity ($V_{\rm lsr}$). 

Therefore, applying the simple two-slab model described above
to the two sources with inverse P-Cygni profiles, we obtain 
the best-fit infall models shown in Fig.~\ref{fig:pcygni}, with the corresponding best-fit parameters 
listed in Table~\ref{Table:pcygni}. In the case of source l59-444, the broad outflow 
component (and an additional narrow absorption feature, blue-shifted with respect 
to the inverse P-Cygni profile, see Table~\ref{Table:gaussian}) 
have been subtracted to the original spectrum before applying the model.
For both sources the value of $T_c$ has been fixed to that shown in Table~\ref{Table:sources}. 
We have also fixed the value of $V_{\rm lsr}$ to the velocity of N$_2$H$^{+}$ 
(see Table~\ref{Table:gaussian}). In section~\ref{sec:narrow} we suggested that the 
gas front layer may have a low excitation temperature and thus we fixed it to the approximate 
intermediate value ($T_f \simeq 4\,$K) shown by \citet{myers1996} in their Fig.1.
As a comparison Hajigholi et al. (submitted) find an excitation temperature of $\simeq 7-8\,$K
in the absorbing envelope of source G34.3+0.15.


By comparing different infall model profiles to the observations we find that $V_{\rm in}$
and $\sigma $ can vary by as much as $0.2 - 0.5\,$km\,s$^{-1}$, and are not very
sensitive to $T_f$. However, these tests do not take into account the uncertainty
on the beam-filling factor, used to convert the antenna temperature to line
brightness temperature (see Section~\ref{sec:radex}), which may also affect the 
best-fit results. Thus, our results show that while the infall velocity toward l59-444 
is comparable to the values found by \citet{difrancesco2001} and \citet{kristensen2012}
(who analyze low-mass protostars), the values of $\sigma$ (and the value of $V_{\rm in}$ 
toward l30-304) are somewhat higher, since these authors find 
$V_{\rm in}$, $\sigma \la 1\,$km\,s$^{-1}$. Our estimated values of $\sigma$ are consistent with those 
found by Hajigholi et al. (submitted) toward  G34.3+0.15.
Some of the high-mass protostars studied by
\citet{vandertak2013} also show regular and inverse P-Cygni profiles, but they do not 
analyze the line profiles and thus we cannot directly compare our results.
Instead, we can compare our results with those of \citet{herpin2012} 
toward the massive protostar W43-MM1, where they derive infall velocities as 
high as 2.9\,km\,s$^{-1}$ and turbulent velocities $\ga 2\,$km\,s$^{-1}$. These
values are much much more similar to our estimates of $\sigma$ and of the infall 
velocity toward l30-304. 

As far as the only source, l30-512, with a regular P-Cygni profile in our sample, 
this source shows a narrow absorption component at a velocity of 64\,km\,s$^{-1}$.
Therefore, assuming that this additional absorption feature is not related to the 
overall P-Cygni profile, and in order to reconstruct the hypothetical P-Cygni profile
we added back this component to the water spectrum. The two-layers model described
earlier can also be applied to the case of expanding gas, the only change in the model 
is that the infall velocity becomes negative. We show our best-fit result in the bottom
panel of Fig.~\ref{fig:pcygni}, obtained by fixing the value of $T_r$, which
would otherwise diverge. Clearly, the fit is not very good and it is difficult to determine
whether this may be due to contamination of the line profile, up to the level of possibly
resulting in a fake P-Cygni profile. For this reason our results in 
Table~\ref{Table:pcygni} are labelled as tentative.

While the infall and turbulent velocities are distance-independent and, at least
in this toy model, do not depend on the mass distribution within the source, determining
the mass infall rate would require the knowledge of the characteristic distance of the
infall zone or the mass undergoing infall. If the envelopes are in free fall, the velocity 
at a radius $R_{\rm in}$ in the envelope is given by $V_{\rm in} = \sqrt{2GM_{\rm enc}/R_{\rm in}}$
where $M_{\rm enc}$ is the mass enclosed within the radius $R_{\rm in}$. Assuming spherical infall
symmetry, the mass infall rate can be estimated as (e.g., \citealp{kristensen2012}): 
\beq
\dot{M} = 4 \pi \mu m_{\rm H} n_{\rm in} (2 G M_{\rm enc} )^2 V_{\rm in}^{-3}
\eeq
with $n_{\rm in}$ the density at the infall radius, $m_{\rm H}$ and $\mu = 2.8$ the hydrogen
atom mass and mean molecular weight, respectively. Therefore, in order to estimate 
$\dot{M}$ it is necessary to be able to determine either $M_{\rm enc}$ or 
the infall radius $R_{\rm in}$. We do not have a reliable distance determination for source
l59-444 (see Table~\ref{Table:sources}), and thus we cannot estimate any mass.
In the case of source l30-304 we do have a distance estimate and a mass, but 
the large mass and relatively large distance (see Table~\ref{Table:sources})
suggest that the estimated mass represents a much larger mass than that 
enclosed within the infalling region.
As an example, if we used a typical infall radius estimated by \citet{herpin2012} 
($R_{\rm in} \sim 6 \times 10^{17}\,$cm) then in source l30-304 we would get 
a mass infall rate $\dot{M} \sim 5\times 10^{-2}\,$M$_\odot$\,yr$^{-1}$, using 
as $n_{\rm in}$ the density given in Table~\ref{Table:sources} and $V_{\rm in}$
from  Table~\ref{Table:pcygni}. Not surprisingly, $\dot{M}$ is comparable 
to the values estimated by \citet{herpin2012} toward the massive proto-stellar
clump W43-MM1, and is much higher than the infall rates determined by 
\citet{kristensen2012} toward low-mass proto-stellar sources.


%
%

\subsection{Association with maser emission}
\label{sec:maser}

We searched for masers associated with the detected sources (within a $30\arcsec$ radius)
and found a total of six sources associated with maser emission: all four detected sources
in the $\ell=59\deg$ region and two more in the $\ell=30\deg$ field. Sources l59-441 and l59-445,
being very close, are actually associated with the same methanol maser.
All sources, except for l59-442, are associated with a methanol 6.7\,GHz maser.
Source l59-442 is the only one associated with a 22\,GHz water maser (B193852.5+235736,
\citealp{palla1991}), while source l30-304 is also associated with an OH maser.

The methanol masers associated with sources l59-441/l59-445 and l59-444 were found
by comparing our observations with those carried out by \citet{olmi2014b} toward
the $\ell=30\deg$ and $\ell=59\deg$ regions in search of 6.7\,GHz CH$_3$OH and 6.0\,GHz OH masers.
The masers associated with l59-441/l59-445 and l59-444 are G59.63$-$0.19 and G59.78+0.63,
respectively, and they have peak flux densities of 0.58 and 0.03\,Jy.
As discussed by \citet{olmi2014b}, these two methanol masers belong to a new class of
low-brightness masers that could represent an earlier stage of evolution.

The methanol masers have velocities which agree well with the velocities observed in
our detected sources (differences are $\la 1-2$\,km\,s$^{-1}$). However, the velocity
of the 22\,GHz water maser ($V_{\rm lsr} \simeq 50$\,km\,s$^{-1}$) associated with source
l59-442 significantly differs from the central velocity of any of the Gaussian components
of its H$_2$O line profile ($V_{\rm lsr} \simeq 35-37$\,km\,s$^{-1}$). If we assume that
the physical association between these two sources is indeed real, given the small angular
separation between them ($\simeq 4\,$arcsec), then one might speculate that the water maser
is originated somewhere along the path of the outflow. 
In fact, the broad Gaussian component in l59-442 has a line width that encompasses the
velocity of the water maser.

\subsection{Evolutionary trends }
\label{sec:evolution}

The relative intensities of the various components used to decompose the line profiles
of the water spectra have already been used in an attempt to identify possible
evolutionary sequences. For example, 
\citet{kristensen2012} associate the disappearance of the broad component with
later evolutionary stages in their low-mass protostars sample.
\citet{vandertak2013} then propose that the appearance of the narrow component
(or ``envelope'' component) in emission is an indicator of a warmer envelope, and thus
associated to later evolutionary stages. We note that while in all $\ell=59\deg$ 
sources the narrow component appears in absorption, toward the
$\ell=30\deg$ region sources l30-43 and l30-376 are characterized by a narrow component
in emission (see Table~\ref{Table:gaussian}; see also Section~\ref{sec:parspec} regarding 
the water detection toward l30-43).  In source l30-512 the narrow component is in
absorption, but this is a more complex source with a regular P-Cygni profile, 
as discussed in Section~\ref{sec:PCygni}.

An alternative method to analyze the evolutionary stages of our sources makes use of
the evolutionary tracks previously shown in Fig.~\ref{fig:LvsM}. These tracks were
determined following a method similar to that discussed by \citet{molinari2008},  
who proposed an evolutionary
sequence for high-mass protostars in terms of two parameters: the envelope mass and the
bolometric luminosity. These authors suggest an evolutionary sequence which is concentrated
in two main phases: protostars first accrete mass from their envelopes, and later disperse
their envelopes by winds and/or radiation. In Fig.~\ref{fig:LvsM} the calculated evolutionary
tracks thus rise upward almost vertically in luminosity 
during the accretion phase, and then proceed horizontally
to the left (to lower masses) during the envelope dispersal phase. 
The positions of our sources in this plot suggest
that some (but not all) of the sources with at least one line detection are, 
for similar ranges of mass, in a
more advanced stage of evolution. This is certainly the case for the detected $\ell=59\deg$ sources
with a well defined distance (l59-441 and l59-445), while for sources detected toward
the $\ell=30\deg$ region a preferential evolutionary phase is not apparent. None of these
sources, however, seem to belong to the late phase of envelope dispersal.

There are other indicators that can be used to analyze the relative evolutionary
phase of the $\ell=30\deg$ and $\ell=59\deg$ regions. For example, 
from the discussion about $T_{\rm c}$ in Section~\ref{sec:overview} and 
Table~\ref{tab:median}, there is a trend\footnote{We use the word ``trend'' because 
of the large scatter shown by the $T_{\rm c}$ values in Table~\ref{tab:median}.} 
of higher $T_{\rm c}$ toward the detected $\ell=30\deg$ sources,
suggesting they may be in a more advanced stage of proto-stellar evolution,
compared to sources in the $\ell=59\deg$ region.
Furthermore, as already noted in Section~\ref{sec:corr}, 
the typical water line luminosity toward the $\ell=30\deg$ sources is lower
compared to that of the $\ell=59\deg$ region. \citet{kristensen2012} find a similar
decrease in water emission between their Class 0 and Class I low-mass protostars.
Finally, as stated in Section~\ref{sec:maser} two of the $\ell=59\deg$ sources are
associated with low-brightness methanol masers that could indicate an
earlier phase of evolution. Source l30-304 is the only one 
associated with an OH maser, which is sometimes described as characterizing 
later evolutionary stages (\citealp{breen2010}) compared to 6.7\,GHz methanol masers.
Therefore, a firm conclusion about the specific evolutionary phase of the observed
sample is limited by the small number of sources. However, various indicators 
tentatively suggest that the detected sources toward the $\ell=30\deg$ region are 
in a later evolutionary phase compared to those in the $\ell=59\deg$ region.

\section{Summary and conclusions} 
\label{section:conclusions}

Using the HIFI instrument of {\it Herschel} we have observed the
{\it o}-NH$_3(1_0-0_0)$ (572\,GHz), {\it o}-H$_2$O$(1_{10}-1_{01})$ (557\,GHz) and
N$_2$H$^+(6-5)$ (559\,GHz) lines toward a sample of 52 high-mass starless and proto-stellar
clumps selected from the  Hi-GAL survey. The target sources are concentrated in
two galactic fields, at $\ell=30\deg$ and $\ell=59\deg$, and were observed 
during the Science Demonstration Phase of {\it Herschel}.
At least one of the three selected molecular lines was detected in 4 out of 35 sources
toward the $\ell=59\deg$ region, and 7 out of 17 objects toward the $\ell=30\deg$ region.

Almost all water spectra can be described as the sum of a narrow
(FWHM$ \sim 3-5$\,km\,s$^{-1}$), a medium (FWHM$ \sim 6-10$\,km\,s$^{-1}$) and a broad
(up to FWHM$\simeq 30$\,km\,s$^{-1}$) velocity component.
All detected sources of the $\ell=59\deg$ region have a strikingly similar water line-profile, 
characterized by the broad Gaussian component associated with the presence of molecular outflows,
which are also observed in the NH$_3$ line, at least toward one source. 
Therefore, in these sources there is strong evidence that water line emission is linked to the
presence of shocked molecular gas.

The situation is very different in the $\ell=30\deg$ region, with no two profiles 
being identical, and with no clear detection of a broad component in the water spectra.
We find that the continuum temperature, $T_{\rm c}$, is correlated with 
the source bolometric luminosity. We also find an indication that 
the water line luminosity increases with $T_{\rm c}$. The detected
sources in the $\ell=59\deg$ region have both a higher water line luminosity
and water abundance, compared to sources detected in the $\ell=30\deg$ region. 
Although we have few sources, this trend constitutes additional support
to a scenario where the water abundance is linked to shocks.

In terms of the gas kinematics, the general appearance in absorption of the narrow 
Gaussian component indicates a high H$_2$O column density and a low excitation 
temperature, suggesting that the absorption is being caused by the outer envelope 
and ambient cloud.  The correlation between the line widths of the broad and medium 
components in the $\ell=59\deg$ sources suggests an outflow or shock origin for the 
medium component. In addition, we also detect in a few sources 
inverse and regular P-Cygni profiles, though
we do not find any link with other source parameters. 
A firm conclusion about the specific evolutionary phase of the observed
sample is limited by the small number of sources. However, various indicators
tentatively suggest that the detected sources toward the $\ell=30\deg$ region are
in a later evolutionary phase compared to those in the $\ell=59\deg$ region.

\begin{acknowledgements}
HIFI has been designed and built by a consortium of institutes and university departments from across Europe, Canada and the USA under the leadership of SRON Netherlands Institute for Space Research, Groningen, The Netherlands and with major contributions from Germany, France and the USA. Consortium members are: Canada: CSA, U.Waterloo; France: CESR, LAB, LERMA, IRAM; Germany: KOSMA, MPIfR, MPS; Ireland, NUI Maynooth; Italy: ASI, IAPS-INAF, Osservatorio Astrofisico di Arcetri-INAF; Netherlands: SRON, TUD; Poland: CAMK, CBK; Spain: Observatorio Astronomico Nacional (IGN), Centro de Astrobiologia (CSIC-INTA). Sweden: Chalmers University of Technology - MC2, RSS \& GARD; Onsala Space Observatory; Swedish National Space Board, Stockholm University - Stockholm Observatory; Switzerland: ETH Zurich, FHNW; USA: Caltech, JPL, NHSC.
CMP  acknowledges generous support from the Swedish National Space Board.

\end{acknowledgements}

\bibliographystyle{aa}
\bibliography{refs}



\Online
\appendix

\section{Spectra of sources in the $\ell=30\deg$ region}
\label{sect:app1}

All spectra are in $T_\mathrm{A}^*$ units and have been smoothed to a spectral resolution of 1~km\,s$^{-1}$,
unless noted otherwise.


\clearpage

%
\begin{figure}
\begin{center}
\includegraphics[scale=0.62]{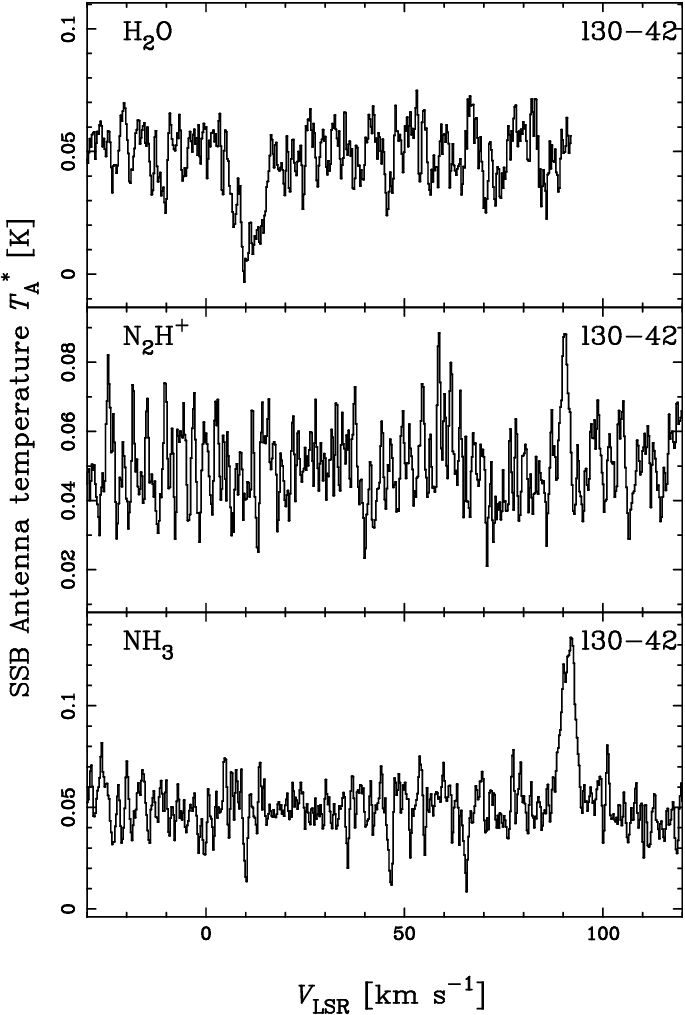}
\caption{Original channel width (0.27 km/s).}   
\end{center}
\end{figure} 


%
\begin{figure}
\begin{center}
\includegraphics[scale=0.62]{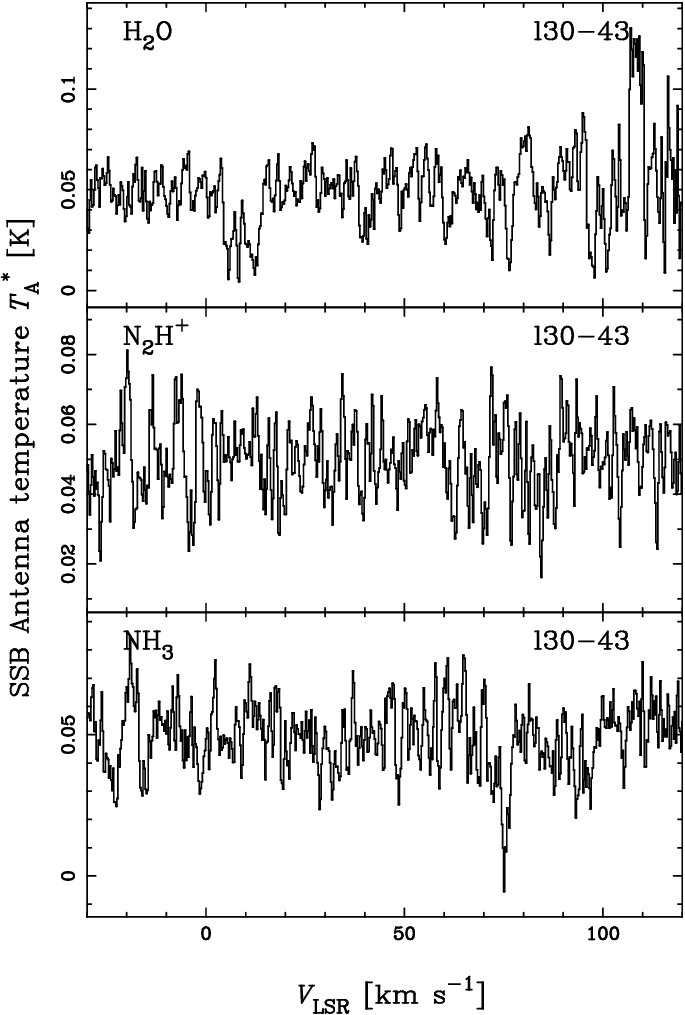}
\caption{Original channel width (0.27 km/s).}   
\end{center}
\end{figure} 

       
%
\begin{figure}
\begin{center}
\includegraphics[scale=0.62]{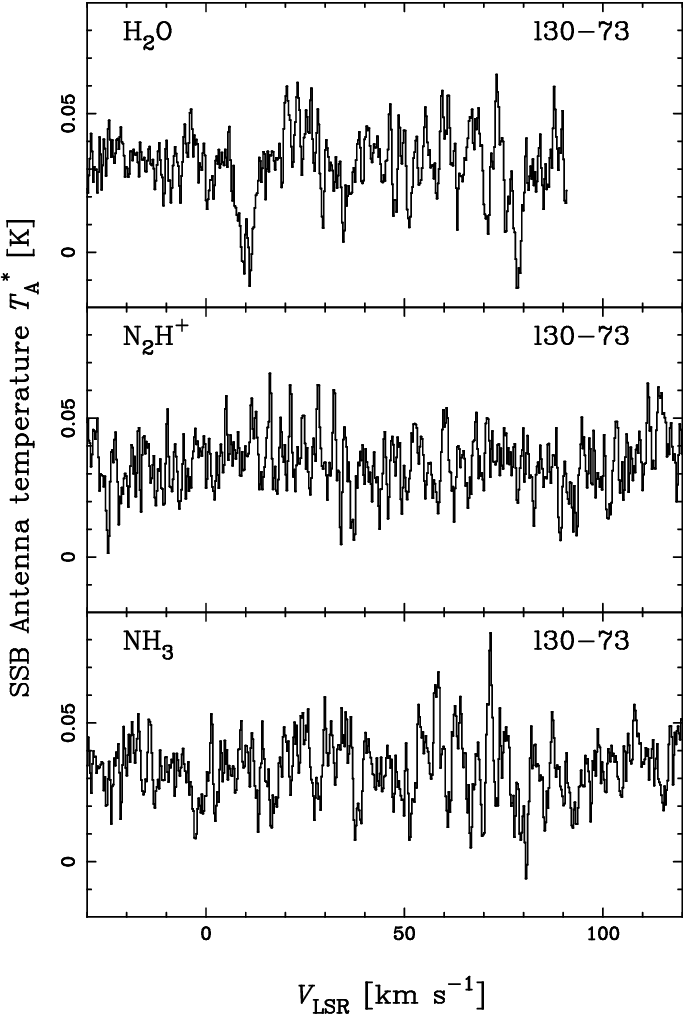}
\caption{Original channel width (0.27 km/s).}   
\end{center}
\end{figure} 
     

%
\begin{figure}
\begin{center}
\includegraphics[scale=0.62]{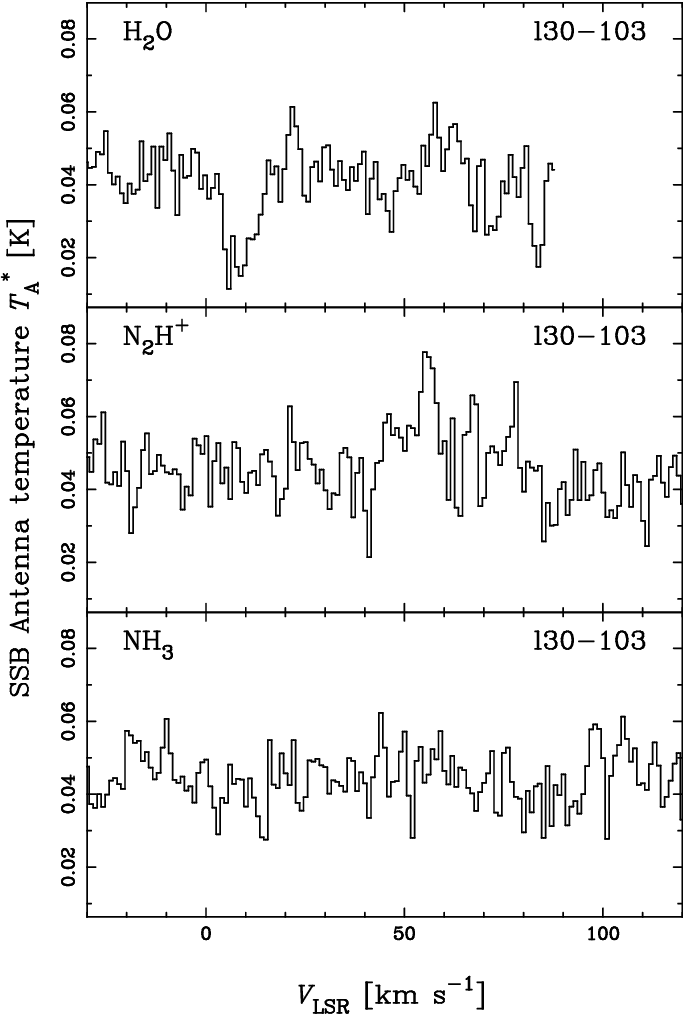}
\caption{}   
\end{center}
\end{figure} 


\clearpage

%
\begin{figure}
\begin{center}
\includegraphics[scale=0.62]{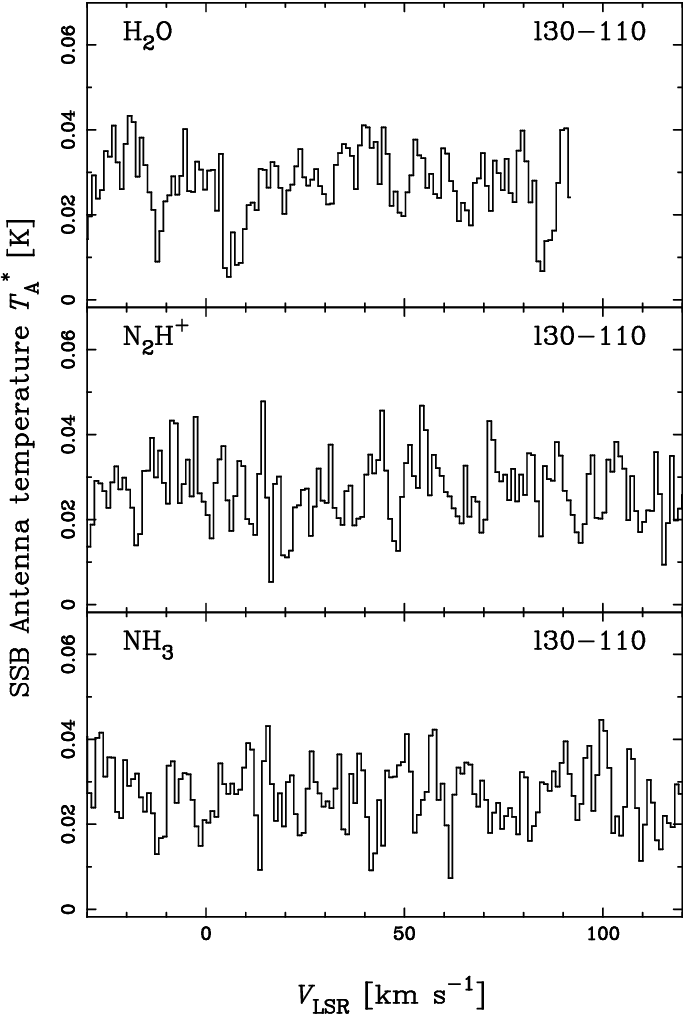}
\caption{}   
\end{center}
\end{figure}


%
\begin{figure}
\begin{center}
\includegraphics[scale=0.62]{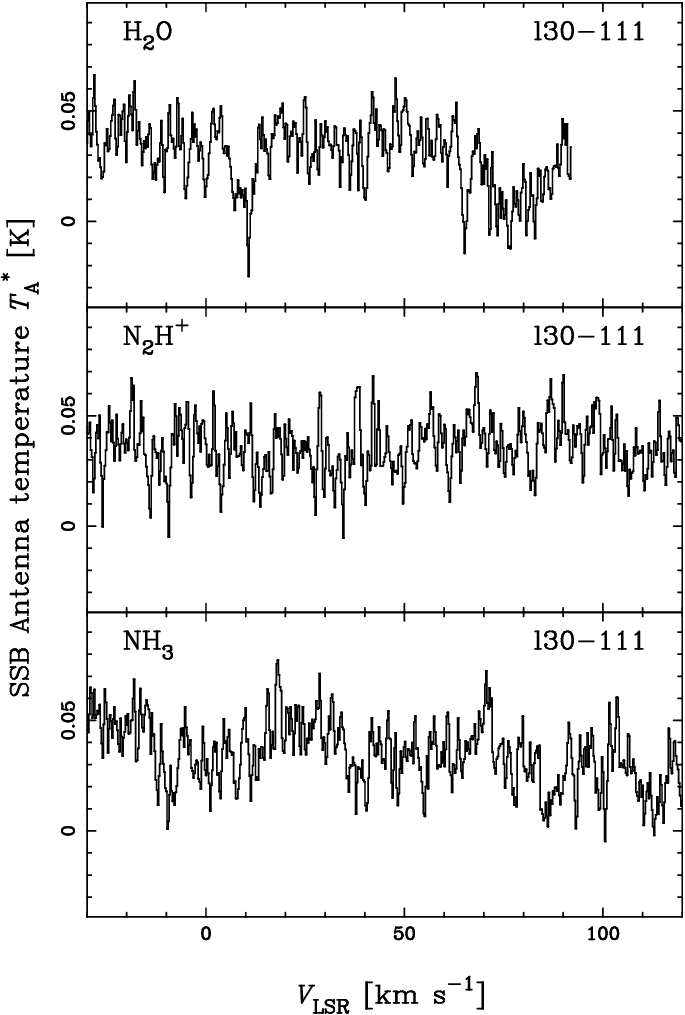}
\caption{Original channel width (0.27 km/s).}   
\end{center}
\end{figure}



%
\begin{figure}
\begin{center}
\includegraphics[scale=0.62]{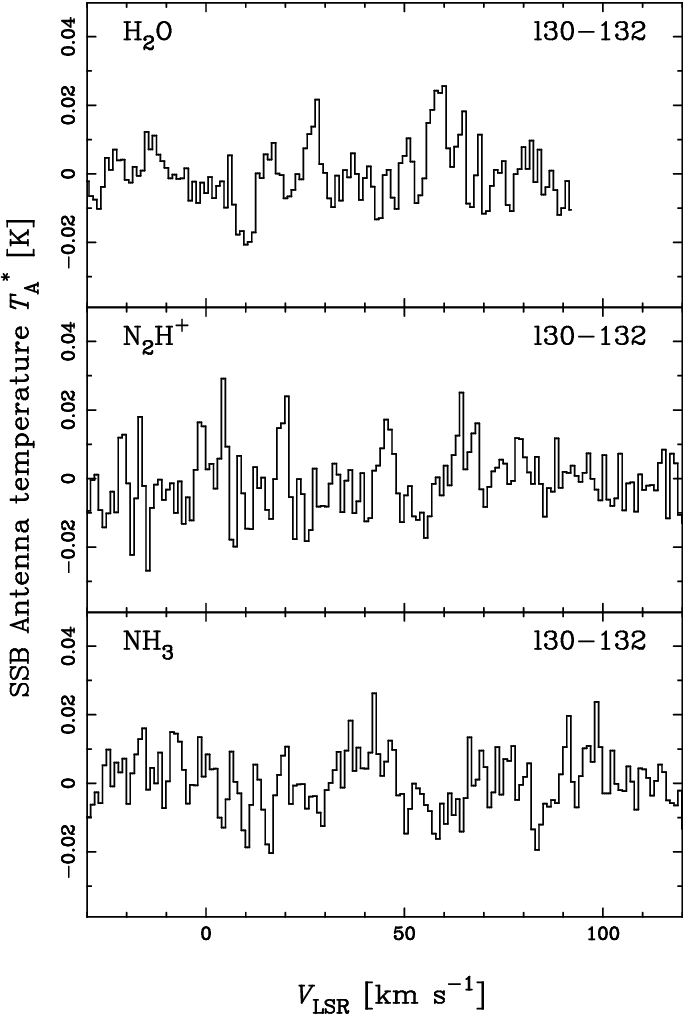}
\caption{}   
\end{center}
\end{figure}


%
\begin{figure}
\begin{center}
\includegraphics[scale=0.62]{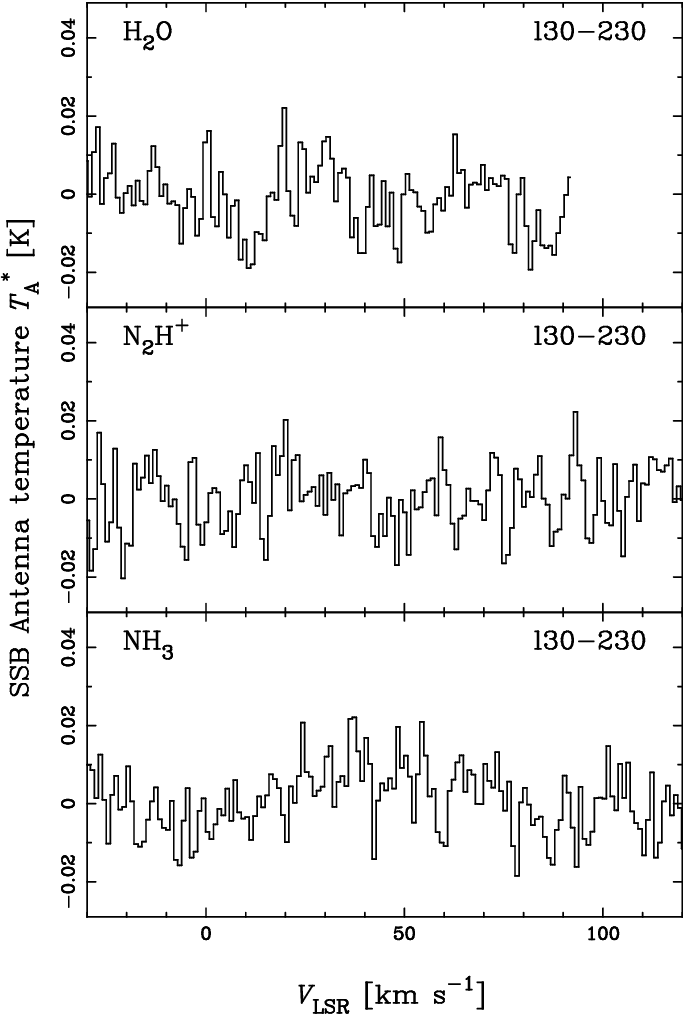}
\caption{}   
\end{center}
\end{figure}


\clearpage

%
\begin{figure}
\begin{center}
\includegraphics[scale=0.62]{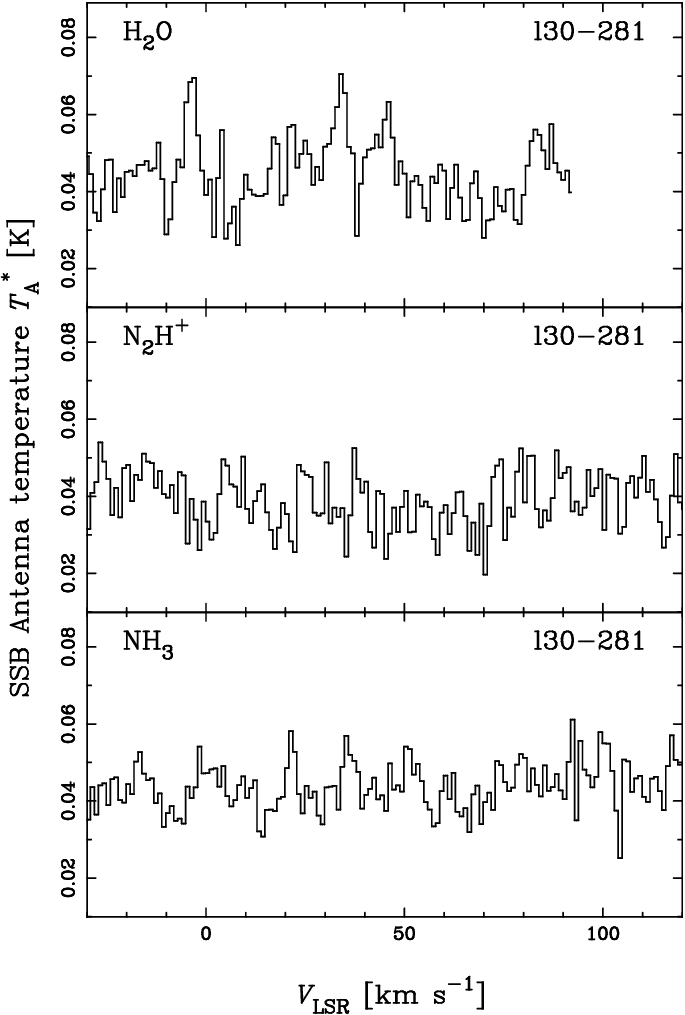}
\caption{}   
\end{center}
\end{figure}


%
\begin{figure}
\begin{center}
\includegraphics[scale=0.62]{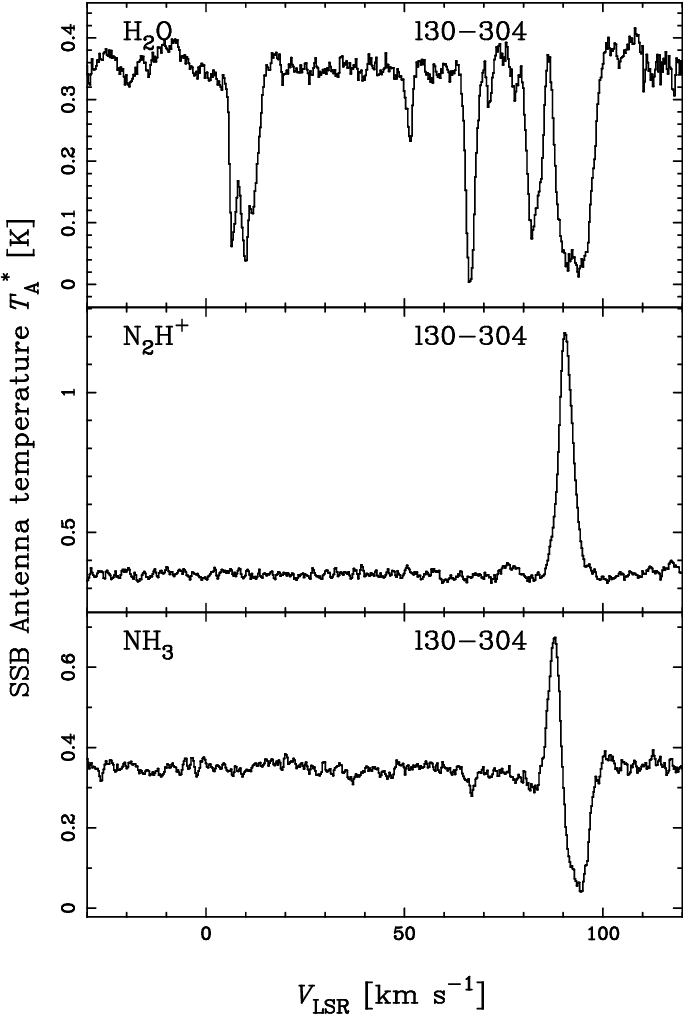}
\caption{Original channel width (0.27 km/s).}   
\end{center}
\end{figure}



%
\begin{figure}
\begin{center}
\includegraphics[scale=0.62]{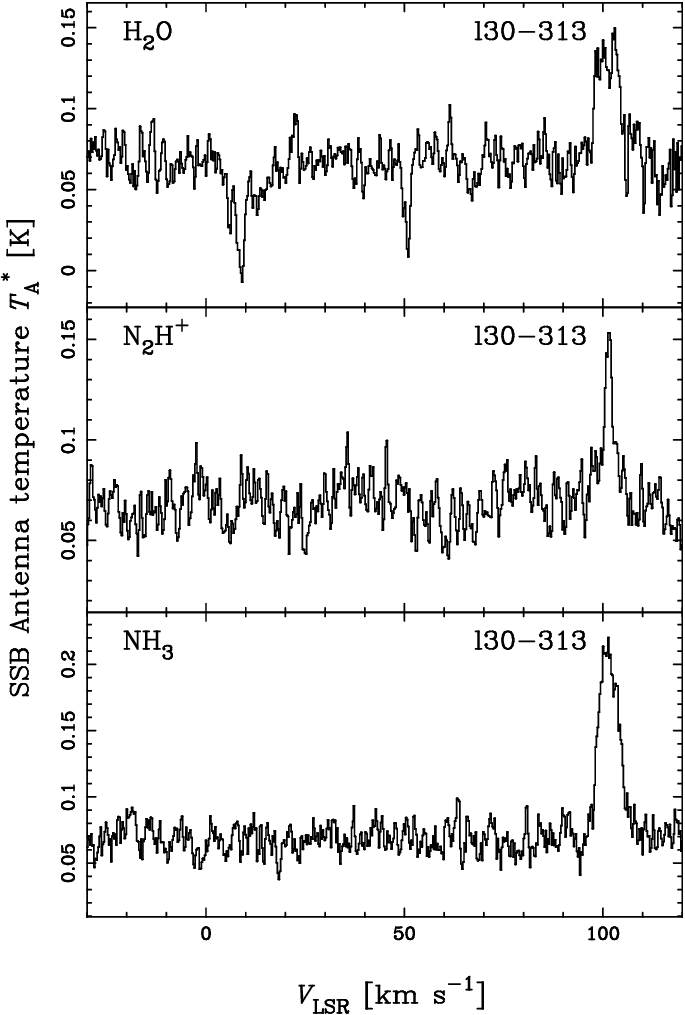}
\caption{Original channel width (0.27 km/s).}   
\end{center}
\end{figure}
       
 
%
\begin{figure}
\begin{center}
\includegraphics[scale=0.62]{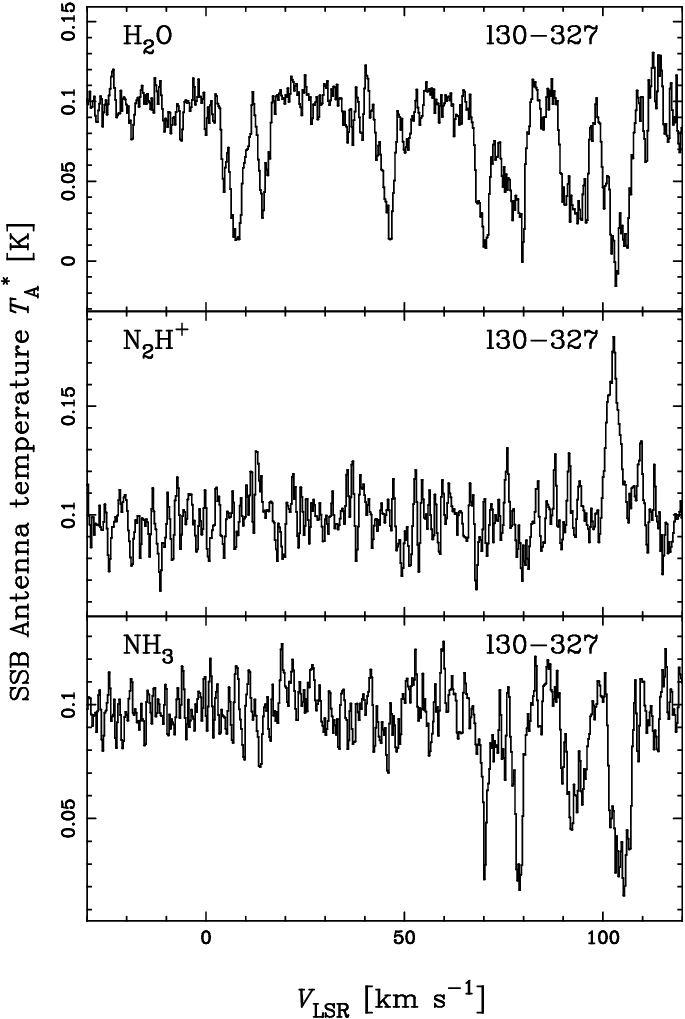}
\caption{Original channel width (0.27 km/s).}   
\end{center}
\end{figure}


%
\begin{figure}
\begin{center}
\includegraphics[scale=0.62]{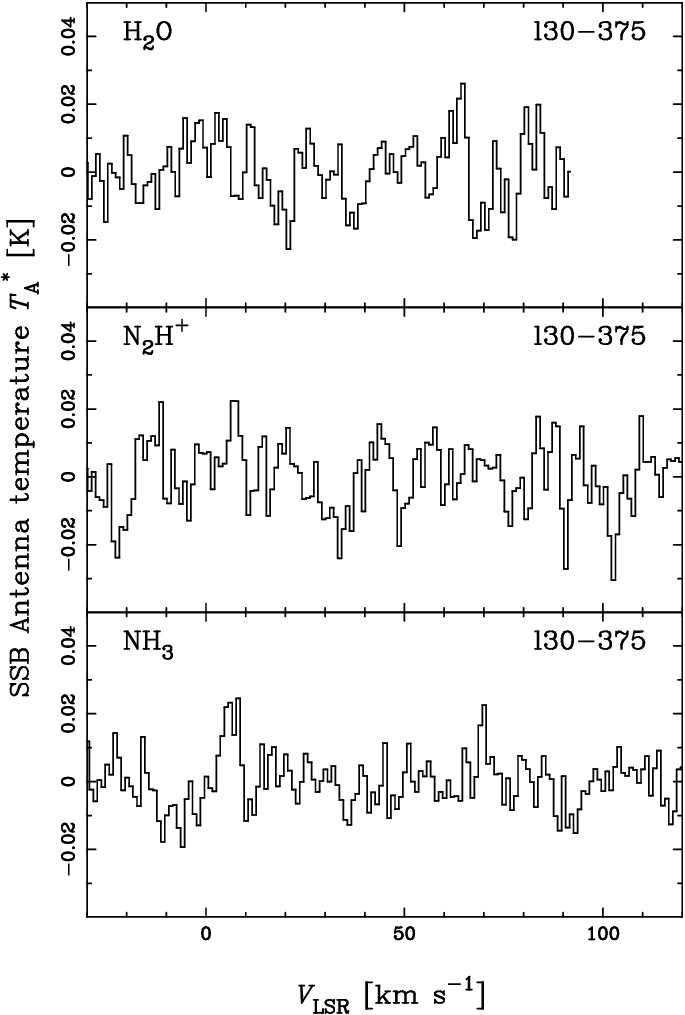}
\caption{}   
\end{center}
\end{figure}
      
      
%
\begin{figure}
\begin{center}
\includegraphics[scale=0.62]{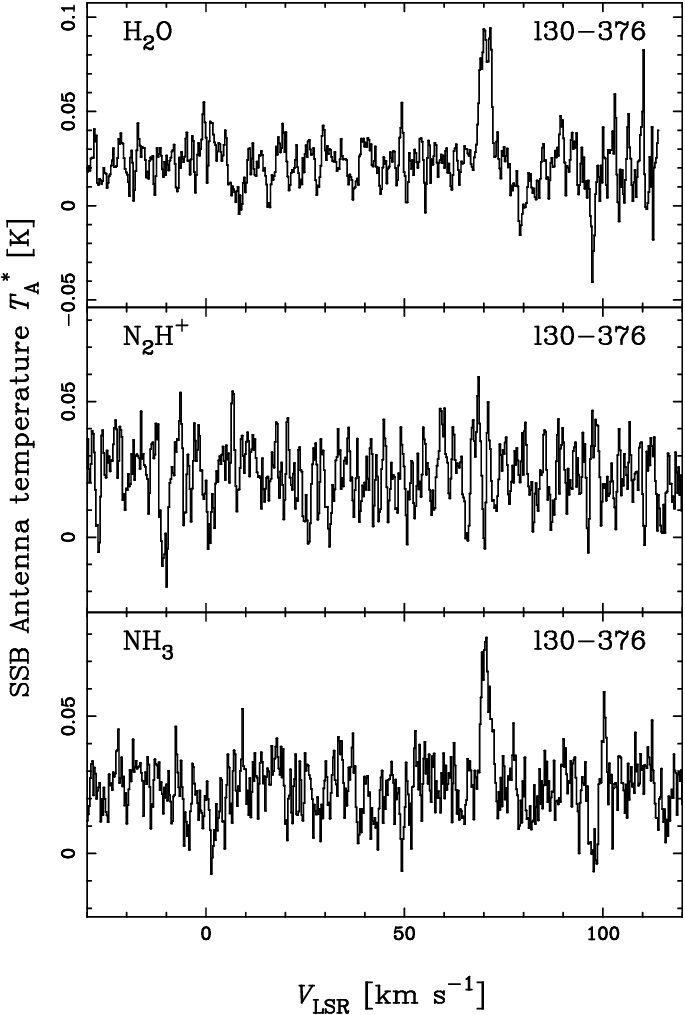}
\caption{Original channel width (0.27 km/s).}   
\end{center}
\end{figure} 


%
\begin{figure}
\begin{center}
\includegraphics[scale=0.62]{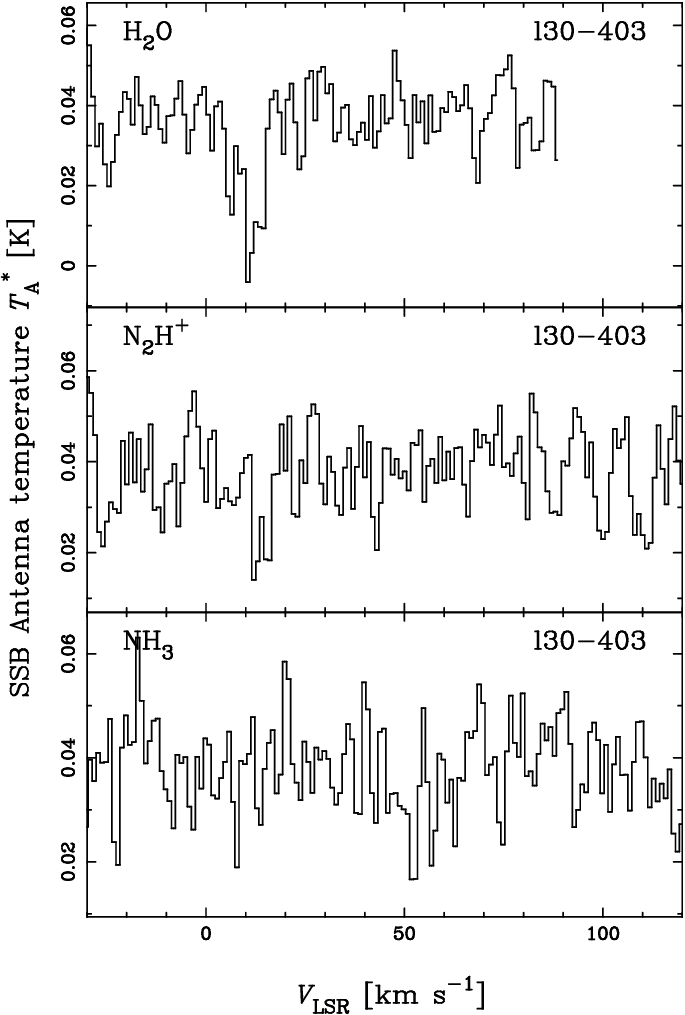}
\caption{}   
\end{center}
\end{figure}
 

%
\begin{figure}
\begin{center}
\includegraphics[scale=0.62]{Figures/l30_432_h2o-n2hp-nh3_1kms.eps}
\caption{}   
\end{center}
\end{figure}


%
\begin{figure}
\begin{center}
\includegraphics[scale=0.62]{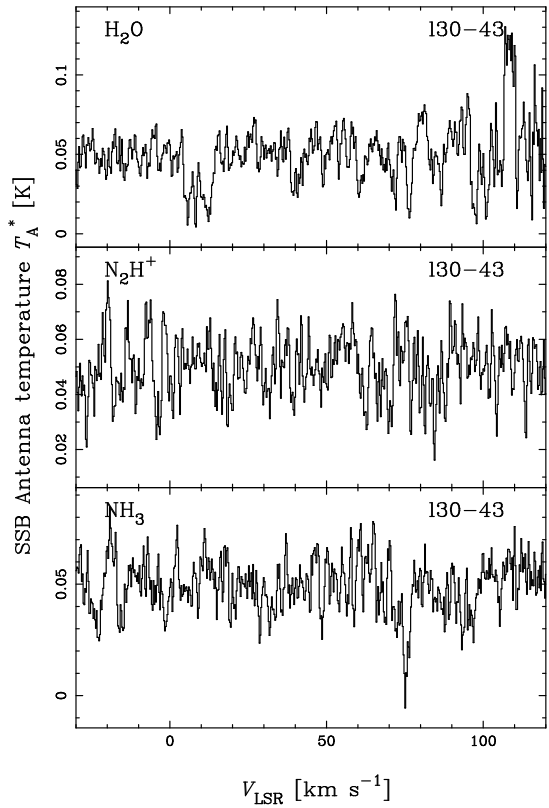}
\caption{Original channel width (0.27 km/s).}   
\end{center}
\end{figure} 



\clearpage

\section{Spectra of sources in the $\ell=59\deg$ region}    
\label{sect:app2}

All spectra are in $T_\mathrm{A}^*$ units and have been smoothed to a spectral resolution of 1~km\,s$^{-1}$,
unless noted otherwise.

\clearpage

\begin{figure}
\begin{center}
\includegraphics[scale=0.62]{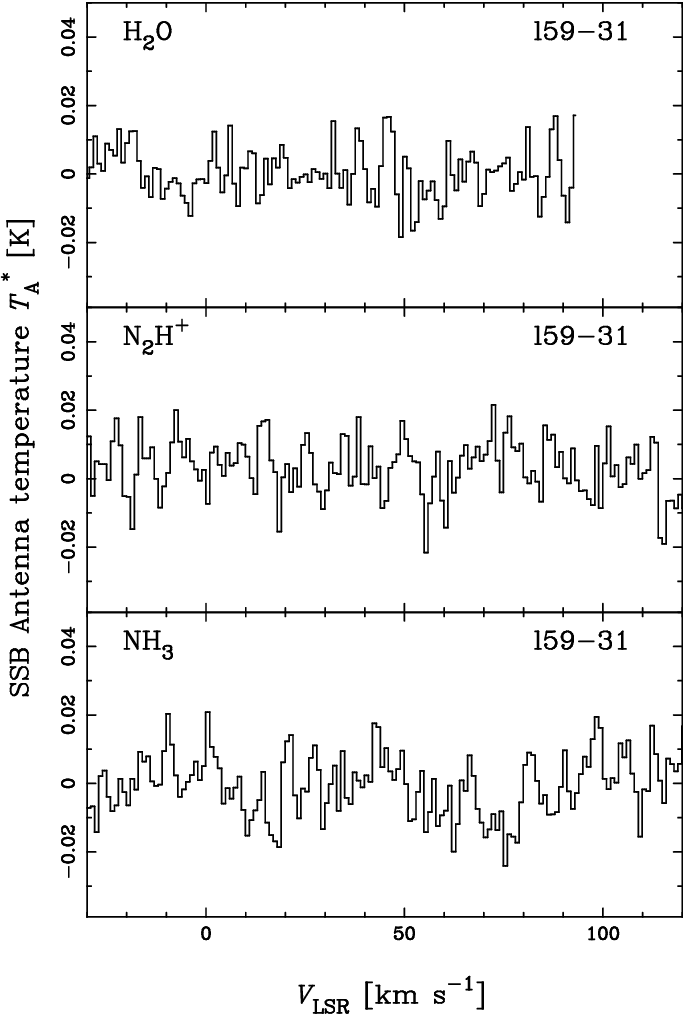}
\caption{}   
\end{center}
\end{figure}

\begin{figure}
\begin{center}
\includegraphics[scale=0.62]{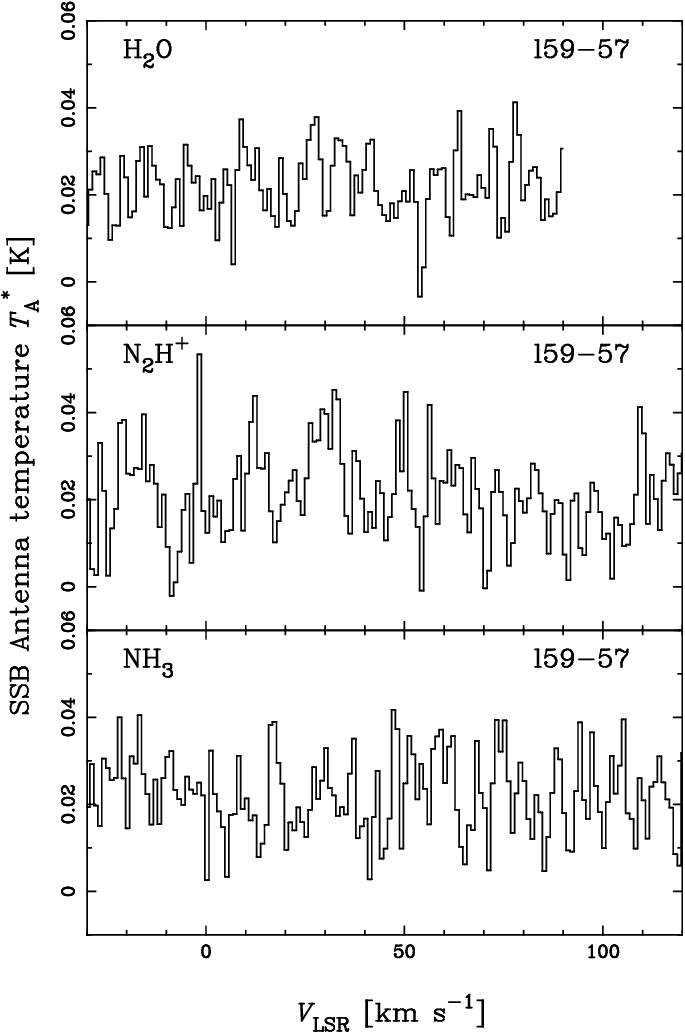}
\caption{}   
\end{center}
\end{figure}

\begin{figure}
\begin{center}
\includegraphics[scale=0.62]{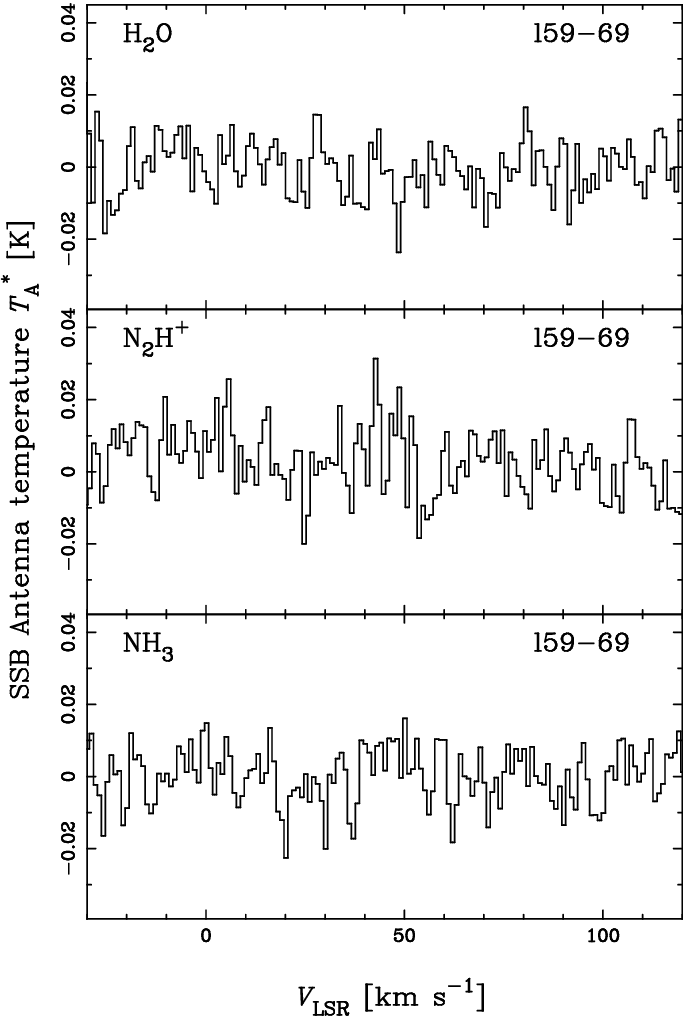}
\caption{}   
\end{center}
\end{figure}

\begin{figure}
\begin{center}
\includegraphics[scale=0.62]{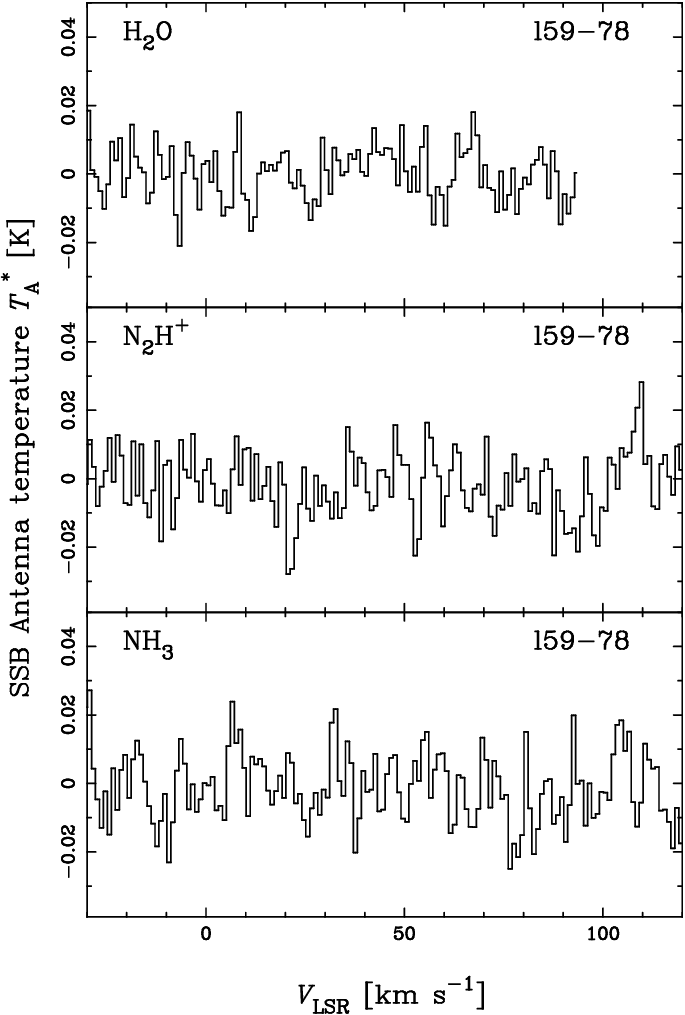}
\caption{}   
\end{center}
\end{figure}

\clearpage

\begin{figure}
\begin{center}
\includegraphics[scale=0.62]{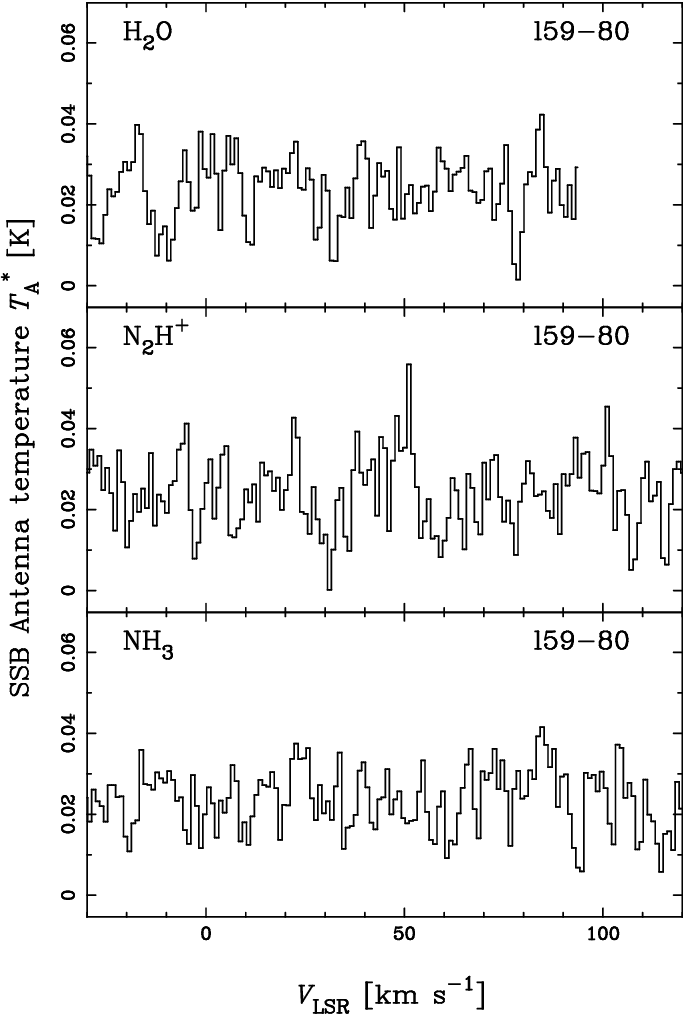}
\caption{}   
\end{center}
\end{figure}

\begin{figure}
\begin{center}
\includegraphics[scale=0.62]{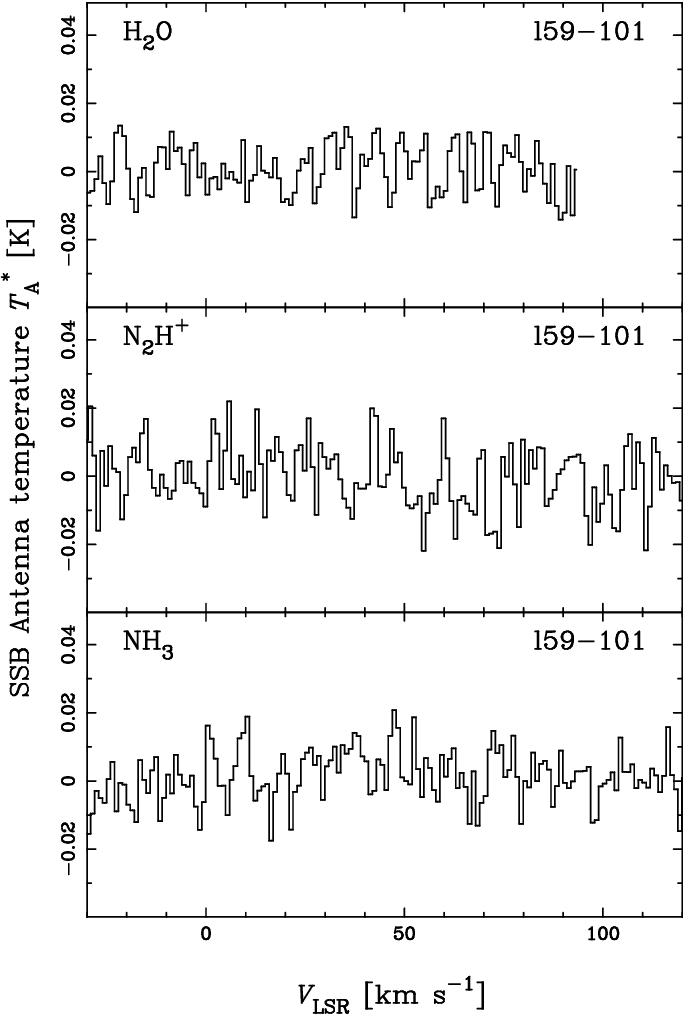}
\caption{}   
\end{center}
\end{figure}

\begin{figure}
\begin{center}
\includegraphics[scale=0.62]{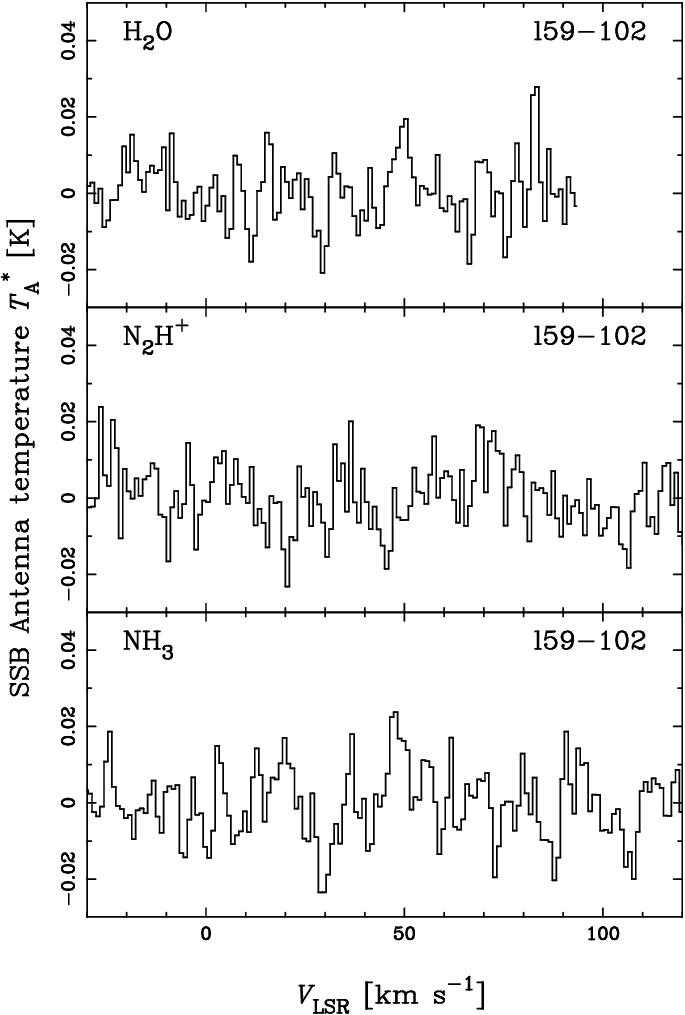} 
\caption{}   
\end{center}
\end{figure}

\begin{figure}
\begin{center}
\includegraphics[scale=0.62]{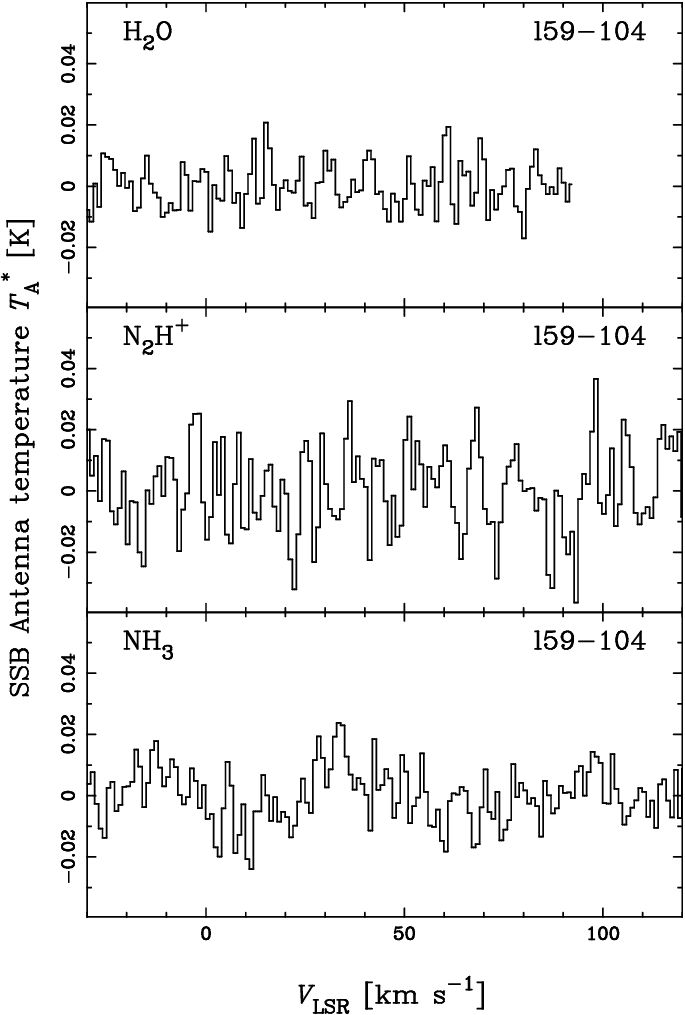}
\caption{}
\end{center}
\end{figure}

\clearpage


\begin{figure}
\begin{center}
\includegraphics[scale=0.62]{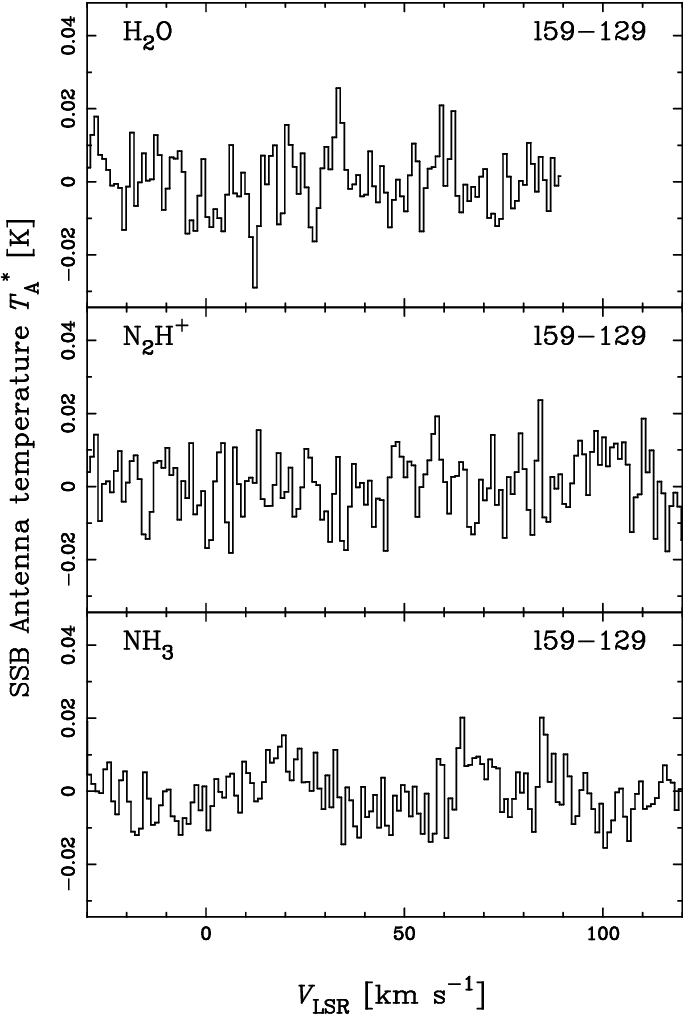} 
\caption{}   
\end{center}
\end{figure}
      

\begin{figure}
\begin{center}
\includegraphics[scale=0.62]{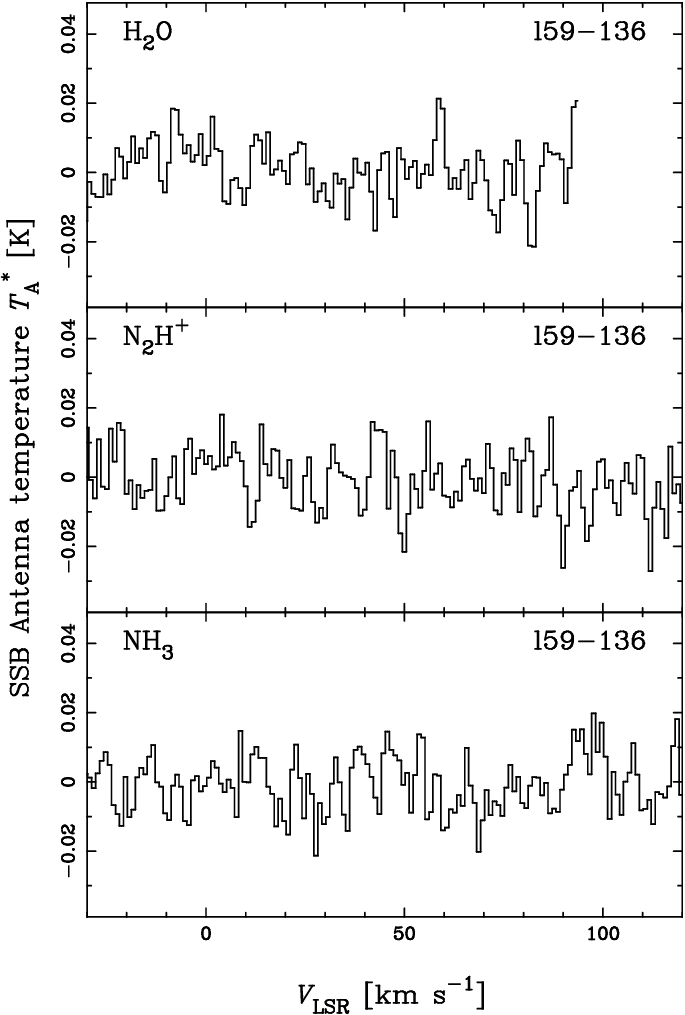}
\caption{}   
\end{center}
\end{figure}

\begin{figure}
\begin{center}
\includegraphics[scale=0.62]{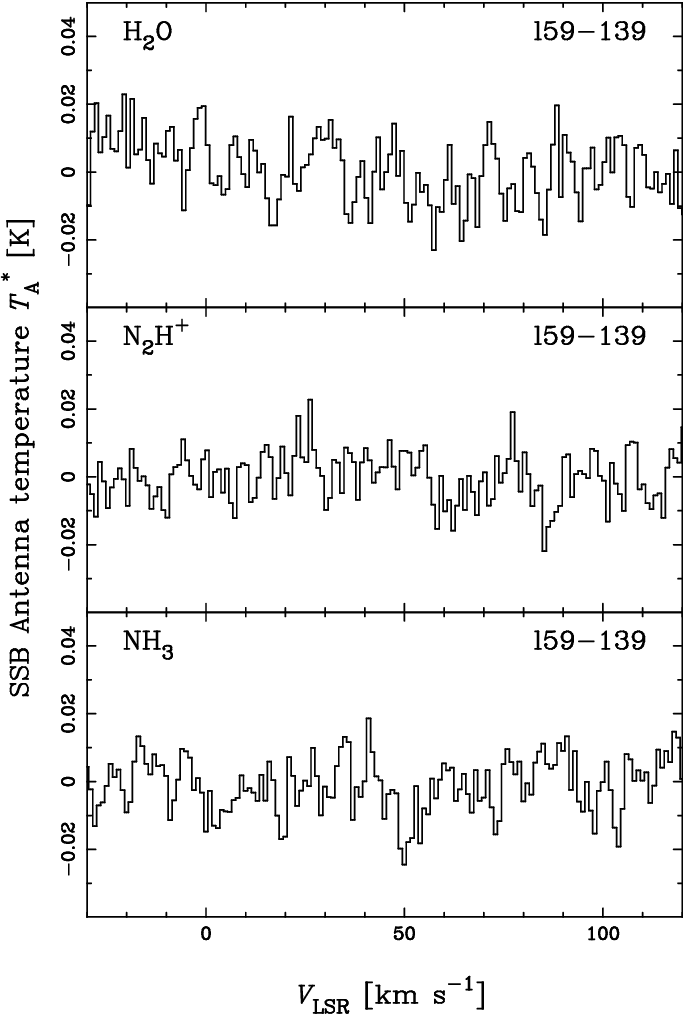}
\caption{}   
\end{center}
\end{figure}

\begin{figure}
\begin{center}
\includegraphics[scale=0.62]{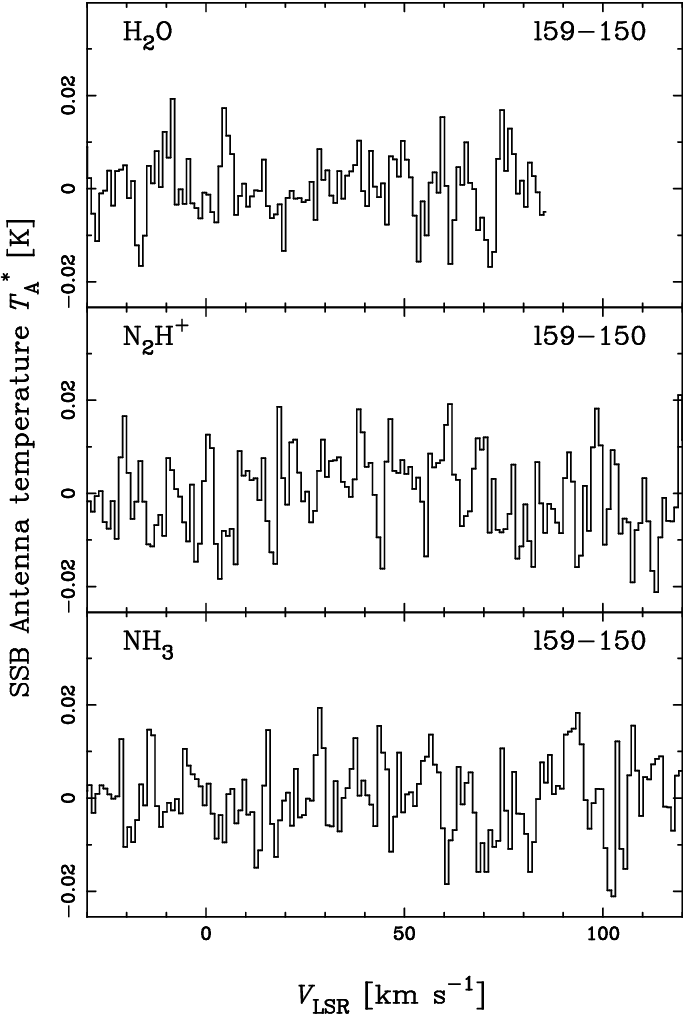}
\caption{}   
\end{center}
\end{figure}

\clearpage

\begin{figure}
\begin{center}
\includegraphics[scale=0.62]{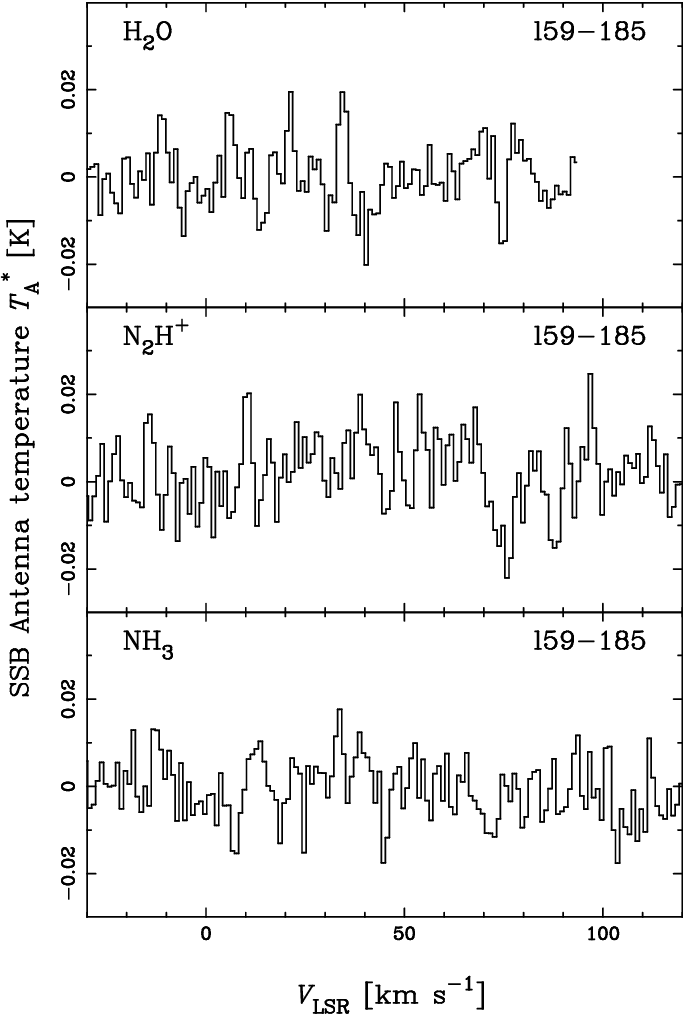}
\caption{}   
\end{center}
\end{figure}


\begin{figure}
\begin{center}
\includegraphics[scale=0.62]{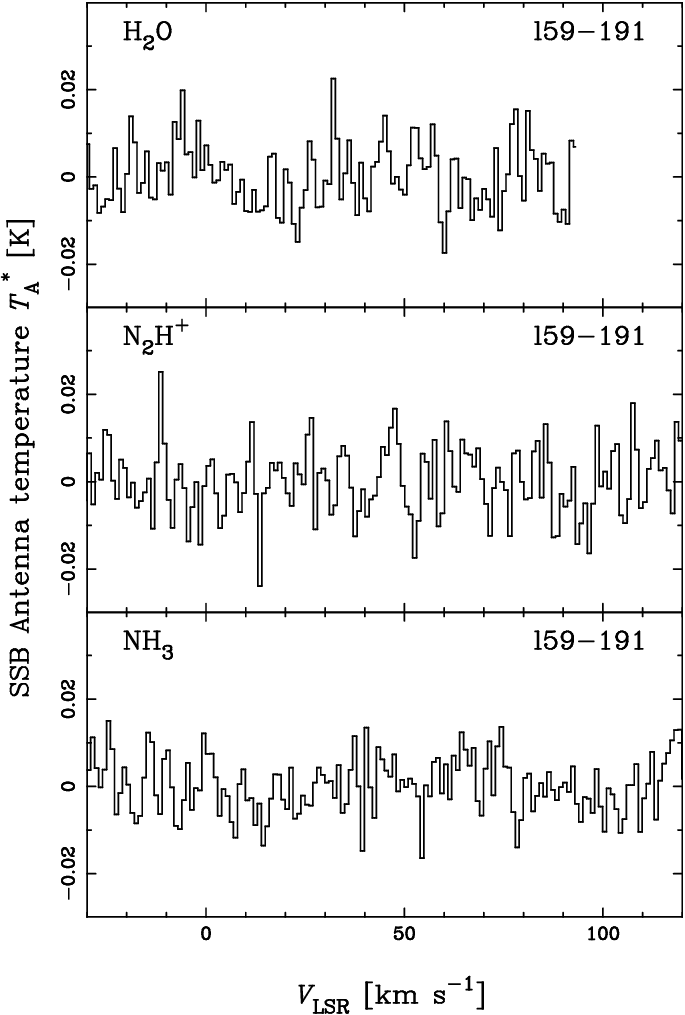}
\caption{}   
\end{center}
\end{figure}

\begin{figure}
\begin{center}
\includegraphics[scale=0.62]{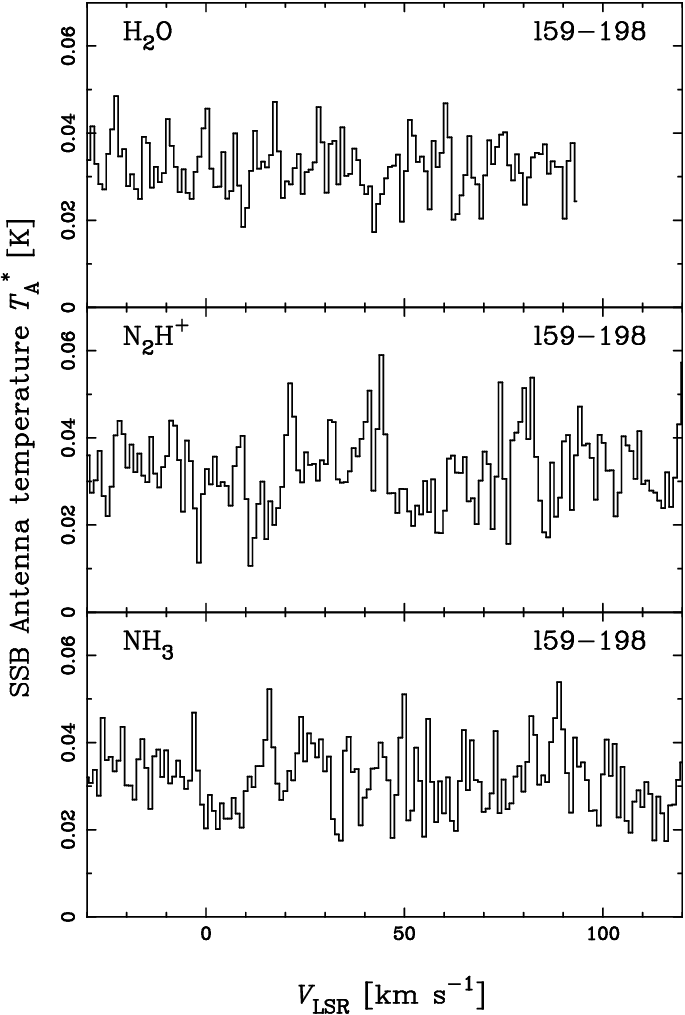}
\caption{}   
\end{center}
\end{figure}

\begin{figure}
\begin{center}
\includegraphics[scale=0.62]{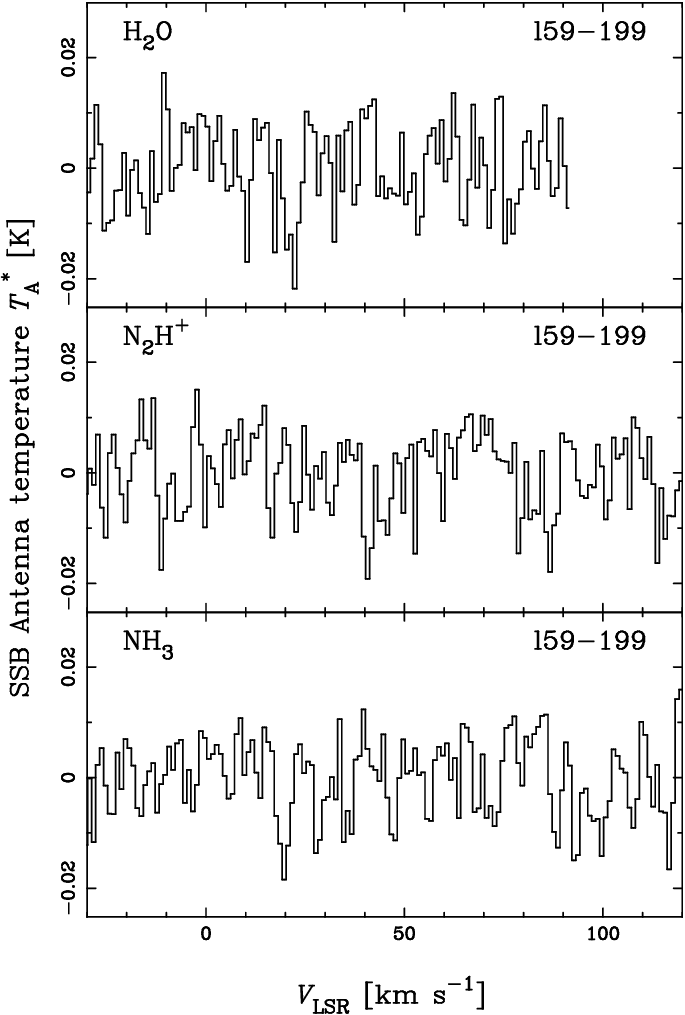}
\caption{}   
\end{center}
\end{figure}

\clearpage

\begin{figure}
\begin{center}
\includegraphics[scale=0.62]{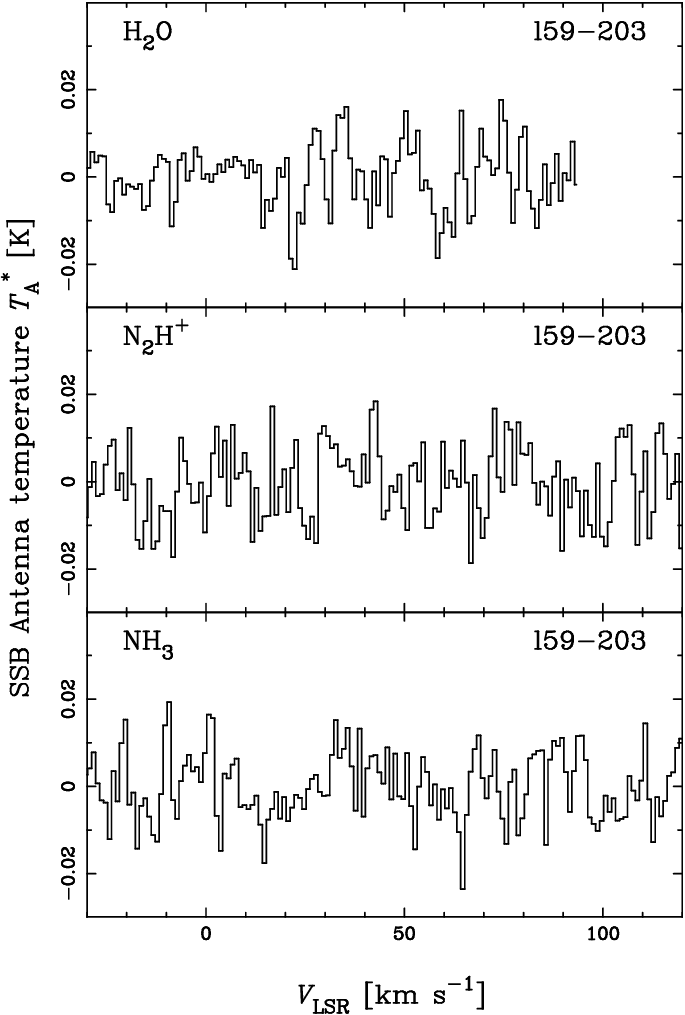}
\caption{}   
\end{center}
\end{figure}


\begin{figure}
\begin{center}
\includegraphics[scale=0.62]{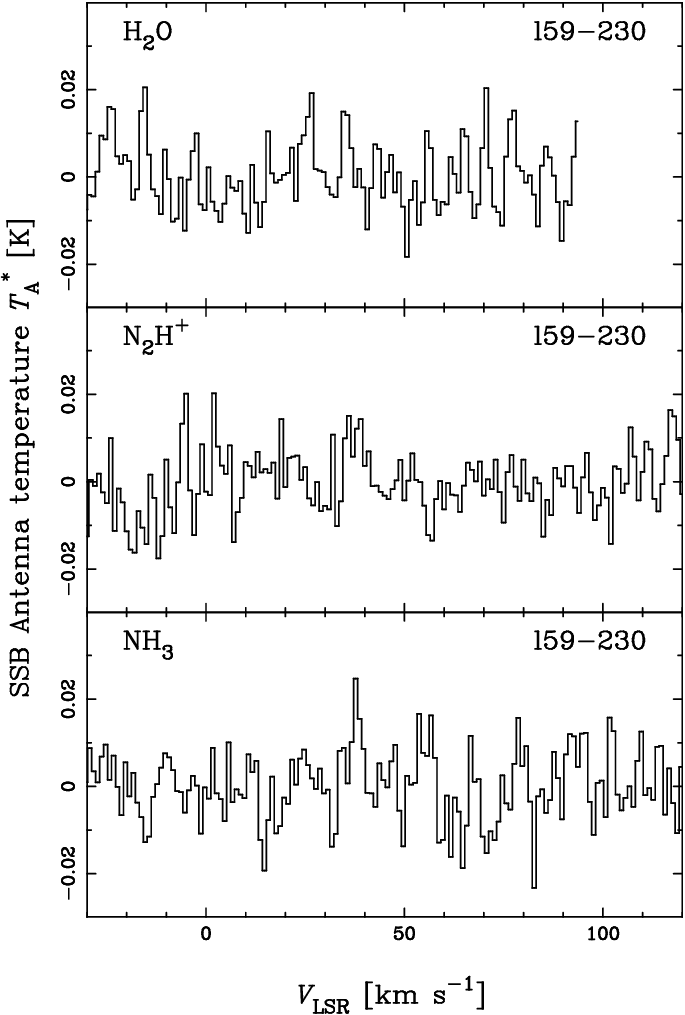}
\caption{}   
\end{center}
\end{figure}

\begin{figure}
\begin{center}
\includegraphics[scale=0.62]{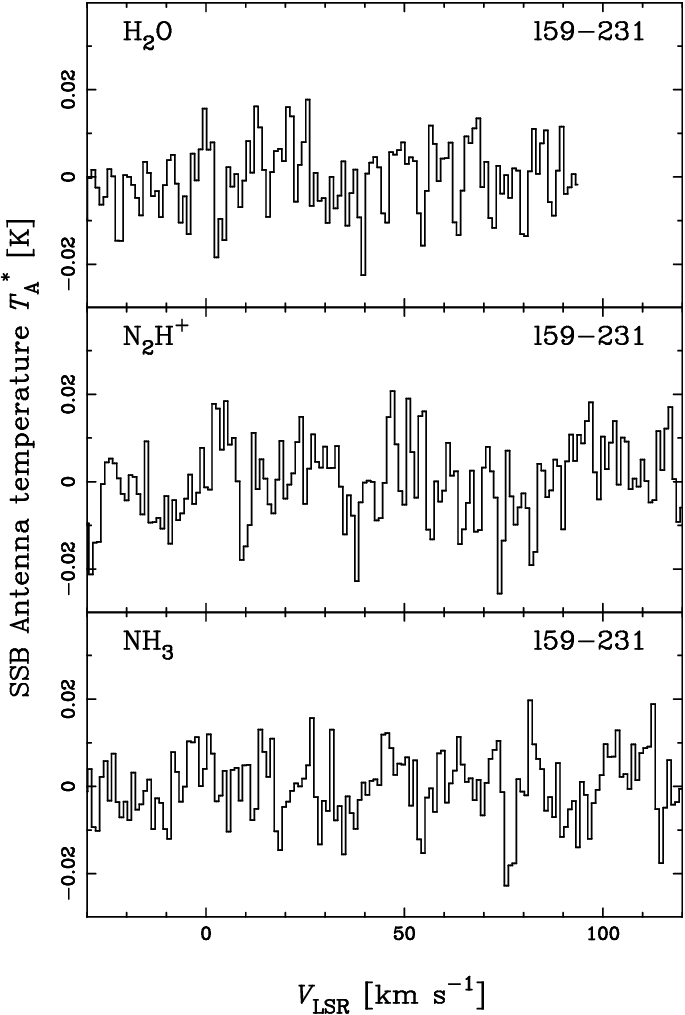}
\caption{}   
\end{center}
\end{figure}

\begin{figure}
\begin{center}
\includegraphics[scale=0.62]{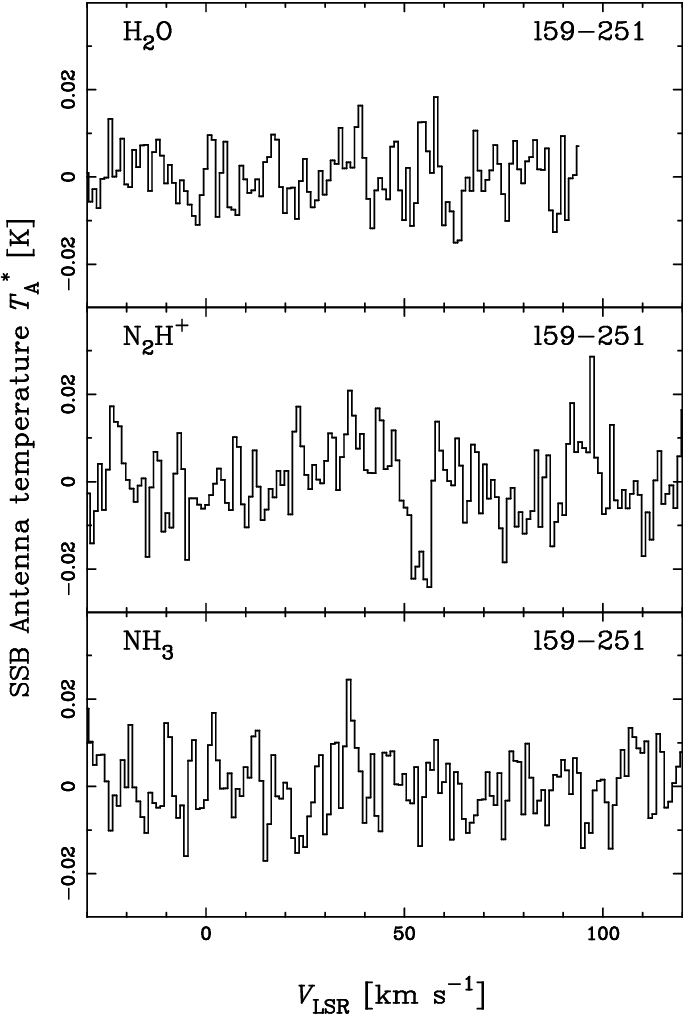}
\caption{}   
\end{center}
\end{figure}

\clearpage

\begin{figure}
\begin{center}
\includegraphics[scale=0.62]{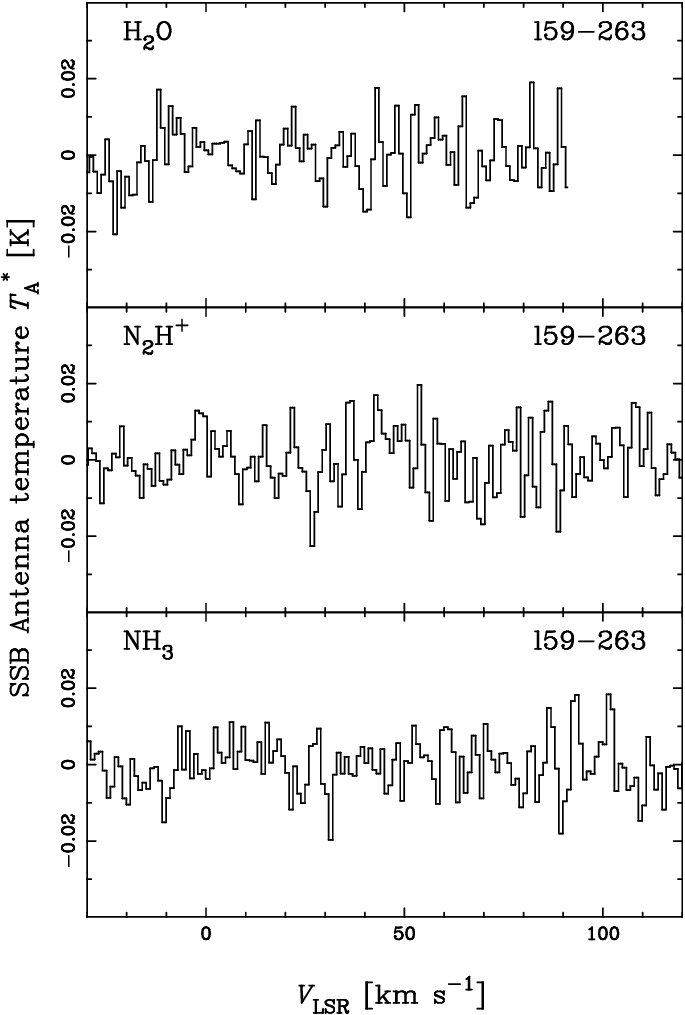}
\caption{}   
\end{center}
\end{figure}

\begin{figure}
\begin{center}
\includegraphics[scale=0.62]{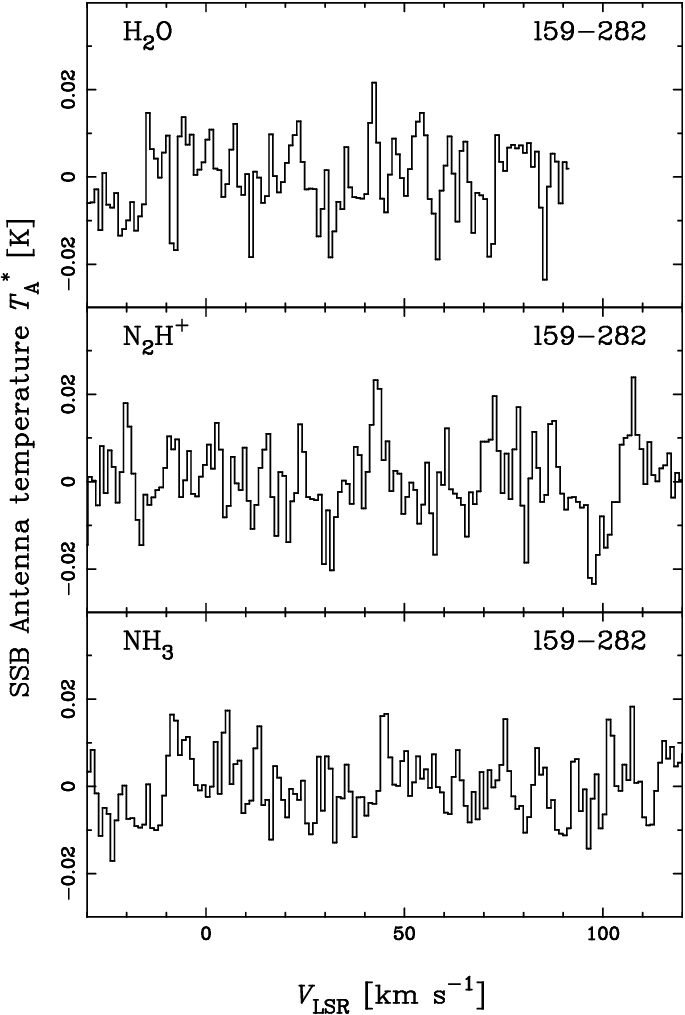}
\caption{}   
\end{center}
\end{figure}

\begin{figure}
\begin{center}
\includegraphics[scale=0.62]{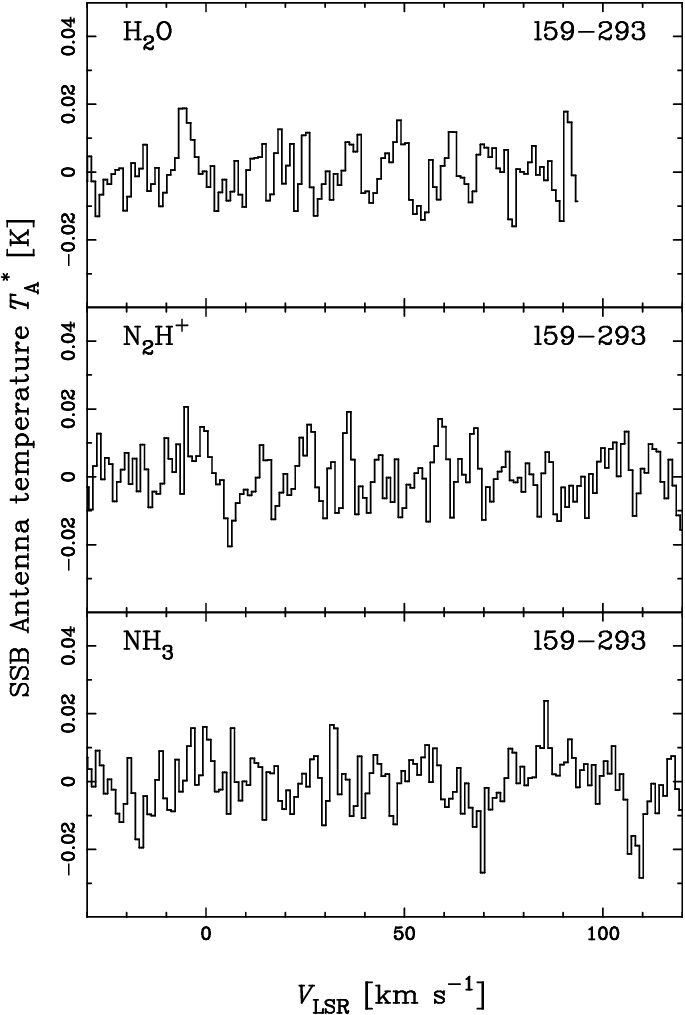}
\caption{}   
\end{center}
\end{figure}

\begin{figure}
\begin{center}
\includegraphics[scale=0.62]{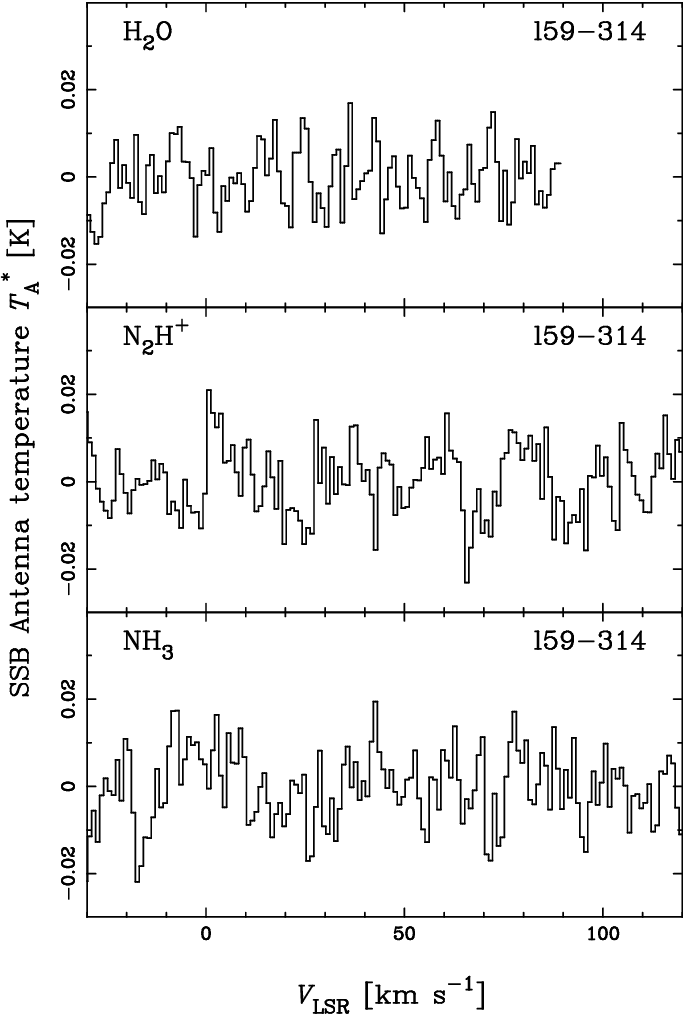}
\caption{}   
\end{center}
\end{figure}

\clearpage

\begin{figure}
\begin{center}
\includegraphics[scale=0.62]{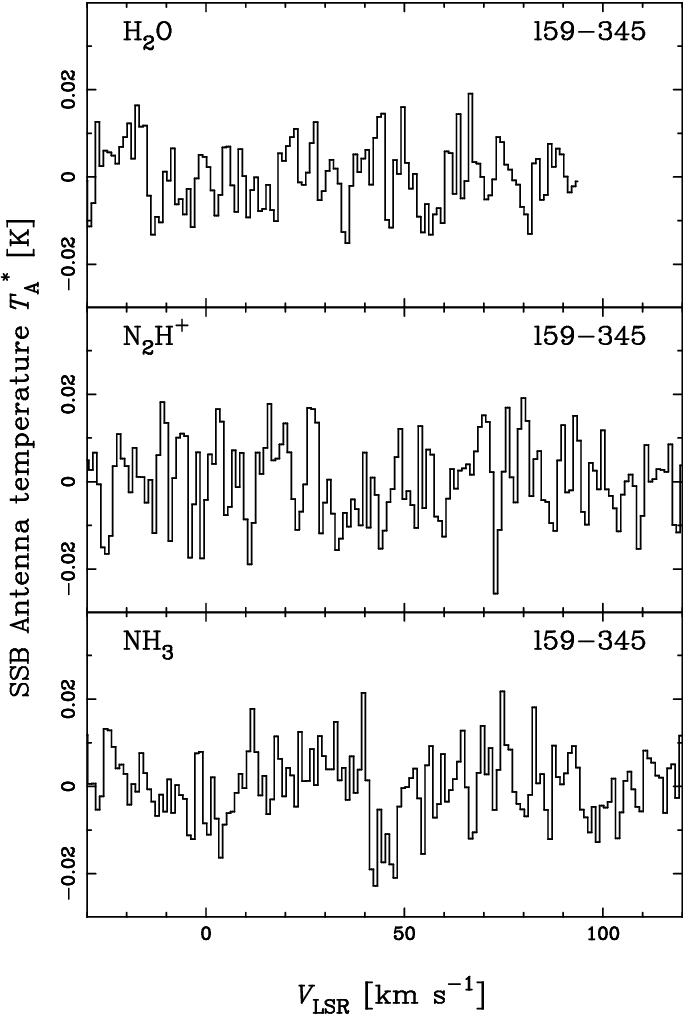}
\caption{}   
\end{center}
\end{figure}

\begin{figure}
\begin{center}
\includegraphics[scale=0.62]{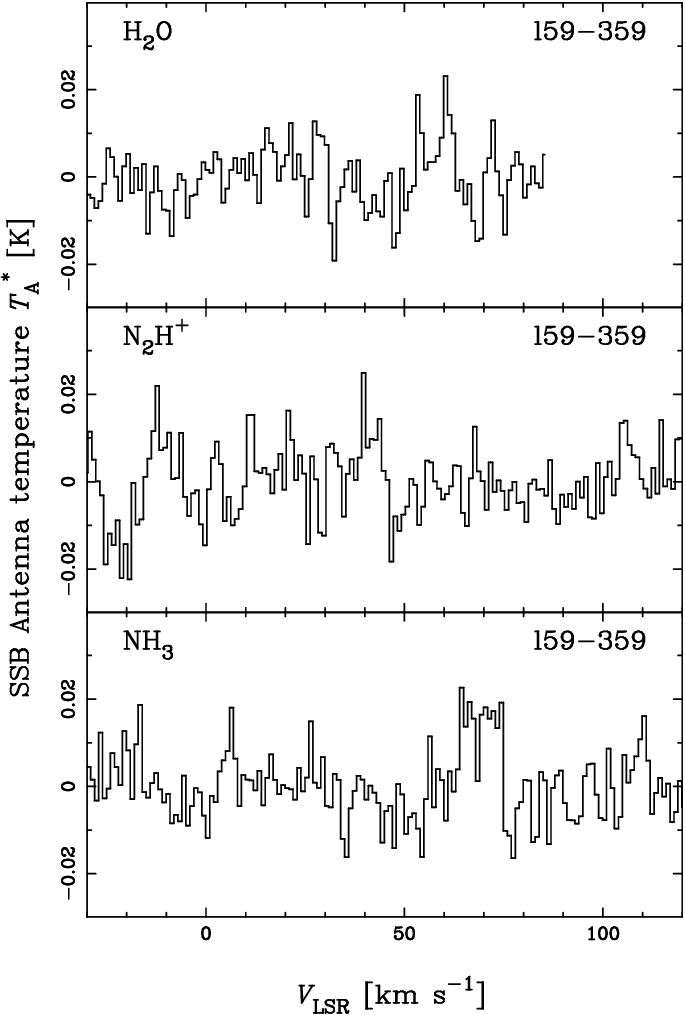}
\caption{}   
\end{center}
\end{figure}

\begin{figure}
\begin{center}
\includegraphics[scale=0.62]{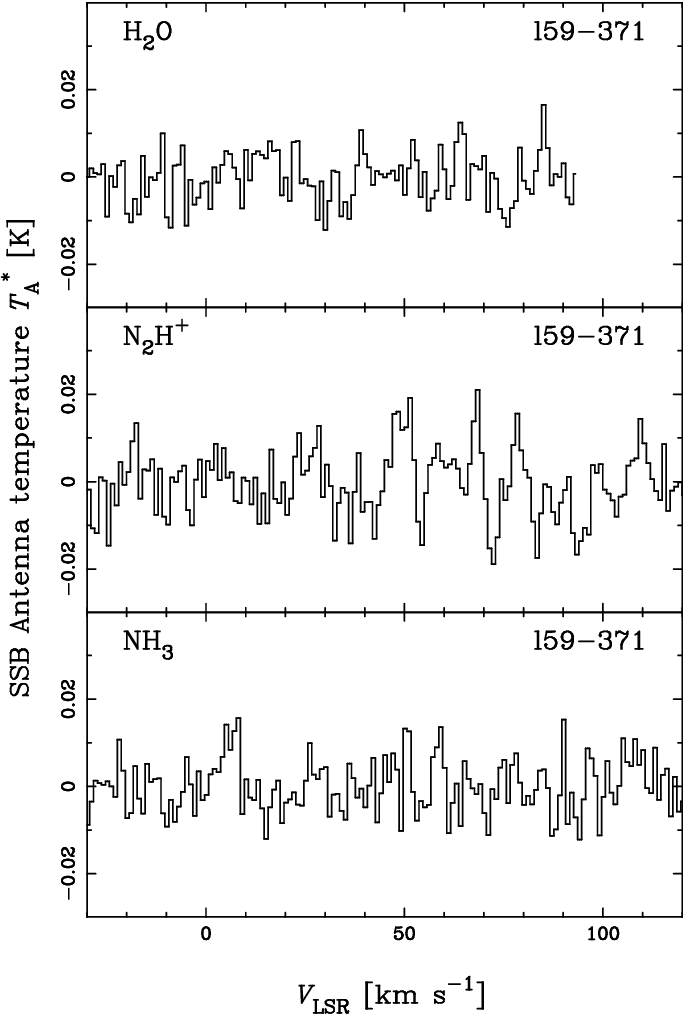}
\caption{}   
\end{center}
\end{figure}

\begin{figure}
\begin{center}
\includegraphics[scale=0.62]{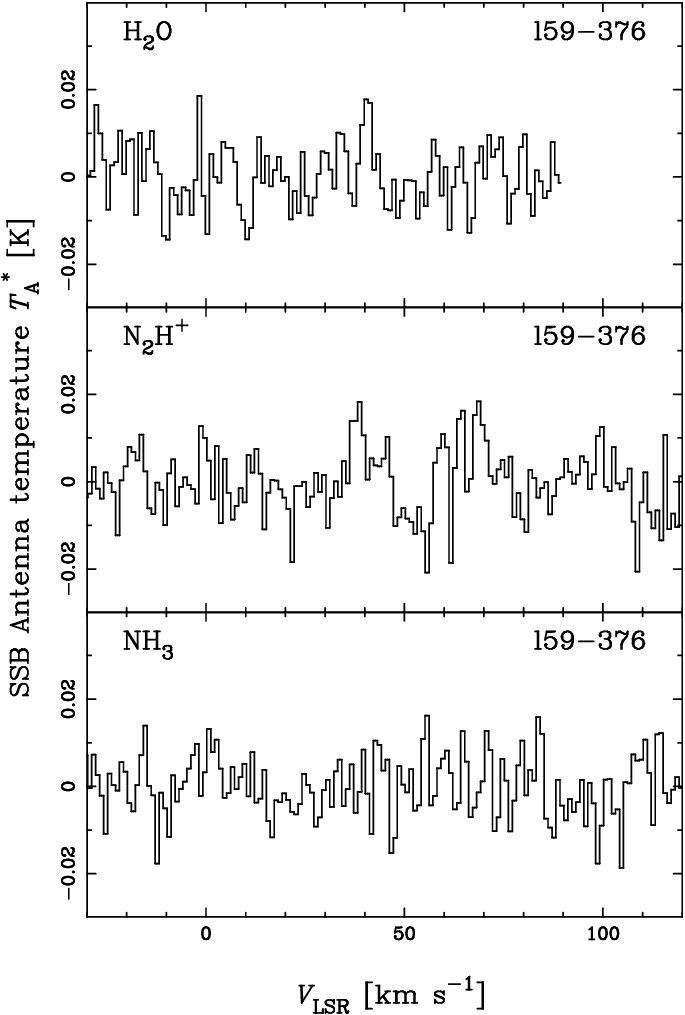}
\caption{}   
\end{center}
\end{figure}

\clearpage

\begin{figure}
\begin{center}
\includegraphics[scale=0.62]{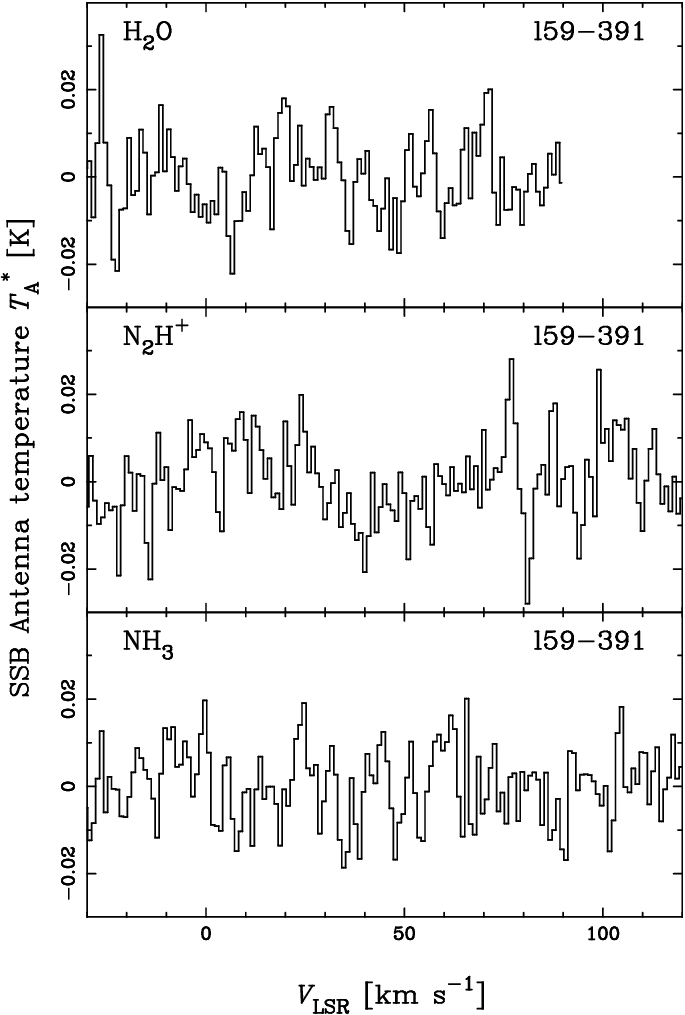}
\caption{}   
\end{center}
\end{figure}

\begin{figure}
\begin{center}
\includegraphics[scale=0.62]{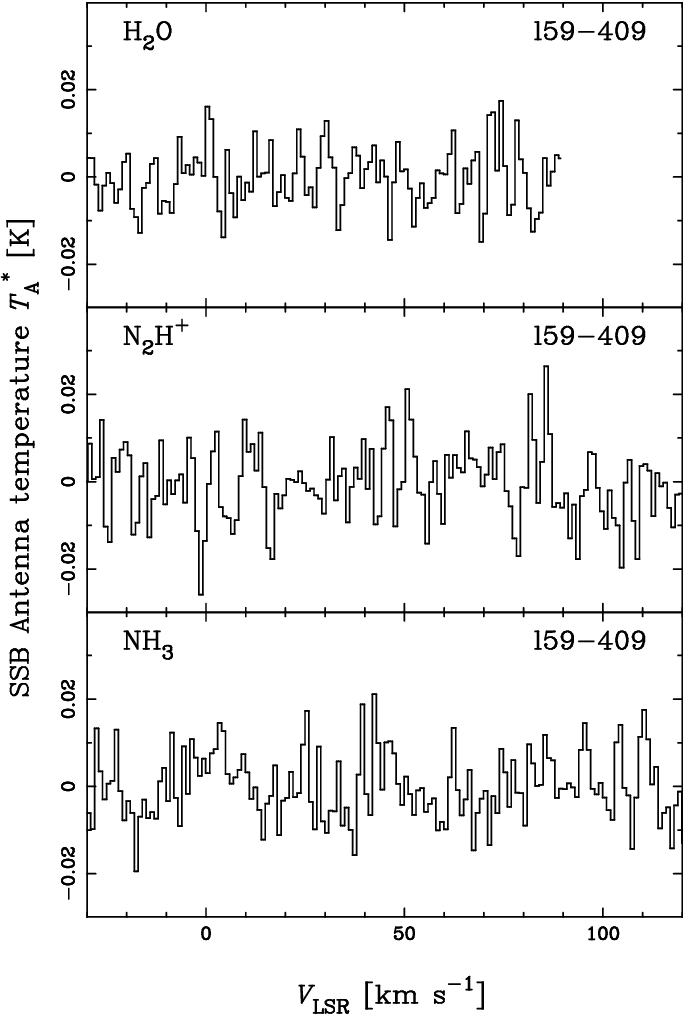}
\caption{}   
\end{center}
\end{figure}

\begin{figure}
\begin{center}
\includegraphics[scale=0.62]{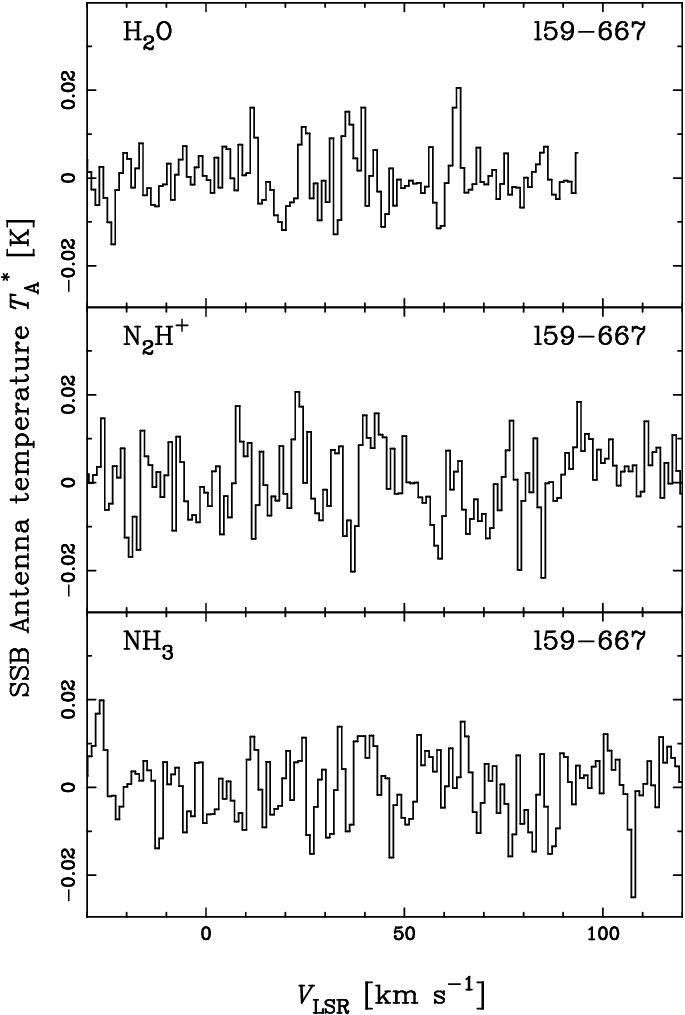}
\caption{}   
\end{center}
\end{figure}


\begin{figure}
\begin{center}
\includegraphics[scale=0.62]{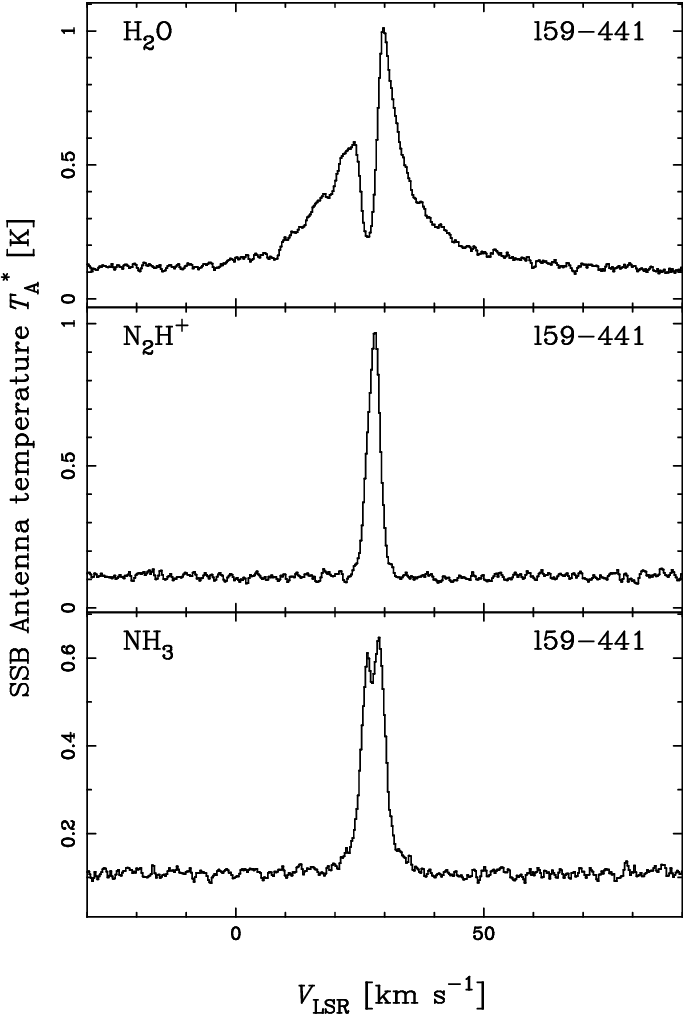} 
\caption{Original channel width (0.27 km/s).}   
\end{center}
\end{figure}

 
\clearpage


\begin{figure}
\begin{center}
\includegraphics[scale=0.62]{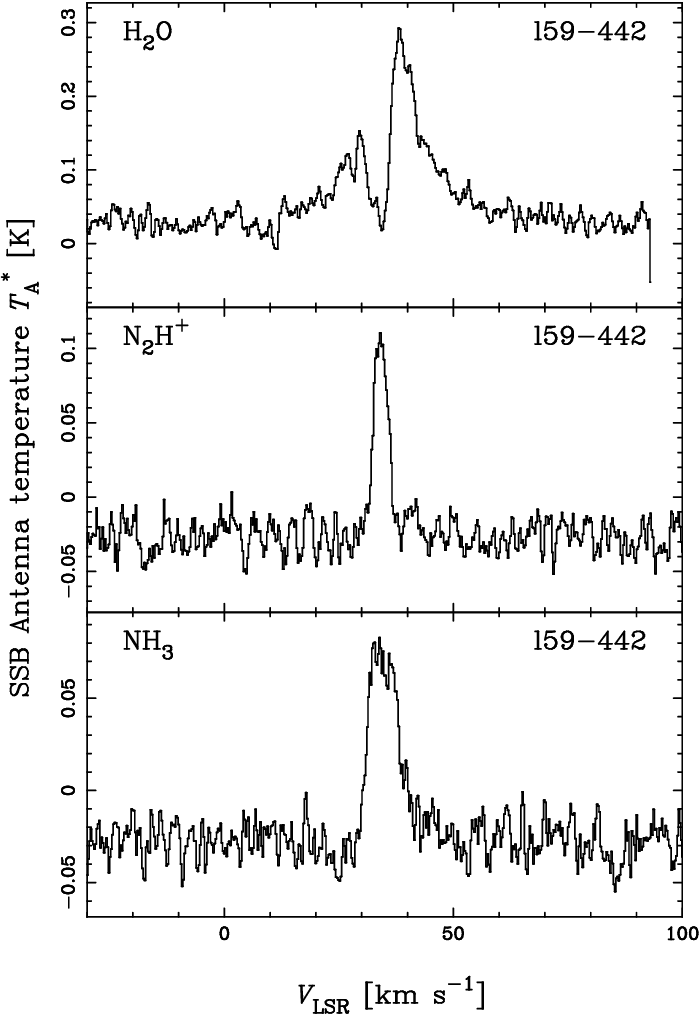} 
\caption{Original channel width (0.27 km/s).}   
\end{center}
\end{figure}

 

\begin{figure}
\begin{center}
\includegraphics[scale=0.62]{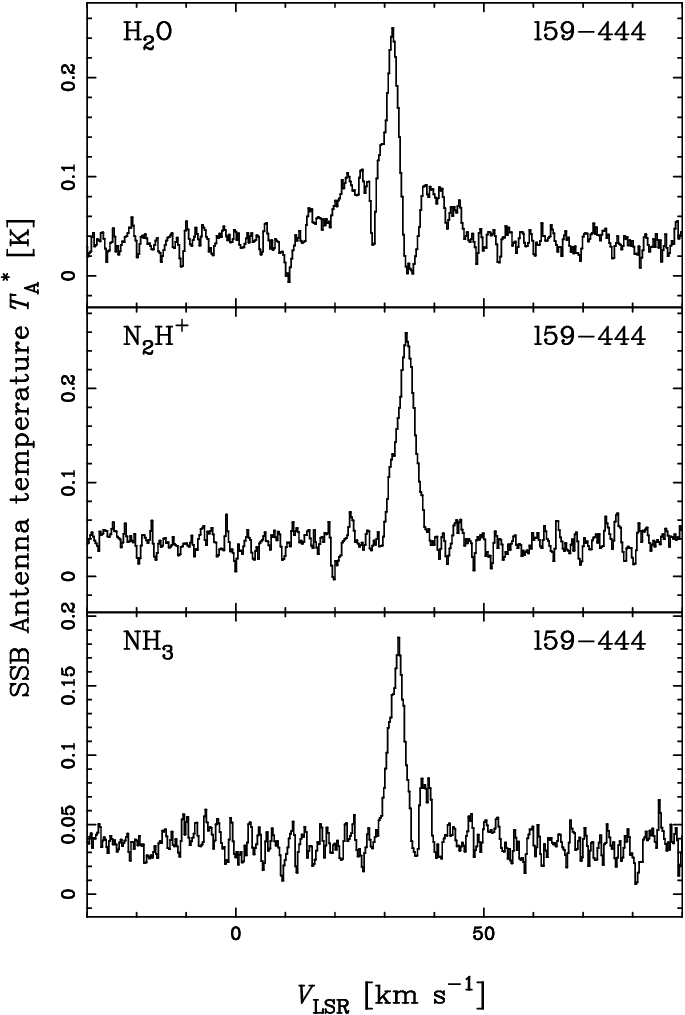} 
\caption{Original channel width (0.27 km/s).}   
\end{center}
\end{figure}



\begin{figure}
\begin{center}
\includegraphics[scale=0.62]{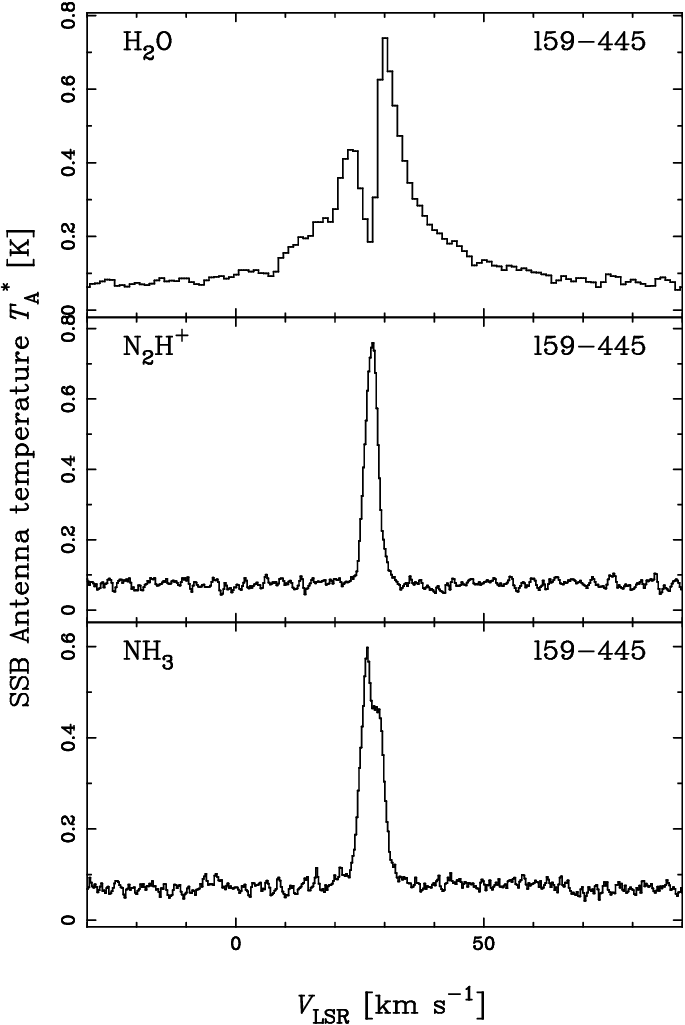} 
\caption{Original channel width (0.27 km/s).}   
\end{center}
\end{figure}




\end{document}